\documentclass[twocolumn,twocolappendix]{aastex631}

\usepackage{soul}
\usepackage{cancel}

\submitjournal{ApJ}

\shorttitle{Uniform Analysis of GPI Debris Disks I}
\shortauthors{Crotts et al.}

\begin{document}

\title{A Uniform Analysis of Debris Disks with the Gemini Planet Imager I: An Empirical Search for Perturbations from Planetary Companions in Polarized Light Images}

\correspondingauthor{Katie A. Crotts}
\email{ktcrotts@uvic.ca}

\author[0000-0003-4909-256X]{Katie A. Crotts}
\affiliation{Physics \& Astronomy Department, University of Victoria, 3800 Finnerty Rd. Victoria, BC, V8P 5C2}

\author[0000-0003-3017-9577]{Brenda C. Matthews}
\affiliation{Herzberg Astronomy and Astrophysics, National Research Council of Canada, 5071 West Saanich Rd., Victoria, BC V9E 2E7, Canada}

\author[0000-0002-5092-6464]{Gaspard Duch\^{e}ne}
\affiliation{Astronomy Department, University of California, Berkeley, CA 94720, USA}
\affiliation{Universit\'{e} Grenoble Alpes/CNRS, Institut de Plan\'{e}tologie et d'Astrophysique de Grenoble, 38000 Grenoble, France}

\author[0000-0002-0792-3719]{Thomas M. Esposito}
\affiliation{Astronomy Department, University of California, Berkeley, CA 94720, USA}
\affiliation{SETI Institute, Carl Sagan Center, 189 Bernardo Ave., Mountain View, CA 94043, USA}

\author[0000-0001-9290-7846]{Ruobing Dong}
\affiliation{Physics \& Astronomy Department, University of Victoria, 3800 Finnerty Rd. Victoria, BC, V8P 5C2}

\author[0000-0001-9994-2142]{Justin Hom}
\affiliation{Steward Observatory and the Department of Astronomy, The University of Arizona, 933 N Cherry Ave, Tucson, 85719, AZ, USA}

\author[0000-0001-7130-7681]{Rebecca Oppenheimer}
\affiliation{American Museum of Natural History, Department of Astrophysics, Central Park West at 79th Street, New York, NY 10024, USA}

\author[0000-0002-7670-670X]{Malena Rice}
\affiliation{Department of Astronomy, Yale University, New Haven, CT 06511, USA}

\author[0000-0002-9977-8255]{Schuyler G. Wolff}
\affiliation{Steward Observatory and the Department of Astronomy, The University of Arizona, 933 N Cherry Ave, Tucson, 85719, AZ, USA}

\author[0000-0002-8382-0447]{Christine H. Chen}
\affiliation{Space Telescope Science Institute (STScI), 3700 San Martin Drive, Baltimore, MD 21218, USA}

\author[0000-0001-5173-2947]{Clarissa R. Do \'{O}}
\affiliation{Center for Astrophysics and Space Sciences, University of California, San Diego, La Jolla, CA 92093, USA}

\author[0000-0002-6221-5360]{Paul Kalas}
\affiliation{Astronomy Department, University of California, Berkeley, CA 94720, USA}
\affiliation{SETI Institute, Carl Sagan Center, 189 Bernardo Ave., Mountain View, CA 94043, USA}
\affiliation{Institute of Astrophysics, FORTH, GR-71110 Heraklion, Greece}

\author[0000-0002-8984-4319]{Briley L. Lewis}
\affiliation{Department of Physics and Astronomy, 430 Portola Plaza, University of California, Los Angeles, CA 90095, USA}

\author[0000-0001-6654-7859]{Alycia J. Weinberger}
\affiliation{Earth and Planets Laboratory, Carnegie Institution for Science, 5241 Broad Branch Rd NW, Washington, DC 20015, USA}

\author[0000-0003-1526-7587]{David J. Wilner}
\affiliation{Center for Astrophysics | Harvard \& Smithsonian, 60 Garden Street, Cambridge, MA 02138, USA}

\author[0000-0001-5172-7902]{Mark Ammons}
\affiliation{Lawrence Livermore National Laboratory, 7000 East Ave, Livermore, CA 94550, USA}

\author[0000-0001-6364-2834]{Pauline Arriaga}
\affiliation{Department of Physics and Astronomy, 430 Portola Plaza, University of California, Los Angeles, CA 90095, USA}

\author[0000-0002-4918-0247]{Robert J. De Rosa}
\affiliation{European Southern Observatory, Alonsode C\'{o}rdova 3107, Vitacura, Santiago, Chile}

\author[0000-0002-1783-8817]{John H. Debes}
\affiliation{Space Telescope Science Institute (STScI), 3700 San Martin Drive, Baltimore, MD 21218, USA}

\author[0000-0002-0176-8973]{Michael P. Fitzgerald}
\affiliation{Department of Physics and Astronomy, 430 Portola Plaza, University of California, Los Angeles, CA 90095, USA} 

\author[0000-0003-4636-6676]{Eileen C. Gonzales}
\altaffiliation{51 Pegasi b Fellow}
\affiliation{Department of Physics and Astronomy, San Francisco State University, 1600 Holloway Ave., San Francisco, CA 94132, USA}
\affiliation{Department of Astronomy and Carl Sagan Institute, Cornell University, 122 Sciences Drive, Ithaca, NY 14853, USA}

\author[0000-0003-4653-6161]{Dean C. Hines}
\affiliation{Space Telescope Science Institute (STScI), 3700 San Martin Drive, Baltimore, MD 21218, USA}

\author[0000-0001-8074-2562]{Sasha Hinkley}
\affiliation{University of Exeter, Astrophysics Group, Physics Building, Stocker Road, Exeter, EX4 4QL, UK}

\author[0000-0002-4803-6200]{A. Meredith Hughes}
\affiliation{Astronomy Department and Van Vleck Observatory, Wesleyan University, 96 Foss Hill Drive, Middletown, CT 06459, USA}

\author[0000-0002-9321-3202]{Ludmilla Kolokolova}
\affiliation{University of Maryland, College Park, MD 20742 ,USA}

\author[0000-0002-1228-9820]{Eve J. Lee}
\affiliation{Department of Physics and Trottier Space Institute, McGill University, 3600 rue University, H3A 2T8 Montreal QC, Canada}

\author[0000-0002-2019-4995]{Ronald A. L\'{o}pez}
\affiliation{Department of Physics and Astronomy, 430 Portola Plaza, University of California, Los Angeles, CA 90095, USA}

\author[0000-0003-1212-7538]{Bruce Macintosh}
\affiliation{Center for Adaptive Optics, University of California Santa Cruz, Santa Cruz, CA 95064, USA}

\author[0000-0002-9133-3091]{Johan Mazoyer}
\affiliation{LESIA, Observatoire de Paris, Universit\'{e} PSL, CNRS, Sorbonne Universit\'{e}, Universit\'{e} de Paris, 5 place Jules Janssen, F-92195 Meudon, France}

\author[0000-0003-3050-8203]{Stanimir Metchev}
\affiliation{Department of Physics and Astronomy, Institute for Earth and Space Exploration, The University of Western Ontario, 1151 Richmond St, London, Ontario, N6A 3K7, Canada}

\author[0000-0001-6205-9233]{Maxwell A. Millar-Blanchaer}
\affiliation{Department of Physics, University of California, Santa Barbara, CA, 93106, USA}

\author[0000-0001-6975-9056]{Eric L. Nielsen}
\affiliation{Department of Astronomy, New Mexico State University, PO Box 30001, MSC 4500, Las Cruces, NM 88003, USA}

\author[0000-0001-9004-803X]{Jenny Patience}
\affiliation{School of Earth and Space Exploration, Arizona State University, PO Box 871404, Tempe, AZ 85287, USA}

\author[0000-0002-3191-8151]{Marshall D. Perrin}
\affiliation{Space Telescope Science Institute (STScI), 3700 San Martin Drive, Baltimore, MD 21218, USA}

\author{Laurent Pueyo}
\affiliation{Space Telescope Science Institute (STScI), 3700 San Martin Drive, Baltimore, MD 21218, USA}

\author[0000-0002-9667-2244]{Fredrik T. Rantakyr\"{o}}
\affiliation{Gemini Observatory, Casilla 603, La Serena, Chile}

\author[0000-0003-1698-9696]{Bin B. Ren}
\affiliation{Universit\'{e} C\^{o}te d’Azur, Observatoire de la C\^{o}te d’Azur, CNRS, Laboratoire Lagrange, Bd de l’Observatoire, CS 34229, F-06304 Nice cedex 4, France}

\author[0000-0002-4511-5966]{Glenn Schneider}
\affiliation{Steward Observatory and the Department of Astronomy, The University of Arizona, 933 N Cherry Ave, Tucson, 85719, AZ, USA}

\author[0000-0003-2753-2819]{Remi Soummer}
\affiliation{Space Telescope Science Institute (STScI), 3700 San Martin Drive, Baltimore, MD 21218, USA}

\author{Christopher C. Stark}
\affiliation{NASA Goddard Space Flight Center, Greenbelt, MD 20771, USA}



\begin{abstract}

The Gemini Planet Imager (GPI) has excelled in imaging debris disks in the near-infrared. The GPI Exoplanet Survey (GPIES) imaged twenty-four debris disks in polarized $H$-band light, while other programs observed half of these disks in polarized $J$- and/or $K1$-bands. Using these data, we present a uniform analysis of the morphology of each disk to find asymmetries suggestive of perturbations, particularly those due to planet-disk interactions. The multi-wavelength surface brightness, the disk color and geometry permit identification of any asymmetries such as warps or disk offsets from the central star. We find that nineteen of the disks in this sample exhibit asymmetries in surface brightness, disk color, disk geometry, or a combination of the three, suggesting that for this sample, perturbations, as seen in scattered light, are common. The relationship between these perturbations and potential planets in the system are discussed. We also explore correlations among stellar temperatures, ages, disk properties, and observed perturbations. We find significant trends between the vertical aspect ratio and the stellar temperature, disk radial extent, and the dust grain size distribution power-law, $q$. We also confirm a trend between the disk color and stellar effective temperature, where the disk becomes increasingly red/neutral with increasing temperature. Such results have important implications on the evolution of debris disk systems around stars of various spectral types.

\end{abstract}


\section{Introduction} \label{sec:intro}
Similar to our Solar System, exoplanetary systems are comprised of planets as well as planetesimal belts of comets and asteroids, accurately named ``debris disks", though detected debris disks around other stars dwarf our own in size, mass and brightness. These are circumstellar disks of dust and gas formed by collisional evolution within planetesimal belts, which allows us to observe these disks in scattered light, from the optical to near-infrared (NIR), as well as in thermal emission, from the mid-infrared to millimeter (mm) wavelengths \citep{Hughes18, Matthews14, Wyatt08}. In order to sustain collisional evolution and replenish dust in the system, the planetesimals must be stirred, either by planetary companions, Pluto-sized planetesimals within the disk itself, or by other gravitational perturbations \citep{Matthews14}. The substructure of the disk therefore constrains the location and mass of planets, including those comparable to Neptune and Saturn mass on long period orbits which are undetectable via any other planet detection methods (e.g., radial velocity variations, transits, or direct imaging). 

In recent years, advances in direct imaging have enabled high-contrast observations that can resolve smaller and lower surface brightness disks, which are likely to be better analogues to our own Solar System \citep{Michel21}; these observations have revealed that debris disks host a wide variety of substructures and asymmetries, such as gaps, warps, and clumps \citep{Hughes18}. The simplest explanation for many of these features is dynamical interaction with planets, but in many cases, the purported planets are undetected. However, in several debris disk systems with known planets (e.g. $\beta$ Pic and HD 106906, \citealt{Lagrange09, Kalas15, Lagrange16}), these planets have been found to be directly linked with the known asymmetries in the disk \citep{Chauvin12,Nesvold17,Crotts21}. In both scenarios, the disk morphology can be used to help determine whether disk-planet interactions are taking place. Additionally, other mechanisms can leave imprints on debris disks as well. For example, $\beta$ Pic is thought to have experienced a recent giant impact, as a large clump of dust and gas has been observed on the West side of the disk \citep{Telesco05, Dent14}. 

In other words, the more that we study the properties and structures of debris disks, the more that we can start to understand how planets, along with other mechanisms (such as a recent giant impact), can affect the overall debris disk morphology. Multiple studies including n-body and dynamical simulations have attempted to show these effects. For example, \citet{LC16} simulate a disk with an eccentric, 10 M$_{\oplus}$ planet orbiting within the disk, and find that this alone can create many of the disk morphologies observed, depending on viewing orientation, such as the ``Needle" and the ``Moth". These morphologies consist of swept back or extended disk halos, as well as eccentric disks leading to surface brightness asymmetries. Other studies show that recent giant impacts can also create similar type of morphologies, where \citet{Jones23} were able to recreate the structure of several debris disks, such as the aforementioned needle- and moth-like morphologies. As with $\beta$ Pic, giant impacts can leave clumps of gas and small dust grains at the collision point, which may help to differentiate between a planet and a giant impact scenario. 

While dynamical simulations are often inspired by disk observations, we can use these results, along with results from other debris disk studies, to return to observations (both past and new) and compare derived disk structures, which in turn will help determine what mechanisms are shaping the disk. Due to the tailored nature of individual debris disk observations, analyses of observations are typically done on a single disk to disk basis, allowing for a variety of different methods which may lead to different results. Therefore, uniform analyses on a larger sample of debris disks can minimize inconsistent results by analyzing all disks using the same methods. This also allows for comparison between debris disks to better understand how debris disks evolve over time and around different spectral types, as well as study other differences/similarities such as the vertical and radial disk structures. 

GPI, previously located on the Gemini South telescope in Chile, provides the perfect opportunity to perform such a uniform analysis, as the extreme AO instrument has imaged multiple debris disks with excellent resolution. \citet{Esposito20} first introduced these disks as a whole sample, presenting both polarized and total intensity observations of 25 debris disks in the $H$ band, as part of GPIES, \citealt{Macintosh18,Macintosh14,Macintosh08}). The names of these disks, along with information on each system, can be found in Table \ref{tab:star_sum}. Additionally, roughly half of the disks observed were also observed through one of GPI's Large and Long Programs (PID GS-2018A-LP-6) in polarized and/or total intensity using the $J$ and $K1$ bands. This large sample of resolved debris disks allows for a uniform, multiwavelength analysis of debris disk morphologies, which may reveal and/or confirm structures that are consistent with either planet-disk interactions or another mechanism.

In this study, we take a step beyond the work of \citet{Esposito20} by using the multiwavelength GPI disk sample to perform a uniform, empirical analysis with the goal of fully characterizing the disk morphology in the NIR, and identifying disks that are potentially perturbed. We choose to perform solely an empirical analysis, as radiative-transfer modelling can be computationally expensive and often not ideal for fitting asymmetric disks. We also focus primarily on polarized intensity observations. Even though total intensity observations are valuable in their own right and in combination with polarized intensity, these observations are highly subjected to disk self-subtraction due to the PSF-subtraction process. Because PSF-subtraction is not required for polarized intensity, as starlight is inherently unpolarized, these observations better represent the true disk structure, which is an important part of this study. 

Through this analysis, we derive the disk geometry, surface brightness, and disk color for the disks with multiwavelength observations. As part of the disk geometry, we also fit for offsets of the disk along the major- and minor-axis to check whether or not the disk is eccentric or has an asymmetric geometry, such as from a warp. We additionally measure whether or not any surface brightness or disk color asymmetries are present. The methods for deriving these disk properties are laid out in Section \ref{sec:results}, while the results for each individual disk can be found in Appendix \ref{sec:summary}. We then use these derived disk properties to categorize each disk based on similarities in asymmetries and discuss possible sources of perturbation in Section \ref{sec:sum}, along with discussion of broader trends found between disk and stellar properties.  

\begin{table*}[ht!]
	\centering
	\caption{\label{tab:star_sum}Summary of system properties including distance, age, stellar effective temperature/mass and luminosity. Distance measurements are from \citet{Gaia20}, except for $\beta$ Pic which is taken from \citet{Nielson20}. T$_{eff}$, M$_{*}$ L$_{*}$ values are taken from \citet{Esposito20}, and are new measurements done for the GPIES campaign, along with two of the system ages, as described in \citet{Nielson19}. Age References: (1) \citet{Nielson16}, (2) \citet{Bell15}, (3) \citet{Nielson19}, (4) \citet{Zuckerman19}, (5) \citet{Pecaut16}.}
	\begin{tabular*}{\textwidth}{c @{\extracolsep{\fill}} ccccc}
	\\
	    \hline
	    \hline
		Name & distance (pc) & Age (Myr) & T$_{eff}$ (K) & M$_{*}$ (M$_{\odot}$) & L$_{*}$ (L$_{\odot}$) \\
		\hline
		AU Mic & $9.71 \pm 0.00$ & 23-29 (1) & 3500 & $0.64^{+0.03}_{-0.02}$ & $0.06 \pm 0.03$ \\
		$\beta$ Pic & $19.44 \pm 0.05$ & 23-29 (1) & 8200 & $1.73^{+0.00}_{-0.02}$ & $9.33 \pm 3.13$ \\
		CE Ant & $34.10 \pm 0.03$ & 7-13 (2) & 3420 & $0.31^{+0.06}_{-0.06}$ & $0.07 \pm 0.07$ \\
		HD 30447 & $80.31 \pm 0.14$ & 38-48 (2) & 6900 & $1.45^{+0.00}_{-0.01}$ & $3.51 \pm 0.72$ \\
		HD 32297 & $129.73 \pm 0.55$ & 15-45 (3) & 7700 & $1.69^{+0.02}_{-0.02}$ & $8.12 \pm 1.68$ \\
		HD 35841 & $103.08 \pm 0.14$ & 38-48 (2) & 6500 & $1.30^{+0.01}_{-0.01}$ & $2.35 \pm 0.54$ \\
		HD 61005 & $36.45 \pm 0.02$ & 45-55 (4) & 5600 & $0.98^{+0.02}_{-0.07}$ & $0.68 \pm 0.07$ \\
		HD 106906 & $102.38 \pm 0.19$ & 12-18 (5) & 6500 & $2.70^{+0.12}_{-0.11}$ & $5.89 \pm 1.15$ \\
		HD 110058 & $130.08 \pm 0.53$ & 12-18 (5) & 8000 & $1.70^{+0.03}_{-0.02}$ & $9.33 \pm 2.13$ \\
		HD 111161 & $109.37 \pm 0.25$ & 12-18 (5) & 7800 & $1.72^{+0.02}_{-0.03}$ & $9.33 \pm 1.17$ \\
		HD 111520 & $108.05 \pm 0.21$ & 12-18 (5) & 6500 & $1.26^{+0.09}_{-0.07}$ & $2.69 \pm 0.37$ \\
		HD 114082 & $95.06 \pm 0.20$ & 12-18 (5) & 7000 & $1.42^{+0.08}_{-0.11}$ & $4.74 \pm 0.56$ \\
		HD 115600 & $109.04 \pm 0.25$ & 12-18 (5) & 7000 & $1.54^{+0.02}_{-0.10}$ & $5.27 \pm 0.37$ \\
		HD 117214 & $107.35 \pm 0.25$ & 12-18 (5) & 6500 & $1.47^{+0.02}_{-0.01}$ & $5.01 \pm 0.90$ \\
		HD 129590 & $136.32 \pm 0.44$ & 14-18 (5) & 5910 & $1.40^{+0.02}_{-0.01}$ & $3.35 \pm 0.96$ \\
		HD 131835 & $129.74 \pm 0.47$ & 14-18 (5) & 8100 & $1.77^{+0.05}_{-0.04}$ & $10.41 \pm 2.21$ \\
		HD 145560 & $121.23 \pm 0.29$ & 14-18 (5) & 6500 & $1.29^{+0.14}_{-0.05}$ & $3.47 \pm 0.14$ \\
		HD 146897 & $132.19 \pm 0.41$ & 7-13 (5) & 6200 & $1.28^{+0.02}_{-0.01}$ & $3.40 \pm 0.66$ \\
		HD 156623 & $108.33 \pm 0.33$ & 14-18 (5) & 8350 & $1.90^{+0.04}_{-0.05}$ & $13.06 \pm 1.80$ \\
		HD 157587 & $99.87 \pm 0.23$ & 165-835 (3) & 6300 & $1.44^{+0.01}_{-0.01}$ & $2.69 \pm 0.23$ \\
		HD 191089 & $50.11 \pm 0.05$ & 23-29 (1) & 6400 & $1.35^{+0.01}_{-0.01}$ & $2.54 \pm 0.17$ \\
		HR 4796 A & $70.77 \pm 0.24$ & 7-13 (2) & 9600 & $2.23^{+0.04}_{-0.05}$ & $26.44 \pm 5.48$ \\
		HR 7012 & $28.79 \pm 0.13$ & 23-29 (1) & 7700 & $1.70^{+0.01}_{-0.02}$ & $8.13 \pm 1.67$ \\
		\hline
		\hline
	\end{tabular*}
\end{table*}

\section{Observations and Data Reduction} \label{sec:observations}
For this study, we have obtained GPI polarimetric observations in the $J$ ($\lambda_{c} = 1.25 \mu$m), $H$ ($\lambda_{c} = 1.65 \mu$m), and $K1$ bands ($\lambda_{c} = 2.05 \mu$m) for 24 disks total. All 24 disks were observed in the $H$ band as a part of the GPIES survey (PI: B. Macintosh), while 10 of the disks were also observed in the $J$ band and 11 were observed in the $K1$ band as a part of the Debris Disk Large and Long Program (PI: C. Chen). All observations were taken in polarimetric mode, with a field of view (FOV) of $2.8'' \times 2.8''$ and a pixel scale of $14.166 \pm 0.007$ mas per lenslet \citep{Rosa15}. A summary of the observations for each disk and each band can be found in Table \ref{tab:data_sum}. While the HD 143675 disk is included as a part of GPIES, because the disk is so radially small and close to the focal plane mask (FPM), we were unfortunately unable to determine the geometry and therefore do not include it in this study. We direct the interested reader to \citet{Hom20} for an analysis of both the polarized and total intensity observations which are better resolved.

For the $H$-band observations, we use the polarized intensity data presented in \citet{Esposito20}. As for the $J$- and $K1$-band observations, we uniformly reduce these data using the same recipe as the $H$-band data. For a more detailed and technical description of this reduction process, see Section 4 in \citealt{Esposito20}. In short, using the GPI data reduction pipeline (\citealt{Perrin14}, and references therein), we first start with the raw data for each disk, which are reduced into 3D Stokes data cubes. The first two dimensions of these cubes contain the spatial information (x,y), and the third dimension contains the Stokes parameters [$I$,$Q$,$U$,$V$]. Through this process, the raw data are dark subtracted and \textit{destriped} with a Fourier filter \citep{Ingraham14} and bad pixel corrected. A cross-correlation algorithm is also used to match the detector with the expected positions of each lenslet's two PSFs \citep{Draper14} before they are assembled into the 3D cubes. The data are flat-fielded and the position of the central star is measured using fiducial satellite spots \citep{Wang14} which are later used for photometric calibration. To ensure good reductions, we remove any bad frames which appear to be distorted or where the star is not placed correctly behind the coronagraph. 

Once the Stokes cubes are created from the raw data, the cubes are further reduced and combined into a single radial Stokes cube containing $Q_{\phi}$ and $U_{\phi}$. Through this process, the cubes are accumulated, cleaned using a double differencing procedure developed specifically for GPI ADI data \citep{Perrin15}, and then smoothed using a Gaussian kernel with a FWHM of 1 pixel. The mean stellar polarization (which can include both stellar and instrumental polarization) is then subtracted by measuring the flux in an annulus near the FPM edge \citep{MB16b}. This step is particularly important for cleaning the image and better recovering the disk's surface brightness, as success on subtracting the instrumental polarization from the image depends on the user's input of the annulus location and size. In our case, we find that an annulus of 2-5 pixel width and placed typically at a mean radius of 7-11 pixels from the star (although this is somewhat varied per disk) gave the best results, i.e. most effectively removed the instrumental polarization. The cubes are then rotated so that North is up, and are combined into a single radial Stokes cube. Finally, using the satellite spot measurements, the radial Stokes cube is converted from units of ADI coadd$^{-1}$ to real units of mJy arcesc$^{-2}$. 

\begin{figure*}
\centering
	\caption{\label{Fig:hband} Polarized intensity observations of the 12 out of 24 debris disks detected by GPI in the $J$, $H$ and $K1$ bands. The circles represent the size of the FPM for each band ($\sim$0.09$''$, $\sim$0.12$''$ and $\sim$0.15$''$ respectively), and the crosses represent the location of the star. Similar to Figure 5 in \citet{Esposito20}, the data are scaled in units of mJy arcsec$^{-2}$ by the numbers in the lower left corner in order to have similar brightness. Additionally, the disk surface brightness is linear from 0 to 1, and log scale from 1 to 20 mJy arcsec$^{-2}$. The arrows in the lower right corner represent the North and East directions.}
	\includegraphics[width=0.497\textwidth]{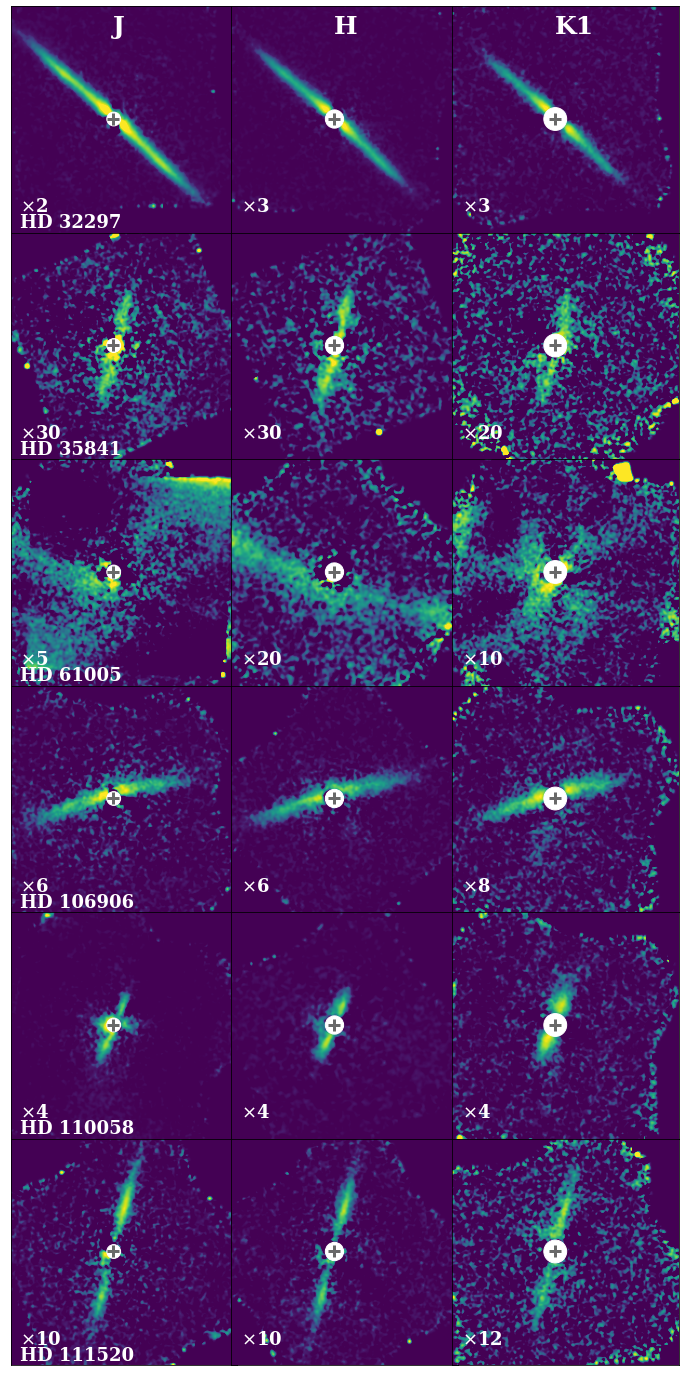}
        \includegraphics[width=0.497\textwidth]{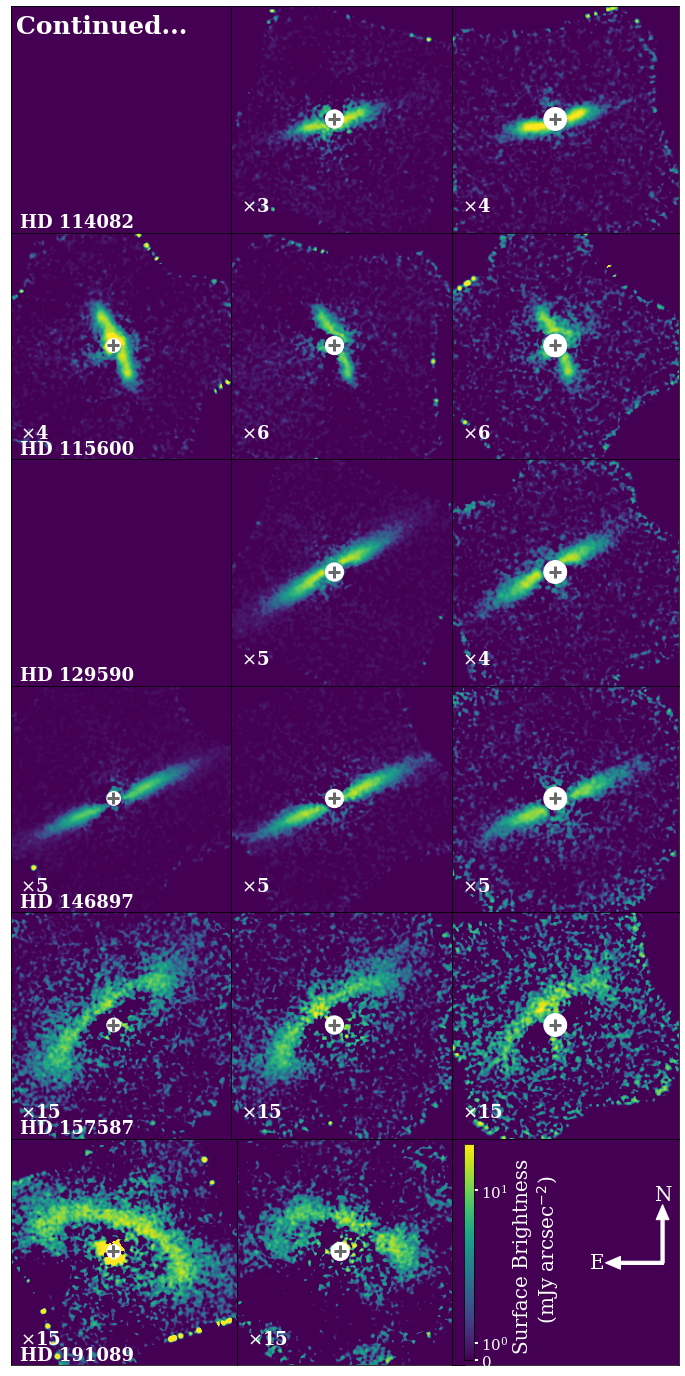}
\end{figure*}

\begin{figure*}[ht!]
\centering
	\caption{\label{Fig:new_jband} Reduced polarized intensity observations of the remaining 11 debris disks resolved by GPI in the $H$ band. The circles represent the size of the FPM in $H$ band ($\sim$0.12$''$), and the crosses represent the location of the star. The data are scaled similarly as Figure \ref{Fig:hband}, where the disk surface brightness is linear from 0 to 1, and log scale from 1 to 20 mJy arcsec$^{-2}$. The arrows in the lower right corner represent the North and East directions.}
	\includegraphics[width=\textwidth]{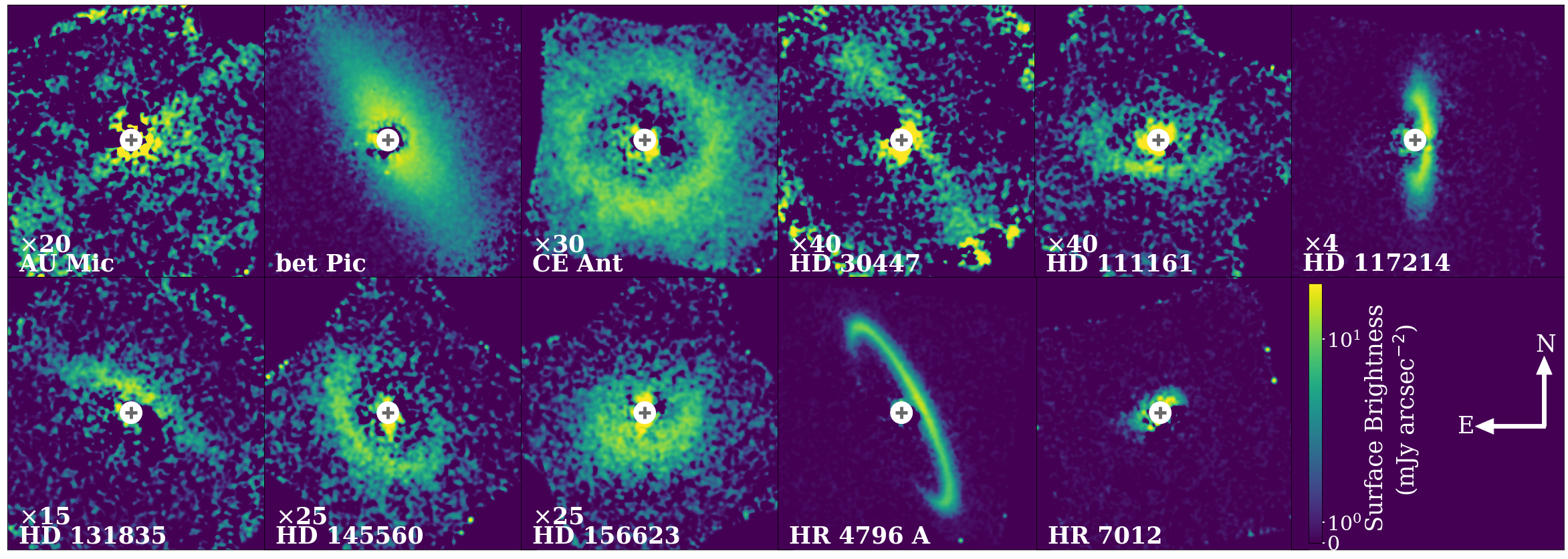}
\end{figure*}

Similar to the $H$ band, we include an extra step for our final $J$ and $K1$ reductions to remove a quadrupole-like noise pattern that often remains in GPI polarized intensity reductions. This is done using the same method as in \citet{Esposito20}, by measuring the contribution and orientation of this quadrupole pattern in $U_{\phi}$ using the function $B = B_{0} I_{r} \text{sin}2(\theta + \theta_{0})$, where $I_{r}$ is the azimuthally averaged total intensity as a function of radius. As described in \citet{Esposito20}, the function is fit by varying the scaling factor $B_{0}$ and offset angle $\theta_{0}$ to minimize the sum of the squared residuals. The best fitting function is then subtracted from the $U_{\phi}$ image, rotated by 45$^{\circ}$, and then subtracted from the $Q_{\phi}$ image. The $H$-band observations can be found in Figure \ref{Fig:hband}, while our final $J$- and $K1$-band reductions can be found in Figure \ref{Fig:new_jband}.

Using the $U_{\phi}$ data, we also create noise maps for each disk. This is under the assumption that $U_{\phi}$ contains no disk signal, as expected for an optically thin debris disk causing single scattering, however, this has not been found to be entirely the case for the $H$-band data (see Appendix A in \citealt{Esposito20}). To create noise maps, we simply calculate the standard deviation at each radius in 1-pixel wide stellocentric annuli of the $U_{\phi}$ image. These noise maps are used to estimate the uncertainty in the surface brightness for each disk, and can also be divided from $Q_{\phi}$ to create signal-to-noise (S/N) maps. Our S/N maps can be seen in Figure \ref{Fig:SN_H} located in the Appendix.

\begin{table*}
	\centering
	\caption{\label{tab:data_sum}Summary of observations. Here, t$_{exp}$ = the integration time for each frame in seconds, t$_{int}$ = the total integration time in seconds, and $\Delta$PA = the total parallactic angle rotation in degrees.}
	\begin{tabular*}{\textwidth}{c @{\extracolsep{\fill}} ccccc}
	\\
	    \hline
	    \hline
		Name & Band & Date & t$_{exp}$ (s) & t$_{int}$ (s) & $\Delta$PA ($^{\circ}$) \\
		\hline
		AU Mic & H & 140515 & 59.65 & 2624.44 & 166.9 \\
		$\beta$ Pic & H & 131212 & 5.82 & 3258.73 & 91.5 \\
		CE Ant & H & 180405 & 119.29 & 3817.37 & 12.8 \\
		HD 30447 & H & 160922 & 59.65 & 3101.61 & 125.8 \\
		HD 32297 & H & 141218 & 59.65 & 2147.27 & 19.1 \\
		... & J & 151206 & 88.74 & 3549.6 & 24.2 \\
		... & K1 & 161118 & 88.74 & 2839.68 & 19.8 \\
		HD 35841 & H & 160318 & 88.74 & 2484.78 & 3.7 \\
		... & J & 180127 & 59.65 & 5726.40 & 19.4 \\ 
		... & K1 & 171228 & 88.74 & 4703.22 & 93.9 \\
		HD 61005 & H & 140324 & 59.65 & 2087.62 & 140.1 \\
		... & J & 151201 & 59.65 & 4891.30 & 164.5 \\
		... & K1 & 180126 & 88.74 & 4969.44 & 150.8 \\
		HD 106906 & H & 150701 & 59.65 & 2564.79 & 20.3 \\
		... & J & 160326 & 59.65 & 3221.10 & 35.2 \\
		... & K1 & 160328 & 88.74 & 3549.60 & 36.5 \\
		HD 110058 & H & 160319 & 59.65 & 2147.27 & 25.2 \\
		... & J & 180126 & 59.65 & 4712.35 & 54.21 \\
		... & K1 & 170420 & 88.74 & 2484.72 & 31.7 \\
		HD 111161 & H & 180310 & 59.65 & 4533.13 & 38.0 \\
		HD 111520 & H & 160318 & 88.74 & 2839.75 & 28.3 \\
		... & J & 160326 & 59.65 & 3519.35 & 39.1 \\
		... & K1 & 160328 & 88.74 & 3194.64 & 35.8 \\
		HD 114082 & H & 170807 & 59.65 & 2087.62 & 12.3 \\
		... & K1 & 170420 & 88.74 & 2839.68 & 23.7 \\
		HD 115600 & H & 150703 & 59.65 & 2624.44 & 24.0 \\
		... & J & 180128 & 29.10 & 2357.10 & 43.4 \\
		... & K1 & 180127 & 88.74 & 4437.0 & 34.3 \\
		HD 117214 & H & 180311 & 59.65 & 1908.68 & 18.5 \\
		HD 129590 & H & 170809 & 59.65 & 2147.27 & 17.9 \\
		... & K1 & 170421 & 88.74 & 2395.98 & 44.3 \\ 
		HD 131835 & H & 150501 & 59.65 & 1908.68 & 74.2 \\
		HD 145560 & H & 180812 & 59.65 & 1670.10 & 17.6 \\
		HD 146897 & H & 160321 & 88.74 & 1774.84 & 28.9 \\
		... & J & 160327 & 59.65 & 4533.40 & 45.6 \\
		... & K1 & 180709 & 88.74 & 4170.78 & 97.6 \\
		HD 156623 & H & 190427 & 88.74 & 2129.81 & 28.2 \\
		HD 157587 & H & 150829 & 88.74 & 2484.78 & 49.9 \\
		... & J & 160326 & 88.74 & 2662.20 & 57.7 \\
		... & K1 & 160327 & 119.29 & 2027.93 & 32.6 \\
		HD 191089 & H & 150901 & 88.74 & 2484.78 & 101.3 \\
		... & J & 170701 & 59.65 & 1908.80 & 11.8 \\
		HR 4796 A & H & 131212 & 29.10 & 640.11 & 2.1 \\
		HR 7012 & H & 180921 & 4.36 & 1117.28 & 19.3 \\
		\hline
		\hline
	\end{tabular*}
\end{table*}

\section{Empirical Analysis \& Results} \label{sec:results}

\subsection{Disk Geometry} \label{sec:geom}
\begin{figure}
        \centering
	\caption{\label{Fig:offset_example} Example of how the FWHM and vertical/radial offset are measured for high inclined disks compared to lower inclined disks. While a Gaussian function is fit to vertical slices along the disk at multiple radial separations (represented by the dotted lines in the left image), for lower inclined disks, a Gaussian function is fit to radial slices (represented by the dotted lines in the right image).}
	\includegraphics[width=0.47\textwidth]{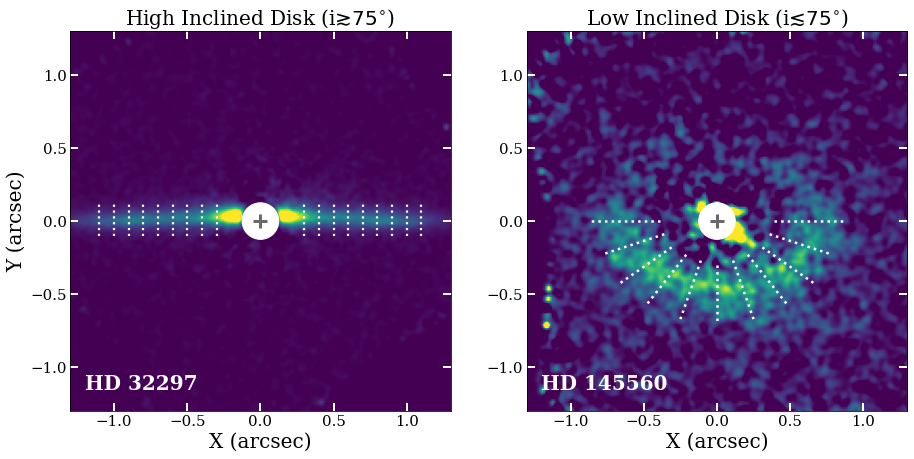}
\end{figure}

To understand the disk morphology as a whole, we first measure the geometry for each disk. For this process, we separate lower inclined disks ($i \lesssim 75^{\circ}$) from higher inclined disks ($i \gtrsim 75^{\circ}$), as a slightly different fitting process is required. The cut off of $\sim$75$^{\circ}$ is chosen because it is at this point that radial structure becomes significant, and the disks are therefore no longer fit well with the method used for higher inclined disks. For higher inclined disks, we fit a Gaussian profile to the surface brightness along vertical slices at multiple radial separations from the star, avoiding noisy regions close to the star. For lower inclined disks which show more radial structure, we instead fit a Gaussian profile to the surface brightness measured along radial slices to more accurately trace the disk geometry. This is done by rotating the image between, at minimum, -90$^{\circ}$ to +90$^{\circ}$ from the given $PA$, and taking vertical slices at each angle (see Figure \ref{Fig:offset_example} for a visual representation). Depending on how much of the disk is visible for the lower inclined disks in our sample, we rotate the image beyond -90 and +90 degrees to also trace the geometry of the back side of the disk. The FWHM and mean of the Gaussian are then extracted, giving us an estimation of the disk width, either vertically or radially depending on the disk inclination, along with either the vertical or radial offset of the disk peak surface brightness from the star. For the majority of our sample we use the $H$-band observations as they tend to have a higher S/N compared to the $J$ and $K1$ band observations, however, for the cases in which the disk is higher S/N in the $J$ or $K1$ bands (this includes HD 114082 and HD 191089), we opt to use these observations instead. 

Using the derived FWHM of the disk, we estimate the vertical or radial aspect ratio by comparing the measured FWHM to R$_{0}$, where R$_{0}$ is defined as the radius of the peak dust density based on scattered light observations, and is derived from modelling the dust density profile. The R$_{0}$ values used are taken from \citet{Esposito20}, which they compiled from their own work and from the literature. To measure the aspect ratio, we calculate the weighted average of the intrinsic disk FWHM. To obtain the intrinsic FWHM, the original measured FWHM from our Gaussian fitting procedure is corrected for the instrumental PSF and any smoothing applied to the image. This is done by subtracting the FWHM of the instrumental PSF and smoothing Gaussian kernels in quadrature from the measured FWHM. Once this is done, we then simply divide R$_{0}$ from the corrected weighted average FWHM. We note that these aspect ratios are significantly higher than those reported for several of the same higher-inclined disks analyzed in \citet{Olofsson22}, including AU Mic, HD 32297, HD 61005, HD 106906, HD 115600, HD 129590 and HR 4796. This discrepancy is mainly due to the difference in measuring the vertical FWHM, where we are empirically measuring the vertical FWHM from the data, compared to \citet{Olofsson22} who determines the vertical FWHM from disk models. By performing this measurement empirically, the vertical width becomes correlated with the disk inclination. Additionally, we are probing the contribution of the small grains in the disk halo, rather than just the planetesimal belt. We therefore do not consider these measurements as true aspect ratios, but use it mainly to compare the vertical or radial width of each disk as a function of inclination. 

\begin{figure*}
\centering
	\caption{\label{Fig:asp_rat} The aspect ratio for each disk, which is defined as the intrinsic FWHM divided by R$_{0}$ given in \citet{Esposito20}, as a function of disk inclination. The red square encapsulates the disks with anomalously high aspect ratios compared to other disks with similar inclinations.}
	\includegraphics[width=\textwidth]{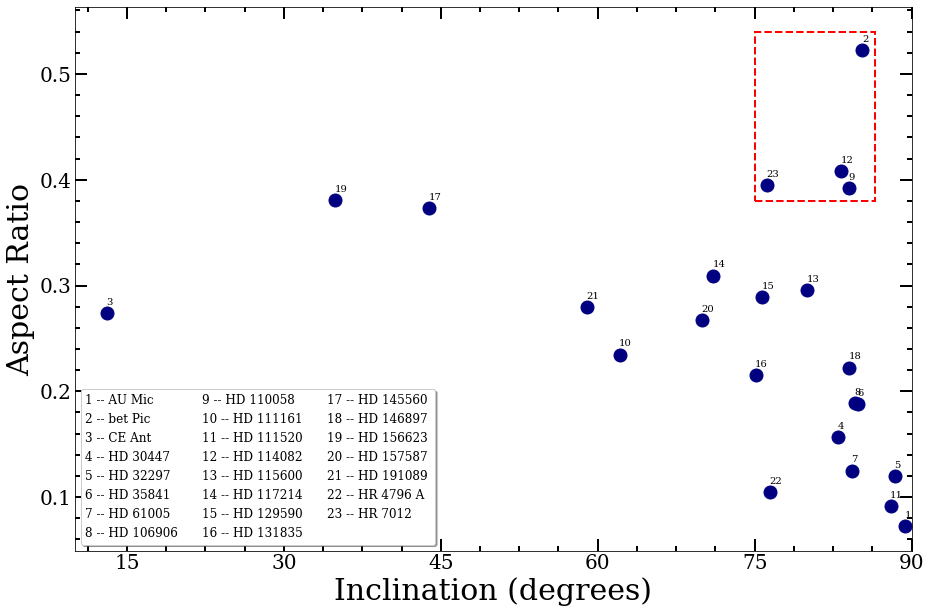}
\end{figure*}

The aspect ratio as a function of inclination is shown in Figure \ref{Fig:asp_rat}. A general trend can be seen from high to low disk inclinations, where the aspect ratio increases with decreasing inclination as we move from probing the vertical aspect ratio alone to probing the radial aspect ratio. We can use this information to also identify disks with large vertical aspect ratios compared to the other disks in our sample at similar inclinations, highlighted in Figure \ref{Fig:asp_rat}. These four disks will be discussed further in Section \ref{sec:sum}.

To constrain the disk geometry, we fit a simple, geometrical inclined ring model to the vertical/radial offset profile, which has also been used in previous debris disk studies \citep{Duchene20, Crotts21, Crotts22}. This model assumes that the disk is radially narrow, although this is unlikely to be the case for many of the disks in our sample (see Section \ref{sec:model_lims} for further discussion on this topic). The reasoning for choosing such a model is its simplicity, allowing us to constrain each disk's geometrical properties in an efficient and empirical manner, without having to rely on more complicated (and often degenerate) radiative-transfer modelling. Our model consists of a circular ring with radius, $R_{d}$, inclination, $i$, position angle, $PA$ (defined as East of North), as well as disk offsets along the major- and minor-axes ($\delta_{x}$ and $\delta_{y}$, respectively). For the lower inclined disks, we fit two ring models simultaneously, the first ring model being a fit to the front side of the disk, while the second ring model is the first model reflected across y-axis to fit the back side of the disk. The best fitting model is found using the MCMC code \textit{emcee} \citep{FM13} by deploying 200 walkers in our defined parameter space over 2000 iterations. 

\begin{figure*}
\centering
	\caption{\label{Fig:vert_prof} The vertical or radial offset from the star for each disk as a function of separation from the star, represented by the dark blue data points. Each disk is rotated by its measured $PA - 90^{\circ}$; therefore negative separations define the East side of the disk, while positive separations define the West side of the disk. Orange curves represent the best fitting narrow, inclined ring.}
	\includegraphics[width=\textwidth]{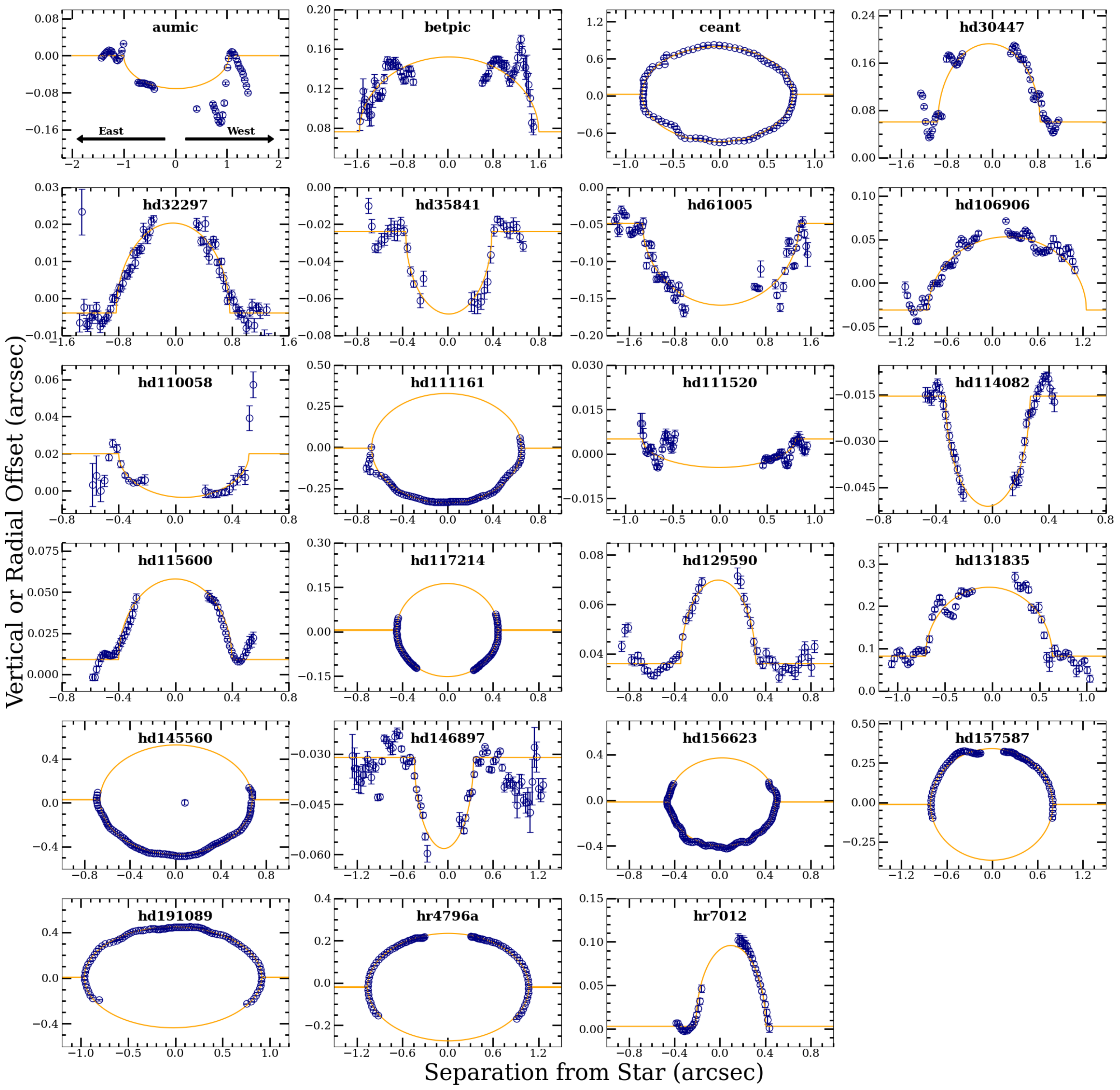}
\end{figure*}

The results for these models can be found in Table \ref{tab:vertoffset_sum} and the vertical/radial offsets with the best fitting ring models for each disk can be found in Figure \ref{Fig:vert_prof}. Additionally, the best fitting models overlaid on the images of each disk can be found in Figure \ref{Fig:spine_H_data} located in the Appendix. We note that in Figure \ref{Fig:vert_prof}, each image is rotated by the measured disk $PA - 90^{\circ}$ so that the major-axis of the disk is horizontal in the image when measuring the vertical/radial offset profile. For simplicity, in this new reference frame, we refer to the disk emission left of the star as the \textbf{East} side/extension, and refer to the disk emission right of the star as the \textbf{West} side/extension. This reference frame and terminology will also be used when measuring the surface brightness, as well as for measuring asymmetries in the surface brightness and disk color. See Table \ref{tab:rotate}, located in the Appendix, for information regarding the degrees of rotation and change in cardinal directions for each disk into the new reference frame. 

While we fit for an offset along the minor-axis ($\delta_{y}$) we do not consider it in our results for the higher inclined disks in our sample, as we find that with this method, $\delta_{y}$ is strongly correlated with other disk properties such as the inclination, vertical width and radial width. In terms of lower-inclined disks, because we are able to fit both the front and back sides of the disk, measurements of $\delta_{y}$ are more robust, and therefore can be useful to determine eccentricity. While we do not find these same correlations significantly for $\delta_{x}$, it is important to take into account that $\delta_{x}$ can be difficult to properly constrain for radially broad disks, as well as for low S/N observations. We also note that a disk offset using this method does not necessarily mean that the disk is eccentric, but can also be the result of other asymmetries in the disk geometry, such as a warp. The uncertainties for both the $\delta_{x}$ and $\delta_{y}$ offsets in Table \ref{tab:vertoffset_sum} include uncertainties in the location of the star for GPI, which has been found to be $\sim$0.05 pixels or 0.7 mas \citep{Wang14}.

\begin{table*}
	\centering
	\caption{\label{tab:vertoffset_sum}Parameters for best fitting inclined ring model using $H$-band data.}
	\begin{tabular*}{\textwidth}{c @{\extracolsep{\fill}} ccccc}
	\\
	    \hline
	    \hline
		Name & $R_{d}$ (AU) & $\delta_{x}$ (AU) & $\delta_{y}$ (AU)& $i$ ($^{\circ}$) & $PA$ ($^{\circ}$) \\
		\hline
		AU Mic & $9.91^{+0.01}_{-0.01}$ & $0.20^{+0.02}_{-0.02}$ & $-0.02^{+0.02}_{-0.02}$ & $86.00^{+0.01}_{-0.01}$ & $126.68^{+0.01}_{-0.01}$ \\
		$\beta$ Pic & $27.06^{+1.18}_{-0.34}$ & $0.26^{+0.83}_{-0.32}$ & $1.82^{+0.05}_{-0.07}$ & $88.90^{+0.09}_{-0.10}$ & $32.24^{+0.04}_{-0.10}$ \\
		CE Ant & $27.13^{+0.01}_{-0.01}$ & $-0.86^{+0.04}_{-0.04}$ & $0.86^{+0.04}_{-0.04}$ & $15.10^{+0.95}_{-1.05}$ & $91.02^{+0.20}_{-0.19}$ \\
		HD 30447 & $75.43^{+0.68}_{-0.72}$ & $-6.53^{+0.23}_{-0.36}$ & $2.92^{+0.19}_{-0.19}$ & $81.47^{+0.05}_{-0.05}$ & $33.56^{+0.39}_{-0.01}$ \\
		HD 32297 & $105.85^{+1.62}_{-1.06}$ & $-4.58^{+0.90}_{-0.95}$ & $-0.58^{+0.13}_{-0.13}$ & $88.26^{+0.04}_{-0.04}$ & $47.63^{+0.02}_{-0.01}$ \\
		HD 35841 & $39.12^{+0.36}_{-0.28}$ & $1.05^{+0.33}_{-0.36}$ & $-2.49^{+0.13}_{-0.13}$ & $83.27^{+0.20}_{-0.24}$ & $167.47^{+0.05}_{-0.05}$ \\
		HD 61005 & $50.21^{+0.17}_{-0.17}$ & $0.68^{+0.19}_{-0.38}$ & $-1.79^{+0.07}_{-0.06}$ & $85.41^{+0.06}_{-0.06}$ & $70.80^{+0.01}_{-0.12}$ \\
		HD 106906 & $107.98^{+0.69}_{-0.79}$ & $20.67^{+1.10}_{-1.10}$ & $-3.20^{+0.20}_{-0.09}$ & $85.34^{+0.05}_{-0.06}$ & $104.00^{+0.03}_{-0.01}$ \\
		HD 110058 & $59.56^{+12.47}_{-0.77}$ & $8.19^{+1.01}_{-13.30}$ & $2.54^{+1.25}_{-0.23}$ & $87.06^{+0.23}_{-0.19}$ & $158.55^{+0.11}_{-1.00}$ \\
		HD 111161 & $72.48^{+0.08}_{-0.09}$ & $-1.44^{+0.16}_{-0.16}$ & $-0.66^{+0.09}_{-0.09}$ & $59.78^{+0.08}_{-0.06}$ & $83.29^{+0.04}_{-0.08}$ \\
		HD 111520 & $91.42^{+13.80}_{-10.1}$ & $-2.24^{+13.38}_{-10.18}$ & $-1.80^{+0.39}_{-0.34}$ & $89.45^{+0.27}_{-0.27}$ & $165.66^{+0.11}_{-0.13}$ \\
		HD 114082 & $28.50^{+1.40}_{-0.19}$ & $-2.95^{+0.20}_{-0.22}$ & $-1.48^{+0.10}_{-0.16}$ & $83.32^{+0.54}_{-0.20}$ & $105.01^{+0.05}_{-0.04}$ \\
		HD 115600 & $44.02^{+7.36}_{-7.32}$ & $-0.04^{+7.19}_{-7.22}$ & $-1.01^{+2.17}_{-0.17}$ & $82.97^{+1.32}_{-1.30}$ & $24.20^{+0.01}_{-0.19}$ \\
		HD 117214 & $42.77^{+0.09}_{-0.10}$ & $-0.19^{+0.18}_{-0.13}$ & $0.41^{+0.20}_{-0.23}$ & $69.57^{+0.46}_{-0.34}$ & $180.51^{+0.15}_{-0.18}$ \\ 
		HD 129590 & $45.50^{+0.48}_{-1.08}$ & $-1.89^{+0.51}_{-0.90}$ & $4.96^{+0.15}_{-0.14}$ & $84.11^{+0.28}_{-0.26}$ & $120.28^{+0.04}_{-0.03}$ \\ 
		HD 131835 & $89.62^{+0.81}_{-0.80}$ & $-4.60^{+0.78}_{-0.84}$ & $11.01^{+0.22}_{-0.22}$ & $75.94^{+0.23}_{-0.23}$ & $60.81^{+0.02}_{-0.18}$ \\ 	
		HD 145560 & $81.23^{+0.06}_{-0.05}$ & $0.86^{+0.13}_{-0.12}$ & $3.33^{+0.13}_{-0.12}$ & $41.91^{+0.49}_{-0.09}$ & $39.51^{+0.01}_{-0.03}$ \\ 
		HD 146897 & $51.84^{+0.19}_{-0.78}$ & $-6.33^{+0.85}_{-0.27}$ & $-4.08^{+0.11}_{-0.12}$ & $85.99^{+0.01}_{-0.01}$ & $114.62^{+0.02}_{-.01}$ \\
		HD 156623 & $52.56^{+0.69}_{-0.25}$ & $2.10^{+0.17}_{-0.55}$ & $1.68^{+0.09}_{-0.09}$ & $34.70^{+0.46}_{-0.96}$ & $102.86^{+0.03}_{-0.49}$ \\ 
		HD 157587 & $81.24^{+0.05}_{-0.04}$ & $-0.65^{+0.12}_{-0.11}$ & $1.32^{+0.19}_{-0.17}$ & $64.02^{+0.04}_{-0.02}$ & $127.71^{+0.10}_{-0.08}$ \\ 	
		HD 191089 & $46.96^{+0.01}_{-0.03}$ & $-1.20^{+0.05}_{-0.09}$ & $0.35^{+0.09}_{-0.09}$ & $61.85^{+0.09}_{-0.08}$ & $71.40^{+0.10}_{-0.07}$ \\ 
		HR 4796 A & $77.71^{+0.05}_{-0.04}$ & $0.58^{+0.10}_{-0.09}$ & $-1.56^{+0.09}_{-0.09}$ & $76.15^{+0.06}_{-0.07}$ & $26.43^{+0.03}_{-0.03}$ \\ 
		HR 7012 & $8.77^{+0.08}_{-0.05}$ & $2.76^{+0.07}_{-0.14}$ & $0.08^{+0.04}_{-0.04}$ & $72.42^{+0.35}_{-0.16}$ & $113.80^{+0.23}_{-0.19}$ \\ 		
		\hline
		\hline
	\end{tabular*}
\end{table*}

\subsection{Surface Brightness} \label{sec:sb}
Once the disk vertical or radial offset and FWHM are measured, we can use these values to measure the surface brightness as a function of stellar separation, as well as measure any brightness asymmetries present between the East and West extension of each disk. The East and West extensions are compared specifically rather than between the front and back side of the disk as brightness asymmetries between the front and back sides are due to preferential forward or backward scattering of dust grains, rather than inherent asymmetries such as an eccentric disk. 

\begin{figure*}
\centering
	\caption{\label{Fig:multi_sb} Disk surface brightness as a function of separation from the star in all three bands. Again, the disk is rotated by measured $PA - 90^{\circ}$ so that negative separations define the East side of the disk, while positive separations define the West side of the disk.}
	\includegraphics[width=\textwidth]{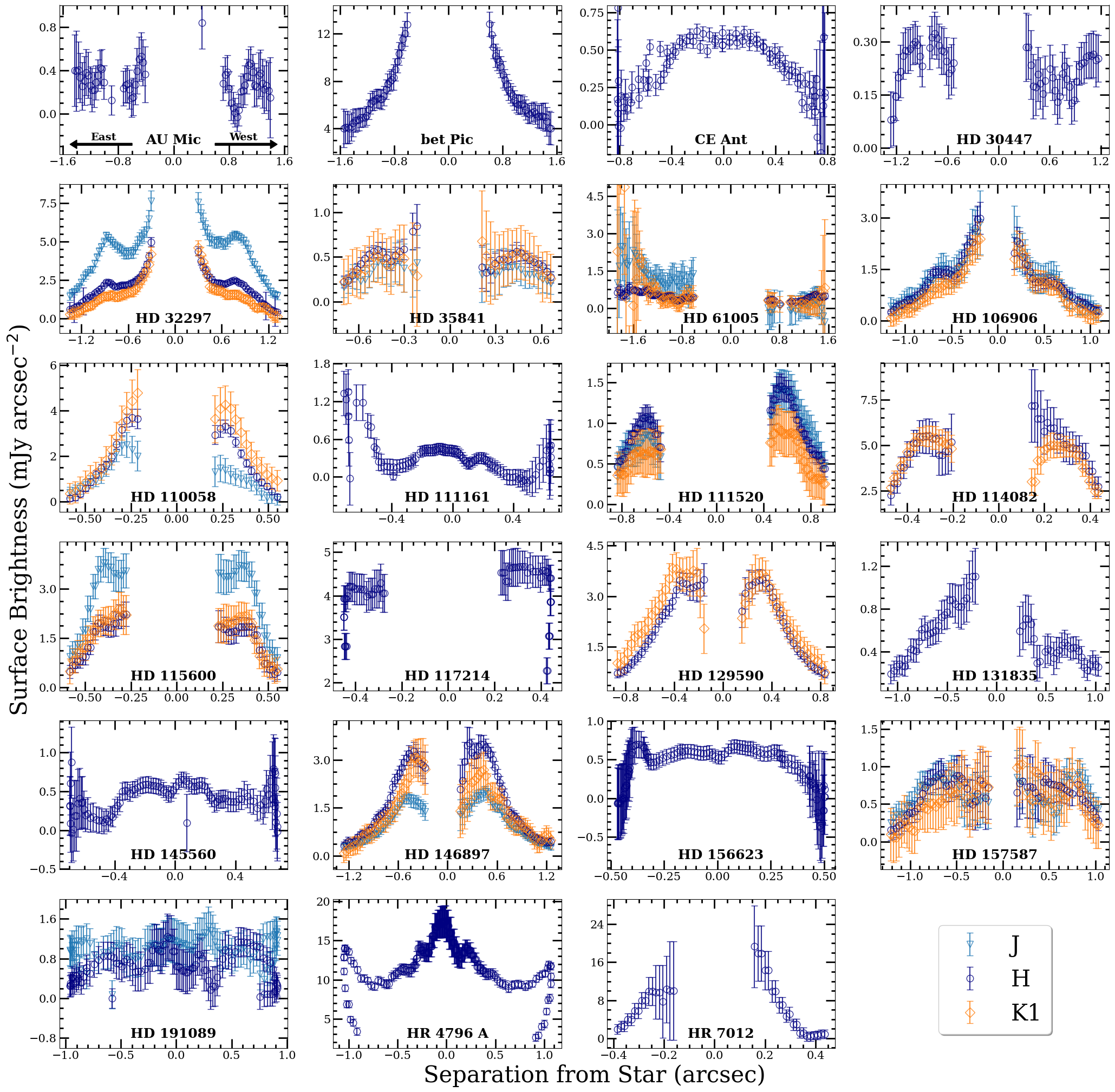}
\end{figure*}

We first measure the surface brightness along each disk for each band. This is done by first rotating the images by their derived $PA$ values (found in Table \ref{tab:vertoffset_sum}) minus 90$^{\circ}$, followed by binning the image into 2$\times$2 pixel bins in order to diminish any correlation between pixels. The vertical/radial offset values are then used to define the location of the peak surface brightness along the disk, where the surface brightness is averaged along several pixels centered around the peak surface brightness location. For the lower inclined disks, the image is rotated between the same angles from the measured $PA$ as done when measuring the vertical/radial offset, followed by averaging the surface brightness around the peak surface brightness location. The resulting surface brightness profiles can be found in Figure \ref{Fig:multi_sb}. 

\begin{figure*}
\centering
	\caption{\label{Fig:sb_asymm} Brightness asymmetry between the East and West extensions for each disk in all three wavelengths. Values of 1 represent no brightness asymmetry. Red data points represent significant asymmetry of $\ge3\sigma$.}
	\includegraphics[width=\textwidth]{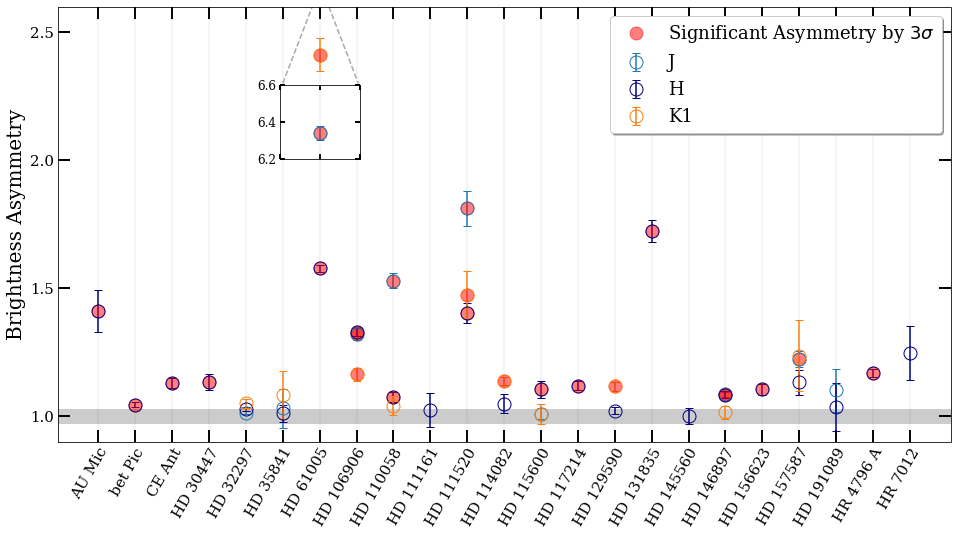}
\end{figure*}

To measure the brightness asymmetry between the East and West extensions, we place apertures at similar separations from the star on either side of each disk. For higher inclined disks, we place a single rectangular aperture on the East and West extensions of the disk, while for lower inclined disks we place two to three square apertures covering from the front of the disk to the disk ansae on either side. In all cases, the height of the aperture is determined by the measured average FWHM of the disk, while the length/placement of the rectangular apertures are determined by the S/N of the disk (i.e. the apertures are placed where the S/N is the highest, again, at a similar separation from the star on either side of the disk). Once the aperture(s) are determined and placed, we then average the flux over the aperture(s) for both our image and uncertainty maps in each band to determine 1$\sigma$ uncertainties. The average surface brightness can then be compared between the East and West extensions to determine whether or not a surface brightness asymmetry is present. The surface brightness asymmetry for each disk can be found in Figure \ref{Fig:sb_asymm}, which is defined as the brighter extension divided by the dimmer extension. We find 16/23 disks have a significant brightness asymmetry (i.e. by 3$\sigma$ in at least one band), which is well over half the disks in our sample.   

\subsection{Disk Color} \label{sec:color}
For the disks in our sample that have multiwavelength observations, we can also measure the disk color between bands. Given that the scattering properties of dust grains determines the disk color, these color measurements can give us information about the dust grain properties in the disk such as dust composition, minimum grain size, and porosity. While it is difficult to untangle these dust grain properties from the disk color alone, we can still use these results to compare the disk color of our sample in NIR wavelengths to look for trends, as well as compare the disk color between the East and West extensions to determine if any asymmetries are present. 

\begin{figure*}
\centering
	\caption{\label{Fig:color_avg} \textbf{Top:} Average disk color for each disk, between all three wavelengths. Disks with a negative value have a blue color, while disks with a positive value have a red color and disks close to zero have a neutral disk color (shown by horizontal grey line). \textbf{Bottom:} Disk color asymmetry between all three wavelengths, measured by taking the absolute value of the East disk color subtracted from the West disk color. Values of 0 represent no disk color present. Red data points represent significant asymmetry by $\ge3\sigma$.}
	\includegraphics[width=\textwidth]{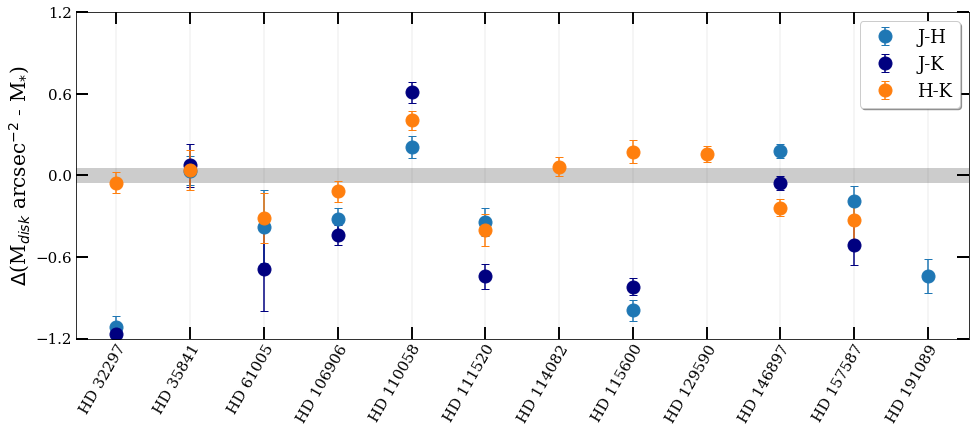}
        \includegraphics[width=\textwidth]{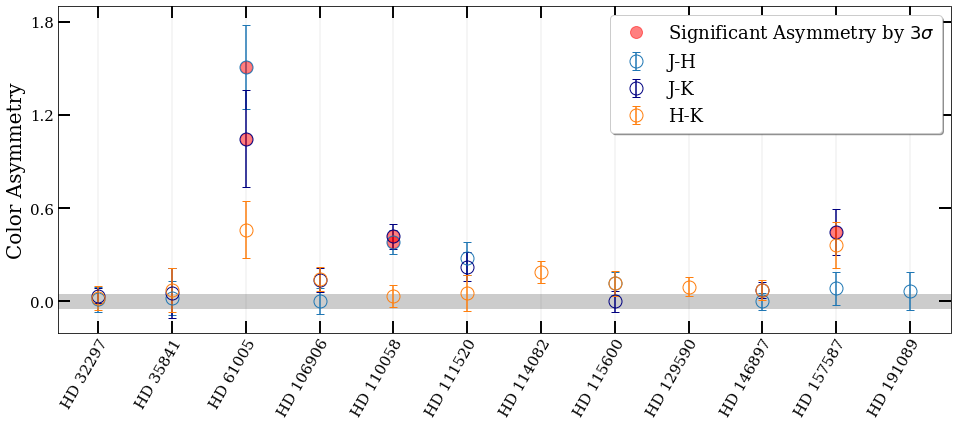}
\end{figure*}

To measure the disk color, we start with the same process as measuring the surface brightness asymmetry, where the flux on either side of the disk is averaged over the same apertures used previously. This averaged flux is then converted to magnitudes and compared between a pair of bands. Finally, the difference in stellar magnitude between the same pair of bands is measured and subtracted from the difference in magnitude of the disk (i.e. $J$-$H$ = $\Delta \text{mag}(J_{disk} - H_{disk}) - \Delta \text{mag}(J_{star} - H_{star})$). This is done to eliminate the bias introduced by the color of the star. The average disk color (averaged across the whole disk) can be found in the top plot of Figure \ref{Fig:color_avg}. In this case, a negative value indicates a blue disk color, meaning that the dust grains scatter more efficiently at shorter wavelengths, while a positive value indicates a red disk color, meaning that the dust grains scatter more efficiently at longer wavelengths. Lastly, a 0 value indicates a grey or neutral disk color, meaning that the scattering efficiency has no preference between short and long wavelengths, and can be the result of a large minimum dust grain size (on the order of a couple of microns or greater; \citealt{Boccaletti03}). 

In addition to the average disk color, we also measure the difference in color between the East and West extensions. The lower plot of Figure \ref{Fig:color_avg} shows the absolute value of the difference in disk color between the East and West extension. Here, a value of 0 means that no asymmetry is present. We find that 3/12 disks have significant color asymmetries of 3$\sigma$ or greater in at least one band (HD 61005, HD 110058 and HD 157587), while 2 additional disks have color asymmetries with a significance between 2$\sigma$ and 3$\sigma$ (HD 111520 and HD 114082). In the case of an axisymmetric disk with a uniform distribution of dust grains, we would expect no difference in disk color between the East and West extensions. Therefore, an asymmetry in the disk color may be the result of an asymmetric distribution of dust grains. For example, a bluer East extension may suggest that a population of small dust grains have been released or redistributed to this area of the disk. Such an event could occur due to recent collisions in the disk or possibly an interaction with the interstellar medium (ISM; \citealt{Debes09}). 

\section{Discussion} \label{sec:sum}

For a discussion of results for each specific disk system, along with comparison to the literature, we refer the reader to Appendix \ref{sec:summary}. Here we discuss the limitations of our model, as well as broader trends found in our sample.

\subsection{Ring Model Limitations} \label{sec:model_lims}
While our ring model for fitting the vertical/radial offset profiles is simplistic and allows us to efficiently derive geometrical properties for our large sample of disks, this simplicity comes with some caveats and limitations.

For one, our ring model assumes a radially narrow ring, which is likely not the case for many of the disks in our sample. This caveat may lead to poor fits, such as for $\beta$ Pic, and may also have led to exaggerated offsets along the major-axis in some cases. For radially narrow disks, such as HR 4796 A, measurements of $\delta_{x}$ are more robust. As mentioned in Section \ref{sec:geom}, our simplistic model also has an effect on the measured offset along the minor-axis, or $\delta_{y}$, where $\delta_{y}$ tends to be exaggerated for disks with higher inclinations ($\gtrsim$75$^{\circ}$) as we are only fitting the front side of the disk. This influenced our decision to not take into account $\delta_{y}$ for the higher inclined disks in our discussion of disk morphologies, as it is difficult to untangle whether these offsets are real, or simply an effect of our chosen model and other properties of the disk.

The S/N of the observations should also be taken into account, as low S/N observations may also lead to poor fits of our ring model, creating small offsets that do not exist, such as the case with AU Mic. We do find that the several disks with the largest $\delta_{x}$ measurements are higher S/N observations which supports the conclusion that these disks are indeed either eccentric or harbour another geometrical asymmetry, such as a warp. However, future followup for these disks with low S/N observations will be needed to confirm our results.

In summary, our simple ring model is most effective for radially narrow disks, and for lower inclination disks where we can fit both the front and back side of the disk. Even in the case of higher inclined disks, and for most radially broad disks in our sample, this method is still successful in confirming inclination, $PA$, and disk radius, while $\delta_{x}$ measurements are also still useful for determining possible asymmetric geometries that may not be fully captured with more complex modelling, especially when taken into consideration with other factors such as surface brightness asymmetries. 

\subsection{Trends in Brightness Asymmetry}
Our large sample size allows us to look at overall trends that may have implications on debris disk properties and evolution. Here, we look at trends seen in the measured brightness asymmetry derived in Section \ref{sec:sb}. 

Comparing the average brightness asymmetry (brighter side/dimmer side) between all disks, three disks have significant brightness asymmetries over 1.5 (HD 61005, HD 111520, and HD 131835), two disks have significant brightness asymmetries between 1.2 and 1.5 (AU Mic and HD 106906), six disks have significant brightness asymmetries between 1.1 and 1.2, and six disks have significant brightness asymmetries $<$1.1. The majority of disks have brightness asymmetries where the brighter side is $<$1.2 times brighter than the dimmer side, while a small handful of disks have particularly large brightness asymmetries $>$1.2. Out of the disks with the largest brightness asymmetries, HD 106906 is the only disk that has strong evidence of planet induced eccentricity (e.g. \citealt{Nesvold17,Crotts21}). It is unclear if the other three disks are eccentric, although all three have complex morphologies (i.e. multiple rings, clumps, warps, and radial asymmetries) suggesting that they are being actively perturbed by some mechanism.

When comparing the average brightness asymmetry between filters for disks with multiwavelength observations, we find that the average asymmetry is 1.75$\pm$0.04 in the $J$ band, 1.16$\pm$0.03 in the $H$ band, and 1.16$\pm$0.05 in the $K1$ band. Excluding HD 61005, which is an outlier in the $J$ and $K1$ bands, changes these values to 1.24$\pm$0.04, 1.12$\pm$0.03 and 1.03$\pm$0.05 respectively. In both cases, the $J$ band has a significantly higher brightness asymmetry on average than the $H$ and $K1$ bands. When excluding HD 61005, the $K1$ band has the lowest brightness asymmetry on average. These results suggest that the brightness asymmetry is strongest in the smallest dust grains and decreases with increasing wavelength/particle size. This result aligns with trends seen between short and long wavelength observations, where disks appear to be more asymmetric at optical/NIR wavelengths and more symmetric at sub-mm/mm wavelengths.

\subsection{Effects of Stellar Age \& Temperature}
In Section \ref{sec:results} we mainly focused on what our analysis showed for each individual disk, however, with such a uniform analysis on a large sample of disks, we can also use our results to look for larger scale trends. In this Section, we focus on debris disk properties, such as asymmetries and disk color, as a function of stellar temperature and age, to see if there are any correlations that may inform us about debris disk environments and evolution. 

\subsubsection{Brightness and Color Asymmetry} \label{sec:asymm_vs_star}
\begin{figure*}[t!]
\centering
	\caption{\label{Fig:bright_asymm_vs_star} \textbf{Top:} Brightness asymmetry vs. stellar age. \textbf{Bottom:} Brightness asymmetry vs. stellar temperature.}
	\includegraphics[width=\textwidth]{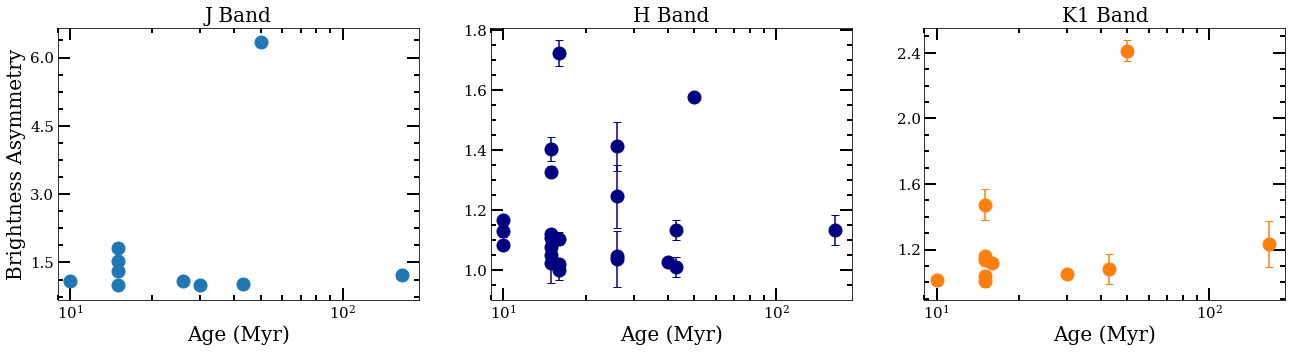}
        \includegraphics[width=\textwidth]{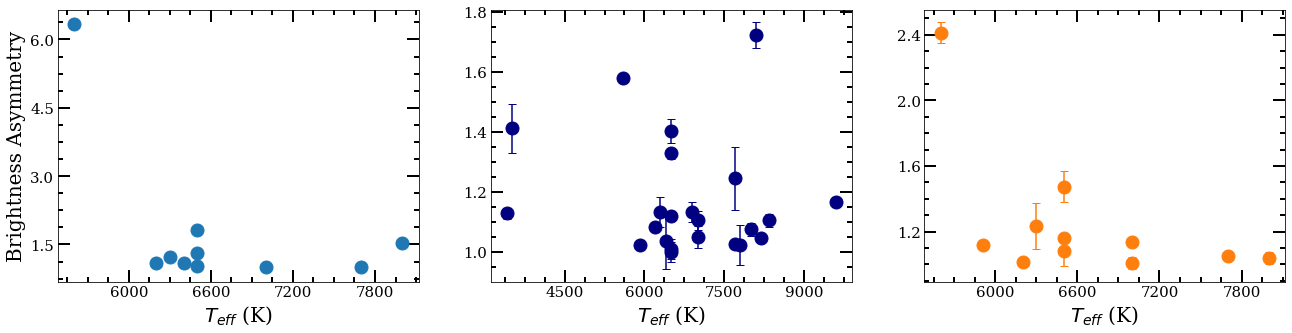}
\end{figure*}

In Figure \ref{Fig:bright_asymm_vs_star}, we plot the measured brightness asymmetry in each band versus the stellar age and temperature. From Figure \ref{Fig:bright_asymm_vs_star}, there does not appear to be a significant trend between the degree of brightness asymmetry with either the stellar age or temperature. While at first glance it may appear as if there is a tentative trend between brightness asymmetry and stellar temperature in the $J$ and $K1$ bands, this is simply due to our small sample size of observations in these bands along with one outlier (HD 61005).

\begin{figure*}
\centering
	\caption{\label{Fig:color_asymm_vs_star} \textbf{Top:} Disk color asymmetry vs. stellar age. \textbf{Bottom:} Disk color asymmetry vs. stellar temperature.}
	\includegraphics[width=\textwidth]{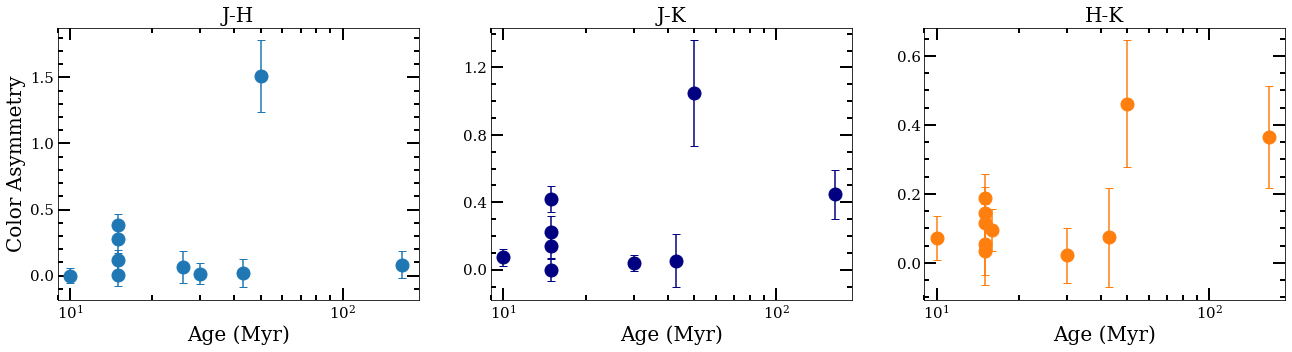}
        \includegraphics[width=\textwidth]{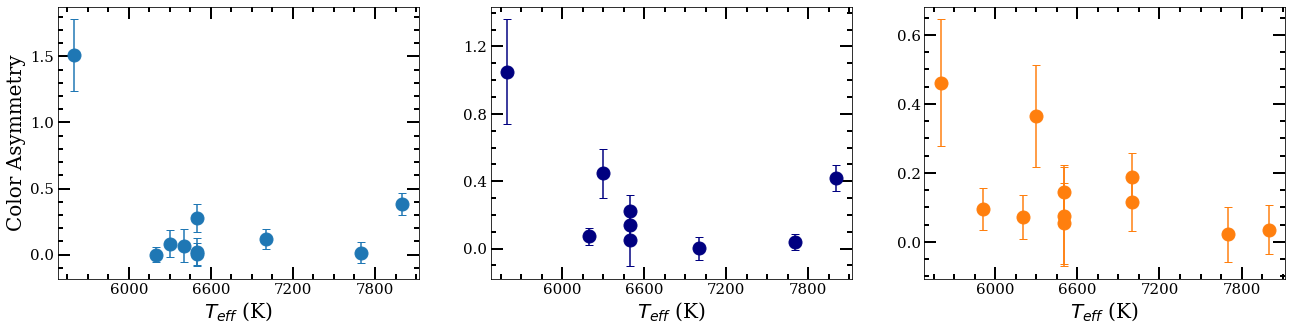}
\end{figure*}

Figure \ref{Fig:color_asymm_vs_star} shows the disk color asymmetry plotted vs. stellar age and temperature. Similar to the brightness asymmetry, no strong trends are seen between disks with a color asymmetry and the age of the system or stellar temperature. This result, along with the brightness asymmetry, suggests that asymmetric disks can be present regardless of the system's age or stellar temperature, although, it should be kept in mind that the average age of our sample is fairly young (less than 100 Myr). Again, it is important to note our small sample size for measured disk colors given the small sample of disks with $J$- and $K1$-band observations, therefore these results may not show the entire picture.

\subsubsection{Average Disk Color} \label{sec:color_vs_star}
In the previous Section, we compare the brightness asymmetry and disk color asymmetry with stellar age and temperature, but we can also compare the disk color itself with these two parameters. As the disk color is the result of dust grain properties in the disk, as described in Section \ref{sec:color}, trends between the disk color and the stellar age or temperature may be informative about the evolution of dust grains in these systems. 

\begin{figure*}
\centering
	\caption{\label{Fig:color_vs_star} \textbf{Top:} Disk color vs. Stellar Age. \textbf{Bottom:} Disk color vs. Stellar Temperature.}
	\includegraphics[width=\textwidth]{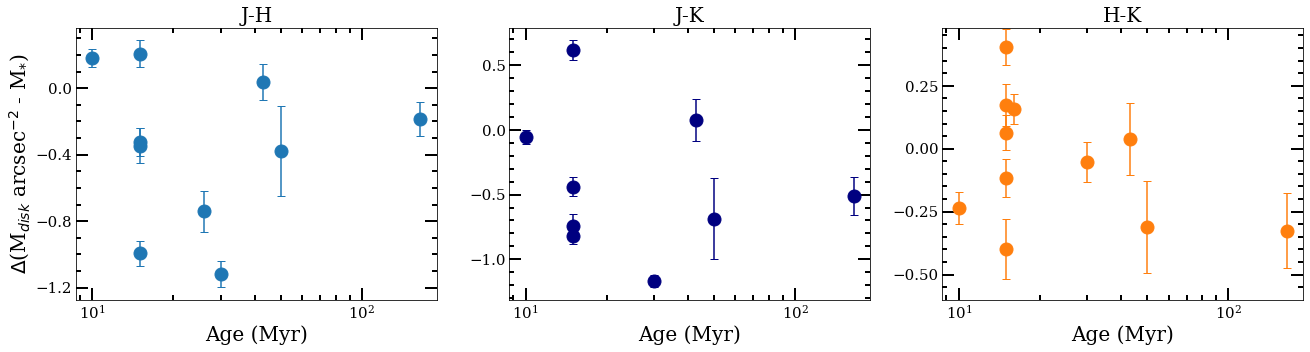}
        \includegraphics[width=\textwidth]{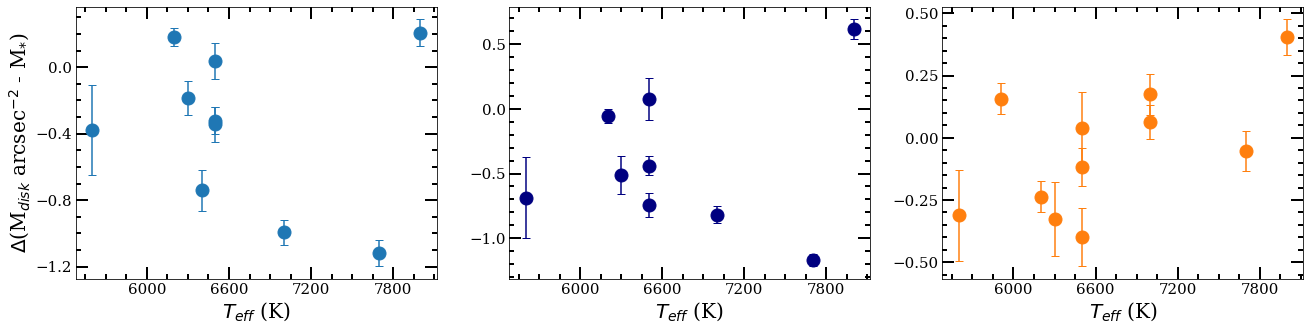}
\end{figure*}

Figure \ref{Fig:color_vs_star} (top) shows, similar to the disk asymmetries, no significant trends between disk color and stellar age, demonstrating once again that age of the system does not have a drastic effect on the properties of debris disks in our sample. This is not the case with respect to the stellar temperature. A trend is suggested in the bottom plot of Figure \ref{Fig:color_vs_star}, strongest in $H$-$K1$, where as we transition from cooler to hotter stellar temperatures, the disk color becomes increasingly grey/red. Calculating the strength of the correlation between the $H$-$K1$ color with temperature, we find a Pearson correlation coefficient of 0.6 with a p-value of 0.05, meaning that the correlation is significant at the 2$\sigma$ (95\%) confidence level. Similar trends have been seen in other color studies, such as with HST \citep{Ren23}. Such a trend is also expected; as the stellar temperature increases, so does the blow-out size of the system, i.e. the dust grain size where the force of radiation pressure is equal to the force of gravity. Because larger dust grains are more efficient at scattering at longer wavelengths compared to small dust grains, this leads to a more red disk color. Additionally, dust grains on the order of several microns or larger can exhibit a grey color, while disks with a larger population of small dust grains will tend toward a blue color. 

While the trend between disk color and stellar temperature is strongest in $H$-$K1$, this trend weakens in $J$-$H$ and $J$-$K1$. However, this may be due to several disks that break this trend. The two most notable disks are HD 32297 and HD 115600, both of which are around hotter stars (7700 K and 7000 K), but have exceptionally strong blue colors in $J$-$H$ and $J$-$K1$. In both cases, the disk color becomes significantly more grey or red in $H$-$K1$, making them more inline with the overall trend. The strong blue color seen at short wavelengths for these two disks suggests that a larger population of small dust grains is present than would be expected for a debris disk orbiting a star of temperature $>$7000 K. One explanation is that these disks may have recently undergone a large/violent collision, producing dust grains smaller than the blow-out size for which radiation pressure has not had enough time to blow out these small grains.

However, a recent large collision may not even be necessary, as studies have shown that bright debris disks ($L_{disk}/L_{*} > 10^{-3}$) around F and A spectral-type stars (as for HD 32297 and HD 115600), with high collisional activity, can naturally produce large amounts of sub-micron sized dust grains that will leave a detectable signature \citep{TK19}. \citet{TK19} show that the halo for these disks can contribute up to $\sim$50\% to the total disk flux at short wavelengths while decreasing towards longer wavelengths. Additionally, small unbound grains can turn the disk color from red to blue. This may explain the strong blue colors in $J$-$H$ and $J$-$K1$, which then becomes significantly less blue in $H$-$K1$. We note that the enhanced blue color is not observed for all bright debris disks around hot stars (i.e., HD 110058). In fact, the HD 110058 debris disk, which has the hottest host star for a disk with multiwavelength observations, is the only disk in our sample that is strongly red between all three wavelengths. Either another factor is affecting the color of this disk (such as composition), or the sub-micron sized grains have been successfully blown out of the system. Either way, these examples show how the disk color is affected by the stellar temperature, and can also be used to help understand the mechanics of a collisional cascade in certain disks.

\subsection{Disks with Large Aspect Ratios} \label{sec:fwhm_discussion}
In Section \ref{sec:geom}, we measured the vertical FWHM (or radial FWHM depending on the inclination) using our Gaussian fitting procedure and used the average FWHM to roughly estimate the aspect ratio. Plotting these aspect ratios vs. inclination showed several debris disks that had a larger vertical aspect ratio compared to other disks of similar inclination (highlighted by the red square in Figure \ref{Fig:asp_rat}). To understand the underlying reason for this discrepancy, we compare the aspect ratio with other disk and system parameters.

In Figure \ref{Fig:aspect_rat_vs_temp}, we show the aspect ratio plotted vs. the stellar temperature for disks with $i > 70^{\circ}$, where the color of each data point represents the reference radius, $R_{0}$, for each disk. The four disks that have particularly high aspect ratios compared to other disks with similar inclinations are highlighted by the red dashed square box, $\beta$ Pic, HD 110058, HD 114082 and HR 7012. One reason these disks may have a high aspect ratio is the combination of their inclinations and the way we measure the aspect ratio, where there may be some back scattering from the far side of the disk that is contributing the vertical width. However, there are two things noticeable in Figure \ref{Fig:aspect_rat_vs_temp}, with respect with the four highlighted disks, that are not related to disk inclination. One, these disks are around relatively hotter stars (7700 K to 8200 K) compared to other disks in our sample with higher inclinations, and two, these four disks are more radially compact in terms of $R_{0}$), with $R_{0} < 40$ au. Additionally, three out of these four disks ($\beta$ Pic, HD 110058 and HR 7012) also have detectable amounts of CO \citep{Dent14,Hales22,Schneiderman21}. HD 114082 has no gas detection, with only an upper limit on the CO mass of $<5\times10^{-6}$ \citep{Kral20}.

The fact that these four disks have multiple factors in common can help us understand what is causing these disk to have a large vertical aspect ratio. While all four disks are around hotter stars, and three out of four have detectable amounts of CO, there are two other disks that also meet this criteria, HD 32297 and HD 131835 (stellar temp = 7700 K and 8100 K), but do not have a high vertical aspect ratio. This suggests that the stellar temperature and the existence of a gas disk, either together or individually, are not the root cause of a disk becoming vertically thick. In fact, \citet{Kral20} found that gas in debris disks should have the opposite effect, making the disk more vertically thin due to the settling of small dust grains. Looking more closely at HD 32297 and HD 131835, one thing that distinguishes these two disks from the other four is that they are both more radially extended in terms of $R_{0}$ ($R_{0}$ = 98.4 au and 107.7 au). In addition to this, AU Mic, which has a smaller $R_{0}$ of 30.2 au and is around an M-dwarf, has a small vertical aspect ratio. This suggests that the combination of a higher stellar temperature and a small $R_{0}$ are requirements for creating a disk that has a particularly large aspect ratio, where dust closer to the star is puffed up (i.e. has a higher inclination dispersion) due to the higher temperatures.  

While this scenario makes physical sense, it is not necessarily the full story. For instance, it is not clear why these disks have small $R_{0}$ values, as \citet{Esposito20} (first reported in \citet{Matra18}) shows that there is a positive correlation between stellar luminosity and $R_{0}$. This means that as the stellar temperature/luminosity increases, we would expect a peak dust density radius farther out from the star, making these four disks outliers. One possible explanation could be that due to the high inclination of these disks, it is difficult to measure the exact peak radius, leading to an underestimation of $R_{0}$. While HR 7012 is undoubtedly compact, this cannot be easily said for the other three disks which extend well beyond their measured values of $R_{0}$ in scattered light. However, in the case of HD 110058 and HD 114082, ALMA observations also show relatively compact disks, with a peak radius of mm-sized grains at 31 au and 24.1 au, respectively \citep{Hales22, Kral20}, consistent with their measured $R_{0}$ in scattered light within uncertainties. The $\beta$ Pic disk is the most uncertain, where both the small grains in scattered light and large grains as seen by ALMA extend way beyond $R_{0}$, which is near the measured inner radius, and is more consistent with being radially broad. While we do measure a consistent disk radius of 27.06 au, this is at the edge of GPI's FOV, and therefore it is possible for the disk radius to lie beyond this distance.

Another explanation for these disks being more compact in terms of $R_{0}$ could be due to shaping from planet or stellar companions. In the case of HR 7012, as mentioned previously, the system has a stellar companion located $>$2000 au from the main star \citep{Torres06}, which has been suspected to be the cause of the disk's significant truncation, however, this has yet to be confirmed. For HD 110058, there is evidence of a warp past 40 au, which suggests perturbation from a planet companion. If there is a planet that is orbiting closely outside of this warp, this could lead to a truncation of the disk. That being said, a planet could cause a similar warp inside of the disk, similar to the $\beta$ Pic system, where \citet{Pearce22} predicts that a sculpting planet of mass $\ge$$0.5 \pm 0.4$ M$_{\text{Jup}}$ with semi-major axis $\le$$8 \pm 8$ au is sufficient to create a warp at 40 au. While a planet is known to exist in the HD 114082 system \citep{Engler22,Zakhozhay22}, this planet is within 2 au of the star, making it dynamically uncoupled from the disk. However as seen in \citet{Engler22}, there is a clear opening within the inner radius of the disk, likely meaning that there are additional planets closer to the disk edge, and given that the disk is radially narrow, this suggests that there may also be a shepherding planet outside of the outer disk edge. Finally, the $\beta$ Pic is also known to have two planets, $\beta$ Pic b and c, \citep{Lagrange10, Lagrange19}. These planets are very likely perturbing the disk and have even been directly linked to the known disk warp located at $\sim$50 au \citep{Mouillet97}. While this may not fully explain the small $R_{0}$ value, \citet{Matra19} found using ALMA observations that the vertical structure of the disk is best fit with two Gaussians rather than one, suggesting the existence of both a cold and hot population of dust grains. The authors state that this distribution of dust grains is not consistent with stirring from $\beta$ Pic b alone, but could be the result of another unseen planet migrating outwards toward the inner disk edge.

\subsubsection{Aspect Ratio \& Particle Size Distribution}
\begin{figure}
\centering
	\caption{\label{Fig:aspect_rat_vs_temp} Aspect ratio for the disks in our sample with $i \gtrsim 70^{\circ}$ as a function of stellar temperature. The color of each point represents $R_{0}$ in au, as indicated by the color bar, taken from \citet{Esposito20}. The four disks highlighted within the red dashed square are the same four disks highlighted in Figure \ref{Fig:asp_rat}.}
	\includegraphics[width=0.46\textwidth]{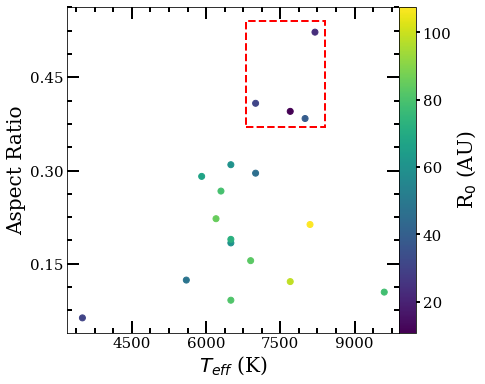}
\end{figure}

\begin{figure*}
\centering
	\caption{\label{Fig:aspect_rat_vs_q} \textbf{Left:} Aspect ratio for the disks listed in Table \ref{disk_q} as a function of their measured dust grain size power-law, $q$. The color of each point represents $R_{0}$ in au. \textbf{Right:} Aspect ratio vs. the stellar temperature, same as Figure \ref{Fig:aspect_rat_vs_temp}, however, the color of each point represents $q$, instead of $R_{0}$.}
	\includegraphics[width=0.96\textwidth]{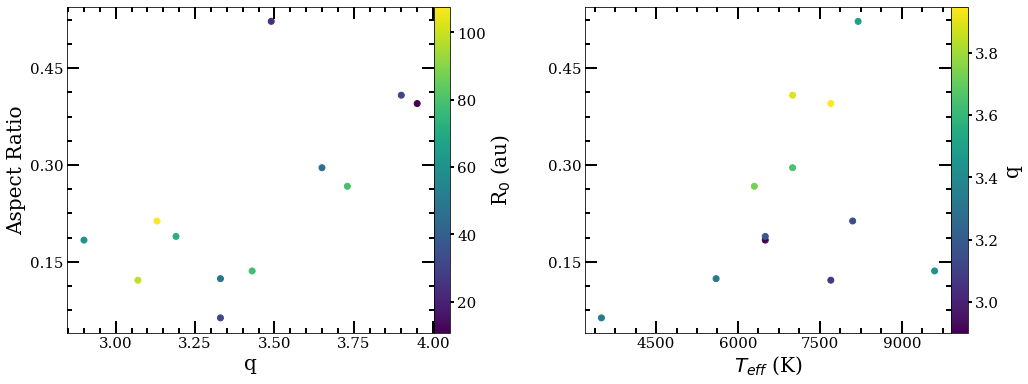}
\end{figure*}

 While an unknown planet may be puffing up the $\beta$ Pic disk, in the case of HR 7012, there is strong evidence of the disk being the result of a high-speed collision between large planetesimals. While the disk is shown to harbour SiO and CO as a result of these collisions \citep{Lisse08,Schneiderman21}, another piece of evidence is the disk's dust grain size distribution power-law ($q$) of 3.95 \citep{Johnson12}. This power law is steeper than the typical power-law for a collisional cascade of $q=3.5$ \citep{Dohnayi69}, where lab work has shown that a high $q$-value is consistent with what is expected for the aftermath of a giant hypervelocity impact \citep{Takasawa11}. This motivates us to look more closely at the affects of $q$ on the vertical aspect ratio, alongside $R_{0}$ and stellar temperature. For disks that have measured $q$-values in the literature (see Table \ref{disk_q}), we plot these values vs. their measured aspect ratios, which can be seen in the left plot Figure \ref{Fig:aspect_rat_vs_q}. We note that most of these $q$ values are measured by extrapolating from millimeter to centimeter observations, however, some disks only have measured $q$ values from radiative transfer modelling of scattered light observations and/or the SED. Plotting aspect ratio vs. $q$, we find a tentative positive trend between $q$ and the vertical aspect ratio, where the average $q$-value for disks with an aspect ratio $\gtrsim$0.25 is $\sim$3.74, while the average $q$-value for disks with an aspect ratio $\lesssim$0.25 is $\sim$3.20. While the disks with $q\gtrsim3.5$ are on average more compact in terms of $R_{0}$, there otherwise does not seem to be a correlation between $q$ and $R_{0}$. The left plot of Figure \ref{Fig:aspect_rat_vs_q} is similar to Figure \ref{Fig:aspect_rat_vs_temp}, however, we replace $R_{0}$ with $q$. Doing so, we find that regardless of the stellar temperature (in contrast to the findings in \citealt{MacGregor16}), $q$ appears to increase with the vertical aspect ratio. Measuring the statistical significance of the correlation between the aspect ratio and $q$, we derive a Pearson correlation coefficient of 0.6 with a p-value of 0.05. When removing $\beta$ Pic, which appears to be an outlier, the Pearson correlation coefficient increases to 0.7 with a p-value of 0.01. These values show that the correlation between aspect ratio and $q$ is significant, however, it is important to keep in mind that our sample size is small.

 A steep $q$-value suggests a large population of the smallest dust grains in the system, and as mentioned before, can be a sign of a giant hypervelocity collision between planetesimals. The two disks with the largest $q$-values are HD 114082 and unsurprisingly, HR 7012, both which have large vertical aspect ratios. While HR 7012 is highly suspected to have a recent giant impact, the same is not true for HD 114082. Unlike the HR 7012 disk, the HD 114082 disk has no significant amount of gas detected \citep{Kral20}. Additionally, past studies of the disk have found a relatively large minimum dust grain size of between 5-10 $\mu$m \citep{Engler22,Wahhaj16}, which is larger than the expected blowout size of 2.4 $\mu$m and is supported by our findings of the disk being neutral in color, again inconsistent with a giant impact scenario. Other studies have shown that a steep $q$-value (between $\sim$3.65 and 4) can simply be the result of collisions between similar sized bodies in the strength-regime \citep{Pan12}, meaning that the collisional bodies are held together by their own material strength rather than by gravity. Analytical and numerical calculations indicate that rocky bodies do not become dominated by self gravity until they reach a size of $\sim$1 km \citep{Wyatt11}, suggesting that collisions in these two disks are primarily between smaller bodies. This is expected for HR 7012, as the fine dust is expected to be from the sub-sequential collisions between sub-mm size dust grains rather than the initial giant impact \citep{Johnson12}. 
 
 When studying the aspect ratio of our sample of debris disks, there are clear trends that have emerged. The stellar temperature, the disk's radial extent, and distribution of dust grain sizes, all appear to affect the vertical aspect ratio. Further study is needed to explore the relationship between the vertical aspect ratio and these other system parameters, in order to help better understand the processes that are occurring in these disks.  

\begin{table}
        \centering
	\caption{\label{disk_q} Measured grain size power law index, $q$, values with uncertainties for each disk listed, taken from the literature.}
	\begin{tabular}{ccc}
	    \hline
	    \hline
		Disk & $q$ & Reference \\
		\hline
            AU Mic & $<$3.33 & \citet{Lohne20} \\
            $\beta$ Pic & $3.49\pm0.06$ & \citet{Lohne20} \\
            HD 32297 & $3.07\pm0.12$ & \citet{Norfolk21} \\
            HD 35841 & $2.90^{+0.10}_{-0.20}$ & \citet{Esposito18} \\
            HD 61005 & $3.33\pm0.04$ & \citet{Lohne20} \\
            HD 106906 & $3.19^{+0.11}_{-0.20}$ & \citet{Crotts21} \\
            HD 114082 & $>$3.9 & \citet{Wahhaj16} \\
            HD 115600 & $3.65\pm0.15$ & \citet{Thilliez17}\\
            HD 131835 & $3.13\pm0.07$ & \citet{Lohne20} \\
            HD 157587 & $3.73^{+0.81}_{-0.08}$ & \citet{Bruzzone18} \\
            HR 4796 A & $3.43\pm0.06$ & \citet{Lohne20} \\
            HR 7012 & $3.95\pm0.10$ & \citet{Johnson12} \\
            \hline
            \hline
        \end{tabular}
\end{table}

\subsection{Polarized Intensity Profiles} \label{sec:pol_profiles}
One interesting observation when comparing the surface brightness profiles shown in Figure \ref{Fig:multi_sb} side by side, is the similarity between the profile shapes for the disks in our sample. For the majority of higher inclined disks, the surface brightness profiles peak at separations closer to the star before gradually decreasing with increasing stellar separation. In the cases of HD 32297 and HD 106906, the surface brightness profile peaks closest to the star, with a second, smaller peak at larger stellar separations. Several disks have more flat surface brightness profiles such as AU Mic, HD 61005 and HD 30447, although AU Mic and HD 61005 extend beyond GPI's FOV. For lower inclined disks, again all the surface brightness profiles are very similar in that the surface brightness gradually decreases from the star before peaking again at the disk ansae. One outlier is HD 191089, where the surface brightness stays fairly flat with separation from the star. However, this disk is relatively low in S/N compared to the other lower-inclined disks in our sample. These surface brightness profiles can provide information about the disk SPF, suggesting that the SPF is very similar between disks. Other studies have also made this observation (e.g., \citealt{Hughes18}) when comparing the SPF of several debris disks, solar system comets, and zodiacal dust. In another example, \citet{Hom23} find that by using the same generic SPF, derived from the SPF of bodies in our solar system (i.e. the rings of Saturn/Jupiter and multiple comets), they were able to achieve low residual models for multiple debris disks a part of our GPI sample, further supporting a universal SPF. Such similarity of the SPF between debris disks and zodiacal dust implies that the dust in the majority of debris disks are porous aggregates, such as with cometary dust.

\subsection{Sources of Disk Morphologies} \label{sec:disk_morph}
While we cannot make any definitive statements of whether or not planets exist in some of these systems without direct detection of said planets, we can take all of our analysis and results for each disk to help determine which scenario the disk morphology is most consistent with, whether that be interaction with a companion or another mechanism. To do so, it is important to understand how different mechanisms affect the disk in different ways.

Planets by themselves can affect the disk morphology in numerous ways. This can be seen in studies such as \citet{LC16}, where they show that a single 10 M$_{\oplus}$ planet on an eccentric orbit can create multiple different morphologies observed in multiple debris disks, such as ``the Needle" and ``the Moth". In addition to these outcomes, planets can create other features such as eccentric disks, brightness asymmetries, gaps, rings, and warps. If a planet lies close the disk edge, it can also effectively stir the disk as discussed in the previous Section. For example, \citet{Pearce22} uses disk stirring along with disk sculpting arguments to predict the masses of potential planets in a large sample of debris disks. It should be noted that interactions with stellar companions (if present), as well as stellar flybys can also perturb debris disks similarly to planets. While planets can effectively sculpt debris disks, it is unclear whether or not a significant color asymmetry would appear solely as a result of planet-disk interactions, although such interactions may result in additional collisions, populating the disk with small grains, and would change the scattering angles in the case of induced eccentricity on the disk. 

For disks with a significant disk color asymmetry, other mechanisms may explain what is happening in the disk. Two mechanisms that have been commonly used to explain perturbed disks are interactions with the ISM \citep{Debes09} and large scale collisions in the disk \citep{Jackson14}. Both these scenarios are able to alter the distribution of dust grains which could cause a disk color asymmetry. In the case of an ISM interaction, if the disk passes through a dense region of the ISM, this can cause preferentially small dust grains to be blown out in the opposite direction of the system's motion. If small grains are redistributed from one side of the disk to the other, this can cause one side of the disk to become brighter and bluer in color than the other, especially at shorter wavelengths. Additionally this can cause the bluer side to also become more radially extended and create a ``Needle" or ``Moth" like morphology depending on the viewing angle. On the other hand, recent large impacts can generate a large amount of small dust grains at the site of collision. These dust grains are put on highly eccentric orbits, making the opposite side of disk more radially extended, while the collision site becomes a pinched point through which the orbits of all the dust grains must pass \citep{Jackson14}. Such an event could cause the side of the disk where the collision occurred to become more blue (due to a concentration of small dust grains) as well as become significantly brighter than the opposite side. For the three disks with brightness asymmetries and color asymmetries (HD 61005, HD 110058 and HD 157587) the brighter side of the disk is also bluer compared to the dimmer extension, as would be expected for either a ISM or large impact scenario.

\begin{figure*}[t!]
\centering
	\caption{\label{Fig:bright_asymm_vs_offset} Here we show the average brightness asymmetry across all bands vs. the measured major-axis offset or $\delta_{x}$ in au. Dark blue data points are disks with multiwavelength observations, while light blue data points are disks with $H$ band observations only. Orange shaded regions represent the parameter space where disks have brightness asymmetries that are consistent with the direction of the major-axis offset.}
	\includegraphics[width=\textwidth]{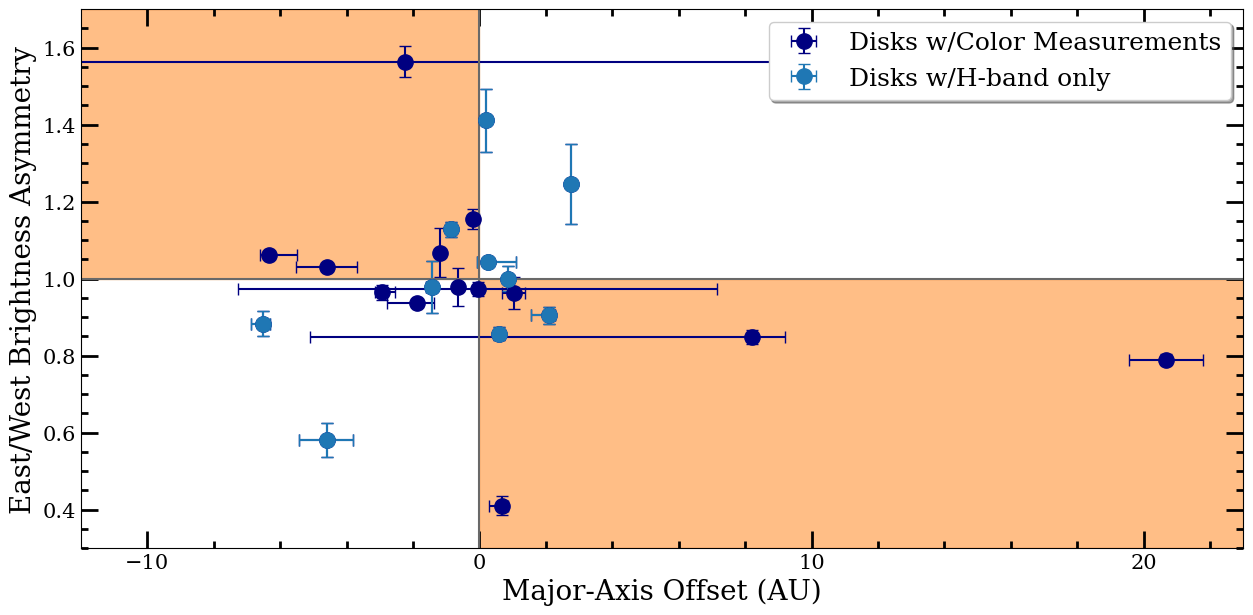}
\end{figure*}

To visually summarize our findings of asymmetries found for each disk, we plot the average brightness asymmetry between all bands, measured in Section \ref{sec:sb}, as a function of the offset found along the major axis shown in Figure \ref{Fig:bright_asymm_vs_offset}. The orange shaded regions represents the area of parameter space where the brightness asymmetry is consistent with the direction of the major-axis offset (i.e. a brighter West side should be closer to the star and vice versa) as in the case of an eccentric disk. We find the majority of disks have brightness asymmetries as expected for an eccentric disk, although there are a handful of disks that have brightness asymmetries that are not consistent with an eccentric disk. A majority of the inconsistent disks do not have multiwavelength observations.  

Going a step further, we place each disk into one of 6 categories based on their brightness asymmetry, major-axis offset, whether or not the brightness asymmetry is consistent with the offset direction, expected brightness asymmetry based on the offset, and the disk color asymmetry. This information can be found in Table \ref{tab:asymm_sum}, and will be discussed further in the subsequent Sections. To calculate the expected brightness asymmetry, we use the relationship between the surface brightness and radius from the star (i.e. $1/r^{2}$). However, given that the scattering angles change when the disk is offset from the star, the disk SPF also affects the expected brightness asymmetry. With this in mind, we also calculate the contribution from the SPF using the generic SPF derived in \citep{Hom23}. This is an approximate estimation as the SPF of the debris disks in our sample may not necessarily conform to this generic SPF, such as the case with HR 4796 A (see Figure 6 in \citealt{Hughes18}), although, as mentioned in Section \ref{sec:pol_profiles}, the similarities between surface brightness profiles suggest this is a fair assumption. In general, the effect of the SPF partially cancels out the expected brightness asymmetry based on $1/r^{2}$ alone, as the opposite side of the disk (apocenter) becomes brighter due to the change in scattering angles. Our approximate estimation of the expected brightness asymmetry in Table \ref{tab:asymm_sum} is represented as a range between the expected brightness asymmetry based on $1/r^{2}$ alone, and when taking into account the contribution from the SPF. For a fair comparison, we recalculate the surface brightness asymmetry for each disk (and each band) at the same radii as we calculate for the expected brightness asymmetry, focusing on stellar separations mid way between the star and the measured disk radius to avoid the effects of limb brightening at the ansae and noise close to the star. The average surface brightness asymmetry can be found in Table \ref{tab:asymm_sum}. For simplicity, we focus on the major-axis offsets to calculate the expected brightness asymmetries.   

\begin{table*}[h!]
	\centering
	\caption{\label{tab:asymm_sum}Summary of asymmetry for each disk. Column three is the brightness asymmetry measured in the $H$ band, while column 5 is the range of expected brightness asymmetries based on the $1/r^{2}$ relationship and the SPF. Column 6 is the average disk color asymmetry for each disk. The table is organized by disks with similar asymmetries or features. See Section \ref{sec:disk_morph} for descriptions of each Category.}
	\begin{tabular*}{\textwidth}{ p{3cm} p{2cm} p{2.8cm} p{2.8cm} p{3.5cm} p{2cm}}
        \\
	    \hline
	    \hline
		Name & $\delta_{x}$ (au) & \centering Brightness Asymmetry & \centering Consistent w/ Offset Direction? & \centering Expected Brightness Asymmetry & \centering Disk Color Asymmetry \cr \\
		\hline
            \textbf{Category 1} \\
            \hline
            HD 32297 & -4.51 & 1.13$\pm$0.05 & Yes & 1.03-1.08 & 0.03$\pm$0.04 \\
            HD 106906 & 20.67 & 1.28$\pm$0.04 & Yes & 1.17-1.43 & 0.09$\pm$0.05 \\
            HD 146897 & -6.31 & 1.20$\pm$0.15 & Yes & 1.10-1.25 & 0.05$\pm$0.03 \\
            HD 156623 & 2.24 & 1.05$\pm$0.02 & Yes & 1.03-1.07 & N/A \\
            HR 4796 A & 0.58 & 1.02$\pm$0.02 & Yes & $\le$1.01 & N/A \\
            \hline
            \hline
            \textbf{Category 2} \\
            \hline
            HD 61005 & 0.69 & 1.82$\pm$0.09 & Yes & 1.01-1.03 & 0.70$\pm$0.15 \\
            HD 110058 & 7.80 & 1.23$\pm$0.03 & Yes & 1.11-1.28 & 0.28$\pm$0.04 \\
            HD 111520 & -2.24 & 1.78$\pm$0.09 & Yes & 1.02-1.05, 1.09-1.25 & 0.15$\pm$0.06 \\
            HD 117214 & -0.19 & 1.14$\pm$0.05 & Yes & $\le$1.01:1 & N/A \\
            \hline
            \hline
            \textbf{Category 3} \\
            \hline
		$\beta$ Pic & 0.0 & 1.04$\pm$0.02 & - & - & N/A \\
		CE Ant & -0.86 & 1.01$\pm$0.02 & Yes & 1.03-1.05 & N/A \\
            HD 115600 & 0.0 & 1.01$\pm$0.05 & - & - & 0.0$\pm$0.04 \\
            \hline
            \hline
            \textbf{Category 4} \\
            \hline
            HD 114082 & -2.95 & 1.14$\pm$0.04 & No & 1.08-1.21 & 0.19$\pm$0.07 \\
            HD 129590 & -1.91 & 1.09$\pm$0.06 & No & 1.03-1.08 & 0.10$\pm$0.06 \\
            HD 157587 & -0.65 & 1.18$\pm$0.05 & No & $\le$1.01 & 0.30$\pm$0.08 \\
            \hline
            \hline
            \textbf{Category 5} \\
            \hline
            AU Mic & 0.19 & 1.95$\pm$0.20 & No & 1.01-1.04 & N/A \\
            HD 30447 & -6.53 & 1.19$\pm$0.08 & No & 1.06-1.17 & N/A \\
            HD 131835 & -4.54 & 1.11$\pm$0.06 & No & 1.04-1.10 & N/A \\
		HR 7012 & 2.74 & 1.94$\pm$0.49 & No & 1.32-1.87 & N/A\\
            \hline
            \hline
            \textbf{Category 6} \\
            \hline
            HD 35841 & 1.03  & 1.08$\pm$0.13 & Yes & 1.02-1.05 & 0.04$\pm$0.08 \\
            HD 111161 & -1.09 & 1.04$\pm$0.07 & No & 1.01-1.03 & N/A \\
		HD 145560 & 0.86 & 1.02$\pm$0.05 & - & $\le$1:1.01 & N/A \\
            HD 191089 & 1.20 & 1.08$\pm$0.12 & Yes & 1.02-1.05 & 0.07$\pm$0.12 \\
		\hline
		\hline
	\end{tabular*}
\end{table*}

\subsubsection{Category 1: Eccentric Disk}
In this first category, the debris disks are consistent with having an eccentric disk. This means that the derived brightness asymmetries are consistent with the direction of the major-axis offset, the expected brightness asymmetry is consistent with the measured asymmetry, and finally these disks do not present a significant disk color asymmetry. There are five disks that fall into this category; HD 32297, HD 106906, HD 146897, HD 156623 and HR 4796 A. 

While HD 32297 is close to axisymmetric, we place it in category 1 as we derive a significant offset of $\sim$4 au, which is present in both the $J$ and $H$ band. The derived offset is also still consistent with the insignificant brightness asymmetry given the large disk radii. On the other hand, the HD 106906 disk is very asymmetric, with a massive disk offset along the major-axis of $\sim$20 au and a significant brightness asymmetry. Despite such a large disk offset, the measured brightness asymmetry is still consistent within the range calculated for the expected brightness asymmetry. If confirmed, the HD 146897 disk also has a large offset relative to the derived disk radius, making it one of the more eccentric disks in our sample. Given the small disk radii, measuring the brightness asymmetry between the star and the disk radius requires us to average the disk surface brightness close to the star, resulting in a high uncertainty measurement of 1.20$\pm$0.15 (i.e. the West side of the disk is 1.20$\pm$0.15 times brighter than the East side). Despite this high uncertainty, the expected brightness asymmetry of $\sim$1.10-1.25 is consistent with the measured brightness asymmetry.

For both the HD 156623 and HR 4796 A disks, we derive brightness asymmetries of 1.05$\pm$0.02 and 1.02$\pm$0.02 between the star and disk radius. For both disks the expected brightness asymmetries derived at the same radii are consistent with these measured brightness asymmetries within 1$\sigma$. These two values are significantly lower than the brightness asymmetries measured across the entire disk in Section \ref{sec:sb} (1.11$\pm$0.02 and 1.17$\pm$0.02, respectively). This is due to the brightness asymmetry being strongest near the disk ansae for both disks, as can be seen in Figure \ref{Fig:multi_sb}. For the HR 4796 A disk, \citet{Olofsson19} found that with the derived eccentricity of $\sim$0.02, their model was unable to match the surface brightness at the ansae, leading to the conclusion that dust may be released preferentially near the East disk ansae due to more frequent collisions. A similar scenario could be the case for HD 156623, although more complex modelling may find that our derived eccentricity is sufficient enough to produce the brightness asymmetry along the entire disk.

A common explanation used to explain an eccentric disk is perturbation from an eccentric planet. For the HD 32297 debris disk, \citet{LC16} have shown that a planet on an eccentric orbit can create the double wing feature seen in the disk halo with HST. This requires the azimuth of the planet to be close to 0$^{\circ}$, which can explain why the disk appears close to axisymmetric. For the planet to sculpt the inner edge of each disk, \citet{Pearce22} derives a minimum planet mass and maximum separation of $1.1^{+0.4}_{-2.0}$ M$_{\text{Jup}}$ and $70^{+8}_{-2}$ au for HD 32297, and $2.0\pm0.4$ and $43^{+7}_{-9}$ for HD 146897, however, this is not including the measured eccentricity from this study. The radially narrow ring of the HR 4796 A disk may be the result of a shepherding planet inside the planetesimal belt, as described in \citet{Olofsson19}. However, in the case of HD 156623, no inner clearing is observed as the polarized intensity is detected down to the FPM, meaning that either a planet is shaping the disk from within the FPM or outside the disk, or another source is causing the disk to become eccentric. Using an orbital separation of 10 au, \citet{Pearce22} finds a planet mass of 0.6 M$_{\text{Jup}}$ is required to sculpt the disk. Given that the HD 156623 disk is gas rich, if the gas disk is eccentric, this can force the dust disk to become eccentric as well \citep{LC19}, a scenario that has also been used to help explain the moth-like wings of HD 32297. However, such a scenario still requires a perturber, such as a planet, to make the gas disk eccentric. 

The HD 106906 system is the only one in this category with a known planet. Past studies have shown that perturbation from the planet HD 106906 b can replicate the observed disk morphology and has a consistent orbit \citep{Nesvold17,Nguyen21,Moore23}. Other studies have show that the disk morphology is also well created by a recent catastrophic collision taking place in the disk's East extension \citep{Jones23}, making it an alternative scenario for the disk asymmetries, although no other evidence of a large collision has been found.

\subsubsection{Category 2: Eccentric Disk + Additional Explanation Needed}
Category 2 consists of debris disks in our sample that are consistent with an eccentric disk, but either have a significant color asymmetry in at least one or more bands and/or have measured brightness asymmetries much larger than expected. The four disks that fall under this category are HD 61005, HD 110058, HD 117214 and HD 111520. 

The HD 61005 and HD 111520 disks have two of the largest brightness asymmetries in our sample; however, the estimated major-axis offsets are too small to explain these large brightness asymmetries. Even if we take the estimated 11 au offset based on the polarized surface brightness profile for HD 111520 \citep{Crotts22}, this only creates a brightness asymmetry of $\sim$1.09-1.25 (compared to the measured 1.78$\pm$0.09 averaged between bands). It is possible that the expected brightness asymmetry would change when taking into account the full eccentricity (i.e. if both the offset along the major- and minor-axis were well constrained), as well as the argument of pericenter, although this is difficult to do empirically for such high inclined disks. In the case of HD 61005, the ``moth"-like halo suggests an argument of pericenter close to 0, meaning that the disk eccentricity would be primarily along the minor-axis, and therefore should not cause a large brightness asymmetry along the major-axis. The halo of the HD 111520 disk shows more radial asymmetry including the warp, ``fork"-like structure, and difference radial extent, suggesting that the argument of pericenter is much farther from zero, and that an offset along the major-axis is required. Dynamical modelling of the system may help to uncover the true orientation of the disk in order to create such asymmetries. 

In \citet{Jones23}, the authors try to explain the morphology for both disks with a recent giant collision; however, neither disk is fully consistent with this scenario. In the case of HD 111520, while a giant collision can create a fork like structure as observed, the orientation of the fork is incorrect, where the micron sized grains align with the lower fork rather than between the two forks as would be expected. We would also expect the site of the collision (i.e. the East extension) to be the brighter side, whereas we observe the opposite. For HD 61005, while a large collision can create the moth-like structure of the disk halo, the authors of \citet{Jones23} note that the brightness ratio between the two sets of disk wings is incorrect, as well as the secondary wings are not as straight as seen in observations. The disk halo morphology may be better explained by interaction with a planet companion on an eccentric orbit, where \citet{LC16} show that such a planet can create the ``moth" and ``bar" like morphologies, although the ``bar" morphology requires a steep dust grain size distribution close to the blow-out size. HD 111520 disk's morphology may also be explained by a planet-disk interaction, as the halo shows a clear 4$^{\circ}$ warp beyond 1.7$''$ \citep{Crotts22}, a planet-disk interaction may also be able to create the fork-like structure \citep{PW14}. 

An interaction with the ISM is another mechanism that may be affecting either disk. This is a scenario that has been used to help explain HD 61005, and has been shown to be able to create both a moth- and needle-like morphology \citep{Maness09, Debes09}. Given that HD 61005 proper motion (corrected for solar reflex motion) is near perpendicular to the disk wings, this may be another explanation for the disk's morphology. Additionally, the proper motion also points slightly more West compared to the major-axis ($\sim$19.2$^{\circ}$ from perpendicular), which may be able to explain the disk color asymmetry, which is found to be significant in the $J$ and $H$ bands, although further study is needed to confirm this. On the other hand, to create the more needle-like morphology of HD 111520's disk halo, the proper motion should be pointing away from the West extension. However, after correcting for solar reflex motion, the proper motion also near perpendicular to the disk, essentially ruling out this scenario.  

Similar to the HD 61005 disk, the HD 117214 disk also has an insignificant offset along the major-axis (along with the minor-axis), while having a significant brightness asymmetry. It is unclear from our data alone what the source of this brightness asymmetry is, as we are unable to perform a multiwavelength analysis for this disk. A deeper analysis of the polarized-intensity data alongside the total-intensity observations (presented in \citealt{Esposito20}) may help shed additional information about the disk morphology as a whole. Finally, the HD 110058 disk is one of the most asymmetric in our sample, with a large brightness asymmetry, possible eccentricity, disk color asymmetry and warp. While the disk offset may be the result of an asymmetric disk geometry due to the warp, especially given that past studies find no eccentricity, the warp itself suggests that the disk is being perturbed by a planet companion. If the disk is eccentric, the expected brightness asymmetry is consistent with the observed brightness asymmetry. However, the disk also has a significant disk color asymmetry in $J$-$H$ and $J$-$K1$, where the East extension is relatively more blue than the West extension, although the strong overall red disk color suggests a larger minimum dust grain size, on the order of $\gtrsim$1 $\mu$m \citep{Boccaletti03}. Further analysis of recently published HST observations \citep{Ren23} may provide additional information.

\subsubsection{Category 3: Additional Geometrical Asymmetries}
Category 3 contains debris disks that have other geometrical asymmetries (rather than an eccentric disk) that may be contributing to their surface brightness asymmetries or are signs of dynamical perturbation from a companion. This category includes $\beta$ Pic, CE Ant, and HD 115600.

In the case of $\beta$ Pic and CE Ant, other morphological asymmetries may be responsible for the brightness asymmetries observed. For $\beta$ Pic, a massive clump of gas and dust resides in the brighter West extension. While this clump mainly resides outside of GPI's FOV, at $\sim$52 au, the brightness asymmetry caused by the clump may extend within GPI's FOV as \citet{Han23} show the clump to extend down to $\sim$35 au (at the edge of GPI's FOV). \citet{Han23} also show that this clump is likely stationary, which is consistent with a recent giant impact scenario. For CE Ant, as mentioned previously and as seen in Figure \ref{Fig:ce_ant_spiral}, the disk contains a spiral arm in the SW quadrant. As our chosen apertures partially cover this area, it is possible that the extra flux from the spiral arm is contributing to the observed brightness asymmetry as measured in Section \ref{sec:sb}. When measuring the surface brightness only at radial separations halfway between the star and the disk radius, we find no significant brightness asymmetry. The existence of this spiral arm also suggests the presence of a planet companion, which hopefully could be imaged in future observations such as with JWST.

The HD 115600 disk is close to axisymmetric with no measured offset, significant brightness asymmetry or significant color asymmetry. However, after measuring the vertical offset profile, we find a tentative warp beyond $\sim$0.45$''$, where the East extension bends downwards and the West extension bends upwards. This makes the HD 115600 disk similar to the HD 110058 disk, which has a confirmed warp, and suggests that a planet on an inclined orbit, relative to the disk, is present in the system. Better resolved observations of the disk, or observations of the disk halo such as with HST, will be useful to confirm the existence of the warp.

\subsubsection{Category 4: Inconsistent with Eccentricity + Color Measurements}
Category 4 includes debris disks that are inconsistent with an eccentric disk, meaning that their brightness asymmetries are not consistent with the direction of the major-axis offset. Additionally, these disks have multiwavelength observations which allow us to perform color measurements to see whether or not any asymmetries in the disk color are present. The three disks that fall in this category are HD 114082, HD 129590, and HD 157587. 

The HD 114082 and HD 129589 disks are similar in that they both present significant brightness asymmetries in the $K1$ band (where the East side is brighter than the right), but not in the $H$ band. Additionally, in both cases, the disk offsets derived from our geometrical fitting support an offset along major-axis in the opposite direction as would be expected to create the observed brightness asymmetries. Given this discrepancy, neither disk has strong disk color asymmetries with significance above 3$\sigma$, making it unclear what is causing the brightness asymmetry specifically in the $K1$ band for either disk. For the overall disk color, both disks exhibit a neutral to red disk color in $H$-$K1$, meaning the disk is brighter at longer wavelengths, and suggests that the minimum dust grain size in these systems are on the order of a few microns or larger. As discussed previously, the estimated minimum dust grain size for HD 114082 is found to be between $\sim$5-10 $\mu$m \citep{Engler22, Wahhaj16}, which is consistent with the near neutral disk color observed based on calculations from \citet{Boccaletti03}. 

Unlike the previous two disks, the HD 157587 disk has a significant color asymmetry, most notably in $J$-$K1$, where the brighter East extension is relatively more blue than the West extension. This color asymmetry is only significant in $J$-$K1$, due to the fact that we only measure a significant brightness asymmetry in the $J$ band, where $H$ and $K1$ observations are consistent with being axisymmetric within 3$\sigma$. This could suggest that the smallest grains in the system may be perturbed, whereas the larger grains are less so. Such a phenomenon could be the result of an ISM interaction, although the proper motion of the system, after correcting for solar reflex motion, is pointing towards the perpendicular relative to the disk major-axis when we would expect it to be pointing more towards the West extension. Shorter wavelength observations, such as with HST, would be useful to determine if there are any structures in the disk halo that could help distinguish the source of the color asymmetry in the HD 157587 disk. 

\subsubsection{Category 5: Inconsistent with Eccentricity + No Color Measurements}
Category 5 contains disks that have measured major-axis offsets and/or brightness asymmetries, but the disk offset is in the opposite direction as expected to create the measured brightness asymmetry. Additionally, these disks do not have multiwavelength observations to show whether or not they have disk color asymmetries. This category includes the following disks: AU Mic, HD 30447, HD 131835, and HR 7012. 

For the disks AU Mic, HD 30447, HD 131835, and HR 7012, all four have either tentative (in the case of HR 7012) or significant brightness asymmetries, although the measured offset is in the opposite direction from what would be expected for an eccentric disk. For AU Mic, the low S/N of the data and spatial scale of the disk make it unfeasible to measure an offset accurately. However, there is evidence that suggests AU Mic may have impacted by a recent catastrophic collision resulting in the fast moving ripples that have been observed \citep{Chiang17}. HR 7012 is another disk that has likely experienced a recent catastrophic collision, as discussed in the previous sections. Surprisingly, with our GPI observations, we derive a disk offset that leads to a very large eccentricity, while there is also the possibility of a large brightness asymmetry, albeit it is still consistent with no asymmetry within 2$\sigma$. If this brightness asymmetry does exist, it would not be consistent with the direction of the derived offset, nor is it consistent with the expected brightness asymmetry. Given that previous observations with SPHERE show the disk to be axisymmetric, the asymmetries seen with GPI may be simply due to unremoved noise close to the FPM.

Both the HD 30447 and HD 131835 disks have significant brightness asymmetries, with a brighter East extension, but both also have derived offsets that suggest that the West extension is closer to the star. The HD 30447 disk is an interesting case, as the peak polarized intensity occurs close to the star in the East compared to the right, suggesting that the East extension could in fact be closer to the star. Given the low S/N of the $H$-band observations, we may not be able to accurately measure the major-axis offset of the disk. It is also possible that the measured offset is due to a geometrical asymmetry other than an eccentric disk. New HST observations show the geometry of the disk halo to also be asymmetric, where the East extension extends farther radially compared to the West extension \citep{Ren23}. A more in depth analysis of the HST observations, and possibly future higher resolution observations of the disk in the NIR will be useful to better constrain the disk geometry. The case for HD 131835 is similar. The GPI observations have fairly low S/N, making it more difficult to measure the disk geometry. Given the evidence for multiple rings and a broad parent disk, it is also possible that the disk geometry is more complicated than can be captured with our simple ring model. Again, higher resolution observations in scattered light will be useful in better constraining the full disk geometry.

\subsubsection{Category 6: Most Axisymmetric}
This final category simply consists of disks that are the most axisymmetric and do not have any strong evidence of harbouring asymmetries. The disks that fall into this category are HD 35841, HD 111161, HD 145560 and HD 191089. While there are small measured offsets for HD 111161 and HD 145560, these are insignificant taking into account the lower S/N of these two observations. The HD 35841 and HD 191089 disks also have measured offsets, but again, these offsets are very small ($\lesssim$1 au). Additionally none of these disks have significant brightness asymmetries, while the HD 35841 and HD 191089 disks also have no significant color asymmetries. Despite being near axisymmetric, three out of four disks have clear cavities within their inner radii, while the HD 35841 disk also appears to have an inner cavity in total-intensity (see \citealt{Esposito20}), meaning that these disks may still be carved by planets, and are worth following up with instruments such as JWST that have the ability to find planets in these systems.

\subsubsection{Summary of Planet-Disk Interactions}
In this Section we briefly summarize which disks may be perturbed and/or shaped by planets based on our findings and results from previous studies. 

Both Category 1 and 2 disks have disk morphologies consistent with planet-disk interactions. All nine disks are consistent with being eccentric, which can be caused by a planet on an eccentric orbit. Several of these disks also exhibit other morphological features that are consistent with planet-disk interactions. For example, both HD 110058 and HD 111520 have confirmed warped disks \citep{Kasper15,Crotts22}, which can be caused by a planet on an inclined orbit relative to the disk. Several of these disks also have disk halos consistent with perturbation from an eccentric planet, in the case of the ``Moth" and ``Needle" like halos of HD 32297, HD 61005, HD 106906 and HD 111520. However, other mechanisms can cause these morphologies as well, such as an ISM interaction or a recent giant impact. There is also the question of whether or not any asymmetries, such as eccentricity, may be carried over from the protoplanetary disk phase, as debris disks are thought to be progenitors of high mass, structured protoplanetary disks \citep{Michel21}. While disks in Category 1 can be solely explained by an eccentric disk, disks in Category 2 (HD 61005, HD 110058, HD 111520, HD 117214) need further investigation to understand their higher than expected surface brightness asymmetries and/or their disk color asymmetries. 

Not every disk in our sample may be eccentric or highly asymmetric, but many disks in our sample still exhibit cavities within their inner radii, showing that the material in this region has been cleared by some mechanism. This includes disks from all categories, including HD 114082, HD 129590, HD 157587, CE Ant, HD 30447, HD 117214, HD 131835, HR 4796 A, HD 111161, HD 145560, and HD 191089, where CE Ant and HD 131835 also harbour multiple rings.. These disks make good candidates for searching for planets within their cavities, such as with JWST. Additionally, the HD 114082 disk is radially compact, suggesting that the disk may be sculpted by a planet orbiting outside of the disk as well. Two disks from the above list with other possible planet-induced features are CE Ant and HD 30447, where CE Ant harbours a spiral arm, tentatively detected in our GPI data, while HD 30447 has a needle-like halo as seen in HST observations presented in \citep{Ren23}. Finally, two additional disks that have evidence of perturbation from a planet are $\beta$ Pic and HD 115600. Both disks are fairly axisymmetric as observed with GPI, although $\beta$ Pic is known to have multiple asymmetries on larger scales, which have been shown to be connected to the known planets in the system. Similar to $\beta$ Pic, as well as HD 110058 and HD 111520, HD 115600 have a tentative warp towards its outer edges as seen with GPI, which if confirmed suggests perturbation from an unseen planet with an inclination relative to the disk. 

In summary, the majority of disks show at least one sign of planet-disk interactions through their morphologies, such as inner gaps, eccentric disks, spiral arms and warps. A number of disks in our sample also have brightness asymmetries, some of which are not consistent with an eccentric disk, along with asymmetries in disk color, showing that other mechanisms may be shaping debris disks as well. Future observations with new and upcoming instruments will hopefully have the potential to detect the disk-perturbing planets in these systems.  

\section{Conclusions}

In this study we present a uniform, empirical analysis of 23 GPI debris disks in polarized intensity, using multiwavelength $J$-, $H$-, and $K1$-band observations. Through this analysis, we fully characterize each disk morphology through measuring the disk geometry, vertical/radial width, multiwavelength surface brightness, and disk color. We also derive any asymmetries present in the disk by measuring disk offsets along the major and minor axes, as well as asymmetries in the surface brightness and disk color. While we analyze each disk individually, we also come to the following broader conclusions;

\begin{itemize}
  \item The majority of our disks present at least one asymmetry. For example, we find 16 out of 23 disks present a significant brightness asymmetry of at least 3$\sigma$ between the East and West extensions in at least one band. Out of the disks with multiwavelength data, 3 out of 12 disks also present a significant disk color asymmetry between the East and West extensions between at least one pair of bands. Additionally, we confirm the warp for the HD 110058 disk, while finding a tentative warp for the HD 115600 disk.
  
  \item Comparing the surface brightness and disk color asymmetries with stellar temperature and age, we find no significant trends. We do, however, find a tentative trend between the overall disk color and stellar temperature, where the disk color becomes increasingly red/grey as the stellar temperature increases. We find this trend strongest in $H$-$K1$, where several disks around hotter stars are strongly blue in $J$-$H$ and $J$-$K1$, breaking the trend. This can be explained by natural collisional evolution, where studies have found that bright debris disks around F and A spectral-type stars can naturally produce a high density of sub-micron sized dust grains. 

  \item We find 4 disks to have significantly higher vertical aspect ratios compared to other debris disks at similar inclinations. This includes the disks $\beta$ Pic, HD 110058, HD 114082 and HR 7012. Comparing the aspect ratios of our sample with other disk/system properties, we find that a combination of stellar temperature and disk radius are correlated with the vertical width, as the four disks mentioned above are all around higher temperature stars \textit{and} are radially compact with $R_{0} < 40$ au. While the estimation of $R_{0}$ is most uncertain for $\beta$ Pic, planet/stellar companions may be responsible for the radially compact nature of HD 110058, HD 114082, and HR 7012. Additionally, we find a positive correlation between the vertical aspect ratio and the dust grain size power law, $q$, where the vertical aspect ratio appears to increase with $q$. While a high $q$-value can be a sign of a giant hypervelocity collision, a high $q$-value can also be the result of collisions between similar sized bodies in the strength regime. Further analysis is needed to better understand the relationship between $q$ and the vertical aspect ratio. 

  \item Categorizing each disk based off their derived asymmetries, we find the following: 5/23 disks are consistent with an eccentric disk, 4/23 disks are consistent with an eccentric disk, but need further explanation for their higher than expected brightness asymmetries and/or disk color asymmetries, 3/23 disks have geometrical/surface brightness asymmetries not necessarily associated with an eccentric disk, 8/23 disks need further followup to determine the source of their brightness asymmetries, and 4/23 are most consistent with being axisymmetric.   
\end{itemize}

Disks that are consistent with an eccentric disk, or harbour another geometrical asymmetry such as a warp or spiral (about half of our sample), may be the result of planet-disk interactions. While not every disk in our sample is significantly asymmetric in terms of surface brightness and/or disk geometry, this does not necessarily mean there are no signs of sculpting by a planet. For example, at least 10 out of 23 disks also show clear gaps within their inner radii, suggesting that possibly one or multiple planets are present and clearing out the material in these regions. To summarize, almost every disk in our sample features at least one asymmetry, while the majority of disks also present an asymmetry or feature that is suggestive of perturbation of a planet companion. These disks provide great candidates to search for planets with current and upcoming instruments, such as JWST NIRCam, in order to better understand exoplanetary architectures and how they evolve over time. 

\begin{acknowledgements}
The authors wish to thank the anonymous referee for helpful suggestions that improved this manuscript. This work is based on observations obtained at the Gemini Observatory, which is operated by the Association of Universities for Research in Astronomy, Inc. (AURA), under a cooperative agreement with the National Science Foundation (NSF) on behalf of the Gemini partnership: the NSF (United States), the National Research Council (Canada), CONICYT (Chile), Ministerio de Ciencia, Tecnolog\'{i}a e Innovaci\'{o}n Productiva (Argentina), and Minist\'{e}rio da Ci\^{e}ncia, Tecnologia e Inova\c{c}\~{a}o (Brazil). This work made use of data from the European Space Agency mission \emph{Gaia} (\url{https://www.cosmos.esa.int/gaia}), processed by the \emph{Gaia} Data Processing and Analysis Consortium (DPAC, \url{https://www.cosmos.esa.int/web/gaia/dpac/consortium}). Funding for the DPAC has been provided by national institutions, in particular the institutions participating in the Gaia Multilateral Agreement. This research made use of the SIMBAD and VizieR databases, operated at CDS, Strasbourg, France. We thank support from NSF AST-1518332, NASA NNX15AC89G and NNX15AD95G/NEXSS. Portions of this work were also performed under the auspices of the U.S. Department of Energy by Lawrence Livermore National Laboratory under contract DE-AC52-07NA27344. KAC and BCM acknowledge a Discovery Grant from the Natural Science and Engineering Research Council of Canada.
\end{acknowledgements}

\facilities{Gemini:South}

\software{Gemini Planet Imager Pipeline (\citealt{Perrin14}, \url{http://ascl.net/1411.018)},
emcee (\citealt{FM13}, \url{http://ascl.net/1303.002}),
corner (\citealt{corner}, \url{http://ascl.net/1702.002}), 
matplotlib (\citealt{Hunter07}; \citealt{droettboom17}), 
iPython (\citealt{perez07}), 
Astropy (\citealt{Collaboration:2018ab}),
NumPy (\citealt{Oliphant_06}; \url{https://numpy.org}),
SciPy (\citealt{Virtanen_20}; \url{http://www.scipy.org/})}

\bibliography{cohesive_analysis.bbl}{}

\begin{thebibliography}{}
\expandafter\ifx\csname natexlab\endcsname\relax\def\natexlab#1{#1}\fi
\providecommand{\url}[1]{\href{#1}{#1}}
\providecommand{\dodoi}[1]{doi:~\href{http://doi.org/#1}{\nolinkurl{#1}}}
\providecommand{\doeprint}[1]{\href{http://ascl.net/#1}{\nolinkurl{http://ascl.net/#1}}}
\providecommand{\doarXiv}[1]{\href{https://arxiv.org/abs/#1}{\nolinkurl{https://arxiv.org/abs/#1}}}

\bibitem[{{Bailey} {et~al.}(2014){Bailey}, {Meshkat}, {Reiter}, {Morzinski},
  {Males}, {Su}, {Hinz}, {Kenworthy}, {Stark}, {Mamajek}, {Briguglio}, {Close},
  {Follette}, {Puglisi}, {Rodigas}, {Weinberger}, \& {Xompero}}]{Bailey14}
{Bailey}, V., {Meshkat}, T., {Reiter}, M., {et~al.} 2014, \apjl, 780, L4,
  \dodoi{10.1088/2041-8205/780/1/L4}

\bibitem[{{Bell} {et~al.}(2015){Bell}, {Mamajek}, \& {Naylor}}]{Bell15}
{Bell}, C. P.~M., {Mamajek}, E.~E., \& {Naylor}, T. 2015, \mnras, 454, 593,
  \dodoi{10.1093/mnras/stv1981}

\bibitem[{{Bhowmik} {et~al.}(2019){Bhowmik}, {Boccaletti}, {Th{\'e}bault},
  {Kral}, {Mazoyer}, {Milli}, {Maire}, {van Holstein}, {Augereau}, {Baudoz},
  {Feldt}, {Galicher}, {Henning}, {Lagrange}, {Olofsson}, {Pantin}, \&
  {Perrot}}]{Browhmik19}
{Bhowmik}, T., {Boccaletti}, A., {Th{\'e}bault}, P., {et~al.} 2019, \aap, 630,
  A85, \dodoi{10.1051/0004-6361/201936076}

\bibitem[{{Boccaletti} {et~al.}(2003){Boccaletti}, {Augereau}, {Marchis}, \&
  {Hahn}}]{Boccaletti03}
{Boccaletti}, A., {Augereau}, J.~C., {Marchis}, F., \& {Hahn}, J. 2003, \apj,
  585, 494, \dodoi{10.1086/346019}

\bibitem[{{Boccaletti} {et~al.}(2018){Boccaletti}, {Sezestre}, {Lagrange},
  {Th{\'e}bault}, {Gratton}, {Langlois}, {Thalmann}, {Janson}, {Delorme},
  {Augereau}, {Schneider}, {Milli}, {Grady}, {Debes}, {Kral}, {Olofsson},
  {Carson}, {Maire}, {Henning}, {Wisniewski}, {Schlieder}, {Dominik},
  {Desidera}, {Ginski}, {Hines}, {M{\'e}nard}, {Mouillet}, {Pawellek}, {Vigan},
  {Lagadec}, {Avenhaus}, {Beuzit}, {Biller}, {Bonavita}, {Bonnefoy},
  {Brandner}, {Cantalloube}, {Chauvin}, {Cheetham}, {Cudel}, {Gry}, {Daemgen},
  {Feldt}, {Galicher}, {Girard}, {Hagelberg}, {Janin-Potiron}, {Kasper}, {Le
  Coroller}, {Mesa}, {Peretti}, {Perrot}, {Samland}, {Sissa}, {Wildi}, {Zurlo},
  {Rochat}, {Stadler}, {Gluck}, {Orign{\'e}}, {Llored}, {Baudoz}, {Rousset},
  {Martinez}, \& {Rigal}}]{Boccaletti18}
{Boccaletti}, A., {Sezestre}, E., {Lagrange}, A.~M., {et~al.} 2018, \aap, 614,
  A52, \dodoi{10.1051/0004-6361/201732462}

\bibitem[{{Bruzzone}(2018)}]{Bruzzone18}
{Bruzzone}, J.~S. 2018, PhD thesis, University of Western Ontario, Canada

\bibitem[{{Cataldi} {et~al.}(2020){Cataldi}, {Wu}, {Brandeker}, {Ohashi},
  {Mo{\'o}r}, {Olofsson}, {{\'A}brah{\'a}m}, {Asensio-Torres}, {Cavallius},
  {Dent}, {Grady}, {Henning}, {Higuchi}, {Hughes}, {Janson}, {Kamp},
  {K{\'o}sp{\'a}l}, {Redfield}, {Roberge}, {Weinberger}, \&
  {Welsh}}]{Cataldi20}
{Cataldi}, G., {Wu}, Y., {Brandeker}, A., {et~al.} 2020, \apj, 892, 99,
  \dodoi{10.3847/1538-4357/ab7cc7}

\bibitem[{{Chauvin} {et~al.}(2012){Chauvin}, {Lagrange}, {Beust}, {Bonnefoy},
  {Boccaletti}, {Apai}, {Allard}, {Ehrenreich}, {Girard}, {Mouillet}, \&
  {Rouan}}]{Chauvin12}
{Chauvin}, G., {Lagrange}, A.~M., {Beust}, H., {et~al.} 2012, \aap, 542, A41,
  \dodoi{10.1051/0004-6361/201118346}

\bibitem[{{Chen} {et~al.}(2020){Chen}, {Mazoyer}, {Poteet}, {Ren},
  {Duch{\^e}ne}, {Hom}, {Arriaga}, {Millar-Blanchaer}, {Arnold}, {Bailey},
  {Bruzzone}, {Chilcote}, {Choquet}, {De Rosa}, {Draper}, {Esposito},
  {Fitzgerald}, {Follette}, {Hibon}, {Hines}, {Kalas}, {Marchis}, {Matthews},
  {Milli}, {Patience}, {Perrin}, {Pueyo}, {Rajan}, {Rantakyr{\"o}}, {Rodigas},
  {Roudier}, {Schneider}, {Soummer}, {Stark}, {Wang}, {Ward-Duong},
  {Weinberger}, {Wilner}, \& {Wolff}}]{Chen20}
{Chen}, C., {Mazoyer}, J., {Poteet}, C.~A., {et~al.} 2020, \apj, 898, 55,
  \dodoi{10.3847/1538-4357/ab9aba}

\bibitem[{{Chiang} \& {Fung}(2017)}]{Chiang17}
{Chiang}, E., \& {Fung}, J. 2017, \apj, 848, 4,
  \dodoi{10.3847/1538-4357/aa89e6}

\bibitem[{{Choquet} {et~al.}(2016){Choquet}, {Perrin}, {Chen}, {Soummer},
  {Pueyo}, {Hagan}, {Gofas-Salas}, {Rajan}, {Golimowski}, {Hines}, {Schneider},
  {Mazoyer}, {Augereau}, {Debes}, {Stark}, {Wolff}, {N'Diaye}, \&
  {Hsiao}}]{Choquet16}
{Choquet}, {\'E}., {Perrin}, M.~D., {Chen}, C.~H., {et~al.} 2016, \apjl, 817,
  L2, \dodoi{10.3847/2041-8205/817/1/L2}

\bibitem[{{Churcher} {et~al.}(2011){Churcher}, {Wyatt}, \&
  {Smith}}]{Churcher11}
{Churcher}, L., {Wyatt}, M., \& {Smith}, R. 2011, \mnras, 410, 2,
  \dodoi{10.1111/j.1365-2966.2010.17422.x}

\bibitem[{{Crotts} {et~al.}(2021){Crotts}, {Matthews}, {Esposito},
  {Duch{\^e}ne}, {Kalas}, {Chen}, {Arriaga}, {Millar-Blanchaer}, {Debes},
  {Draper}, {Fitzgerald}, {Hom}, {MacGregor}, {Mazoyer}, {Patience}, {Rice},
  {Weinberger}, {Wilner}, \& {Wolff}}]{Crotts21}
{Crotts}, K.~A., {Matthews}, B.~C., {Esposito}, T.~M., {et~al.} 2021, \apj,
  915, 58, \dodoi{10.3847/1538-4357/abff5c}

\bibitem[{{Crotts} {et~al.}(2022){Crotts}, {Draper}, {Matthews}, {Duch{\^e}ne},
  {Esposito}, {Wilner}, {Mazoyer}, {Padgett}, {Kalas}, \&
  {Stapelfeldt}}]{Crotts22}
{Crotts}, K.~A., {Draper}, Z.~H., {Matthews}, B.~C., {et~al.} 2022, \apj, 932,
  23, \dodoi{10.3847/1538-4357/ac6c86}

\bibitem[{{Currie} {et~al.}(2015){Currie}, {Lisse}, {Kuchner}, {Madhusudhan},
  {Kenyon}, {Thalmann}, {Carson}, \& {Debes}}]{Currie15}
{Currie}, T., {Lisse}, C.~M., {Kuchner}, M., {et~al.} 2015, \apjl, 807, L7,
  \dodoi{10.1088/2041-8205/807/1/L7}

\bibitem[{{Daemgen} {et~al.}(2017){Daemgen}, {Todorov}, {Quanz}, {Meyer},
  {Mordasini}, {Marleau}, \& {Fortney}}]{Daemgen17}
{Daemgen}, S., {Todorov}, K., {Quanz}, S.~P., {et~al.} 2017, \aap, 608, A71,
  \dodoi{10.1051/0004-6361/201731527}

\bibitem[{{Daley} {et~al.}(2019){Daley}, {Hughes}, {Carter}, {Flaherty},
  {Lambros}, {Pan}, {Schlichting}, {Chiang}, {Wyatt}, {Wilner}, {Andrews}, \&
  {Carpenter}}]{Daley19}
{Daley}, C., {Hughes}, A.~M., {Carter}, E.~S., {et~al.} 2019, \apj, 875, 87,
  \dodoi{10.3847/1538-4357/ab1074}

\bibitem[{{De Rosa} {et~al.}(2015){De Rosa}, {Nielsen}, {Blunt}, {Graham},
  {Konopacky}, {Marois}, {Pueyo}, {Rameau}, {Ryan}, {Wang}, {Bailey},
  {Chontos}, {Fabrycky}, {Follette}, {Macintosh}, {Marchis}, {Ammons},
  {Arriaga}, {Chilcote}, {Cotten}, {Doyon}, {Duch{\^e}ne}, {Esposito},
  {Fitzgerald}, {Gerard}, {Goodsell}, {Greenbaum}, {Hibon}, {Ingraham},
  {Johnson-Groh}, {Kalas}, {Lafreni{\`e}re}, {Maire}, {Metchev},
  {Millar-Blanchaer}, {Morzinski}, {Oppenheimer}, {Patel}, {Patience},
  {Perrin}, {Rajan}, {Rantakyr{\"o}}, {Ruffio}, {Schneider},
  {Sivaramakrishnan}, {Song}, {Tran}, {Vasisht}, {Ward-Duong}, \&
  {Wolff}}]{Rosa15}
{De Rosa}, R.~J., {Nielsen}, E.~L., {Blunt}, S.~C., {et~al.} 2015, \apjl, 814,
  L3, \dodoi{10.1088/2041-8205/814/1/L3}

\bibitem[{{Debes} {et~al.}(2009){Debes}, {Weinberger}, \& {Kuchner}}]{Debes09}
{Debes}, J.~H., {Weinberger}, A.~J., \& {Kuchner}, M.~J. 2009, \apj, 702, 318,
  \dodoi{10.1088/0004-637X/702/1/318}

\bibitem[{{Dent} {et~al.}(2014){Dent}, {Wyatt}, {Roberge}, {Augereau},
  {Casassus}, {Corder}, {Greaves}, {de Gregorio-Monsalvo}, {Hales}, {Jackson},
  {Hughes}, {Lagrange}, {Matthews}, \& {Wilner}}]{Dent14}
{Dent}, W.~R.~F., {Wyatt}, M.~C., {Roberge}, A., {et~al.} 2014, Science, 343,
  1490, \dodoi{10.1126/science.1248726}

\bibitem[{{Dohnanyi}(1969)}]{Dohnayi69}
{Dohnanyi}, J.~S. 1969, \jgr, 74, 2531, \dodoi{10.1029/JB074i010p02531}

\bibitem[{{Donaldson} {et~al.}(2013){Donaldson}, {Lebreton}, {Roberge},
  {Augereau}, \& {Krivov}}]{Donaldson13}
{Donaldson}, J.~K., {Lebreton}, J., {Roberge}, A., {Augereau}, J.~C., \&
  {Krivov}, A.~V. 2013, \apj, 772, 17, \dodoi{10.1088/0004-637X/772/1/17}

\bibitem[{{Draper} {et~al.}(2014){Draper}, {Marois}, {Wolff}, {Perrin},
  {Ingraham}, {Ruffio}, {Rantakyro}, {Hartung}, \& {Goodsell}}]{Draper14}
{Draper}, Z.~H., {Marois}, C., {Wolff}, S., {et~al.} 2014, in Society of
  Photo-Optical Instrumentation Engineers (SPIE) Conference Series, Vol. 9147,
  Ground-based and Airborne Instrumentation for Astronomy V, ed. S.~K.
  {Ramsay}, I.~S. {McLean}, \& H.~{Takami}, 91474Z, \dodoi{10.1117/12.2057156}

\bibitem[{{Draper} {et~al.}(2016){Draper}, {Duch{\^e}ne}, {Millar-Blanchaer},
  {Matthews}, {Wang}, {Kalas}, {Graham}, {Padgett}, {Ammons}, {Bulger}, {Chen},
  {Chilcote}, {Doyon}, {Fitzgerald}, {Follette}, {Gerard}, {Greenbaum},
  {Hibon}, {Hinkley}, {Macintosh}, {Ingraham}, {Lafreni{\`e}re}, {Marchis},
  {Marois}, {Nielsen}, {Oppenheimer}, {Patel}, {Patience}, {Perrin}, {Pueyo},
  {Rajan}, {Rameau}, {Sivaramakrishnan}, {Vega}, {Ward-Duong}, \&
  {Wolff}}]{Draper16}
{Draper}, Z.~H., {Duch{\^e}ne}, G., {Millar-Blanchaer}, M.~A., {et~al.} 2016,
  \apj, 826, 147, \dodoi{10.3847/0004-637X/826/2/147}

\bibitem[{Droettboom {et~al.}(2017)Droettboom, Caswell, Hunter, Firing, \&
  Nielsen}]{droettboom17}
Droettboom, M., Caswell, T.~A., Hunter, J., Firing, E., \& Nielsen, J.~H. 2017,
  Zenodo

\bibitem[{{Duch{\^e}ne} {et~al.}(2020){Duch{\^e}ne}, {Rice}, {Hom}, {Zalesky},
  {Esposito}, {Millar-Blanchaer}, {Ren}, {Kalas}, {Fitzgerald}, {Arriaga},
  {Bruzzone}, {Bulger}, {Chen}, {Chiang}, {Cotten}, {Czekala}, {De Rosa},
  {Dong}, {Draper}, {Follette}, {Graham}, {Hung}, {Lopez}, {Macintosh},
  {Matthews}, {Mazoyer}, {Metchev}, {Patience}, {Perrin}, {Rameau}, {Song},
  {Stahl}, {Wang}, {Wolff}, {Zuckerman}, {Ammons}, {Bailey}, {Barman},
  {Chilcote}, {Doyon}, {Gerard}, {Goodsell}, {Greenbaum}, {Hibon}, {Ingraham},
  {Konopacky}, {Maire}, {Marchis}, {Marley}, {Marois}, {Nielsen},
  {Oppenheimer}, {Palmer}, {Poyneer}, {Pueyo}, {Rajan}, {Rantakyr{\"o}},
  {Ruffio}, {Savransky}, {Schneider}, {Sivaramakrishnan}, {Soummer}, {Thomas},
  \& {Ward-Duong}}]{Duchene20}
{Duch{\^e}ne}, G., {Rice}, M., {Hom}, J., {et~al.} 2020, \aj, 159, 251,
  \dodoi{10.3847/1538-3881/ab8881}

\bibitem[{{Engler} {et~al.}(2018){Engler}, {Schmid}, {Quanz}, {Avenhaus}, \&
  {Bazzon}}]{Engler18}
{Engler}, N., {Schmid}, H.~M., {Quanz}, S.~P., {Avenhaus}, H., \& {Bazzon}, A.
  2018, \aap, 618, A151, \dodoi{10.1051/0004-6361/201832674}

\bibitem[{{Engler} {et~al.}(2017){Engler}, {Schmid}, {Thalmann}, {Boccaletti},
  {Bazzon}, {Baruffolo}, {Beuzit}, {Claudi}, {Costille}, {Desidera}, {Dohlen},
  {Dominik}, {Feldt}, {Fusco}, {Ginski}, {Gisler}, {Girard}, {Gratton},
  {Henning}, {Hubin}, {Janson}, {Kasper}, {Kral}, {Langlois}, {Lagadec},
  {M{\'e}nard}, {Meyer}, {Milli}, {Mouillet}, {Olofsson}, {Pavlov}, {Pragt},
  {Puget}, {Quanz}, {Roelfsema}, {Salasnich}, {Siebenmorgen}, {Sissa},
  {Suarez}, {Szulagyi}, {Turatto}, {Udry}, \& {Wildi}}]{Engler17}
{Engler}, N., {Schmid}, H.~M., {Thalmann}, C., {et~al.} 2017, \aap, 607, A90,
  \dodoi{10.1051/0004-6361/201730846}

\bibitem[{{Engler} {et~al.}(2020){Engler}, {Lazzoni}, {Gratton}, {Milli},
  {Schmid}, {Chauvin}, {Kral}, {Pawellek}, {Th{\'e}bault}, {Boccaletti},
  {Bonnefoy}, {Brown}, {Buey}, {Cantalloube}, {Carle}, {Cheetham}, {Desidera},
  {Feldt}, {Ginski}, {Gisler}, {Henning}, {Hunziker}, {Lagrange}, {Langlois},
  {Mesa}, {Meyer}, {Moeller-Nilsson}, {Olofsson}, {Petit}, {Petrus}, {Quanz},
  {Rickman}, {Stadler}, {Stolker}, {Vigan}, {Wildi}, \& {Zurlo}}]{Engler20}
{Engler}, N., {Lazzoni}, C., {Gratton}, R., {et~al.} 2020, \aap, 635, A19,
  \dodoi{10.1051/0004-6361/201936828}

\bibitem[{{Engler} {et~al.}(2022){Engler}, {Milli}, {Gratton}, {Ulmer-Moll},
  {Vigan}, {Lagrange}, {Kiefer}, {Rubini}, {Grandjean}, {Schmid}, {Messina},
  {Squicciarini}, {Olofsson}, {Th{\'e}bault}, {van Holstein}, {Janson},
  {M{\'e}nard}, {Marshall}, {Chauvin}, {Lendl}, {Bhowmik}, {Boccaletti},
  {Bonnefoy}, {del Burgo}, {Choquet}, {Desidera}, {Feldt}, {Fusco}, {Girard},
  {Gisler}, {Hagelberg}, {Langlois}, {Maire}, {Mesa}, {Meyer}, {Rabou},
  {Rodet}, {Schmidt}, \& {Zurlo}}]{Engler22}
{Engler}, N., {Milli}, J., {Gratton}, R., {et~al.} 2022, arXiv e-prints,
  arXiv:2211.11767, \dodoi{10.48550/arXiv.2211.11767}

\bibitem[{{Esposito} {et~al.}(2016){Esposito}, {Fitzgerald}, {Graham}, {Kalas},
  {Lee}, {Chiang}, {Duch{\^e}ne}, {Wang}, {Millar-Blanchaer}, {Nielsen},
  {Ammons}, {Bruzzone}, {De Rosa}, {Draper}, {Macintosh}, {Marchis}, {Metchev},
  {Perrin}, {Pueyo}, {Rajan}, {Rantakyr{\"o}}, {Vega}, \& {Wolff}}]{Esposito16}
{Esposito}, T.~M., {Fitzgerald}, M.~P., {Graham}, J.~R., {et~al.} 2016, \aj,
  152, 85, \dodoi{10.3847/0004-6256/152/4/85}

\bibitem[{{Esposito} {et~al.}(2018){Esposito}, {Duch{\^e}ne}, {Kalas}, {Rice},
  {Choquet}, {Ren}, {Perrin}, {Chen}, {Arriaga}, {Chiang}, {Nielsen}, {Graham},
  {Wang}, {De Rosa}, {Follette}, {Ammons}, {Ansdell}, {Bailey}, {Barman},
  {Sebasti{\'a}n Bruzzone}, {Bulger}, {Chilcote}, {Cotten}, {Doyon},
  {Fitzgerald}, {Goodsell}, {Greenbaum}, {Hibon}, {Hung}, {Ingraham},
  {Konopacky}, {Larkin}, {Macintosh}, {Maire}, {Marchis}, {Marois}, {Mazoyer},
  {Metchev}, {Millar-Blanchaer}, {Oppenheimer}, {Palmer}, {Patience},
  {Poyneer}, {Pueyo}, {Rajan}, {Rameau}, {Rantakyr{\"o}}, {Ryan}, {Savransky},
  {Schneider}, {Sivaramakrishnan}, {Song}, {Soummer}, {Thomas}, {Wallace},
  {Ward-Duong}, {Wiktorowicz}, \& {Wolff}}]{Esposito18}
{Esposito}, T.~M., {Duch{\^e}ne}, G., {Kalas}, P., {et~al.} 2018, \aj, 156, 47,
  \dodoi{10.3847/1538-3881/aacbc9}

\bibitem[{{Esposito} {et~al.}(2020){Esposito}, {Kalas}, {Fitzgerald},
  {Millar-Blanchaer}, {Duch{\^e}ne}, {Patience}, {Hom}, {Perrin}, {De Rosa},
  {Chiang}, {Czekala}, {Macintosh}, {Graham}, {Ansdell}, {Arriaga}, {Bruzzone},
  {Bulger}, {Chen}, {Cotten}, {Dong}, {Draper}, {Follette}, {Hung}, {Lopez},
  {Matthews}, {Mazoyer}, {Metchev}, {Rameau}, {Ren}, {Rice}, {Song}, {Stahl},
  {Wang}, {Wolff}, {Zuckerman}, {Ammons}, {Bailey}, {Barman}, {Chilcote},
  {Doyon}, {Gerard}, {Goodsell}, {Greenbaum}, {Hibon}, {Hinkley}, {Ingraham},
  {Konopacky}, {Maire}, {Marchis}, {Marley}, {Marois}, {Nielsen},
  {Oppenheimer}, {Palmer}, {Poyneer}, {Pueyo}, {Rajan}, {Rantakyr{\"o}},
  {Ruffio}, {Savransky}, {Schneider}, {Sivaramakrishnan}, {Soummer}, {Thomas},
  \& {Ward-Duong}}]{Esposito20}
{Esposito}, T.~M., {Kalas}, P., {Fitzgerald}, M.~P., {et~al.} 2020, \aj, 160,
  24, \dodoi{10.3847/1538-3881/ab9199}

\bibitem[{{Feldt} {et~al.}(2017){Feldt}, {Olofsson}, {Boccaletti}, {Maire},
  {Milli}, {Vigan}, {Langlois}, {Henning}, {Moor}, {Bonnefoy}, {Wahhaj},
  {Desidera}, {Gratton}, {K{\'o}sp{\'a}l}, {Abraham}, {Menard}, {Chauvin},
  {Lagrange}, {Mesa}, {Salter}, {Buenzli}, {Lannier}, {Perrot}, {Peretti}, \&
  {Sissa}}]{Feldt17}
{Feldt}, M., {Olofsson}, J., {Boccaletti}, A., {et~al.} 2017, \aap, 601, A7,
  \dodoi{10.1051/0004-6361/201629261}

\bibitem[{Foreman-Mackey(2016)}]{corner}
Foreman-Mackey, D. 2016, Journal of Open Source Software, 1, 24,
  \dodoi{10.21105/joss.00024}

\bibitem[{{Foreman-Mackey} {et~al.}(2013){Foreman-Mackey}, {Hogg}, {Lang}, \&
  {Goodman}}]{FM13}
{Foreman-Mackey}, D., {Hogg}, D.~W., {Lang}, D., \& {Goodman}, J. 2013, \pasp,
  125, 306, \dodoi{10.1086/670067}

\bibitem[{{Gaia Collaboration}(2020)}]{Gaia20}
{Gaia Collaboration}. 2020, VizieR Online Data Catalog, I/350

\bibitem[{{Gallenne} {et~al.}(2022){Gallenne}, {Desgrange}, {Milli},
  {Sanchez-Bermudez}, {Chauvin}, {Kraus}, {Girard}, \&
  {Boccaletti}}]{Gallenne22}
{Gallenne}, A., {Desgrange}, C., {Milli}, J., {et~al.} 2022, \aap, 665, A41,
  \dodoi{10.1051/0004-6361/202244226}

\bibitem[{{Gibbs} {et~al.}(2019){Gibbs}, {Wagner}, {Apai}, {Mo{\'o}r},
  {Currie}, {Bonnefoy}, {Langlois}, \& {Lisse}}]{Gibbs19}
{Gibbs}, A., {Wagner}, K., {Apai}, D., {et~al.} 2019, \aj, 157, 39,
  \dodoi{10.3847/1538-3881/aaf1bd}

\bibitem[{{Goebel} {et~al.}(2018){Goebel}, {Currie}, {Guyon}, {Brandt},
  {Groff}, {Jovanovic}, {Kasdin}, {Lozi}, {Hodapp}, {Martinache}, {Grady},
  {Hayashi}, {Kwon}, {McElwain}, {Yang}, \& {Tamura}}]{Goebel18}
{Goebel}, S., {Currie}, T., {Guyon}, O., {et~al.} 2018, \aj, 156, 279,
  \dodoi{10.3847/1538-3881/aaeb24}

\bibitem[{{Greaves} {et~al.}(2016){Greaves}, {Holland}, {Matthews}, {Marshall},
  {Dent}, {Woitke}, {Wyatt}, {Matr{\`a}}, \& {Jackson}}]{Greaves16}
{Greaves}, J.~S., {Holland}, W.~S., {Matthews}, B.~C., {et~al.} 2016, \mnras,
  461, 3910, \dodoi{10.1093/mnras/stw1569}

\bibitem[{{Hales} {et~al.}(2022){Hales}, {Marino}, {Sheehan}, {Ulloa},
  {P{\'e}rez}, {Matr{\`a}}, {Kral}, {Wyatt}, {Dent}, \& {Carpenter}}]{Hales22}
{Hales}, A.~S., {Marino}, S., {Sheehan}, P.~D., {et~al.} 2022, \apj, 940, 161,
  \dodoi{10.3847/1538-4357/ac9cd3}

\bibitem[{{Han} {et~al.}(2023){Han}, {Wyatt}, \& {Dent}}]{Han23}
{Han}, Y., {Wyatt}, M.~C., \& {Dent}, W.~R.~F. 2023, \mnras, 519, 3257,
  \dodoi{10.1093/mnras/stac3769}

\bibitem[{{Han} {et~al.}(2022){Han}, {Wyatt}, \& {Matr{\`a}}}]{Han22}
{Han}, Y., {Wyatt}, M.~C., \& {Matr{\`a}}, L. 2022, \mnras, 511, 4921,
  \dodoi{10.1093/mnras/stac373}

\bibitem[{{Heap} {et~al.}(2000){Heap}, {Lindler}, {Lanz}, {Cornett}, {Hubeny},
  {Maran}, \& {Woodgate}}]{Heap00}
{Heap}, S.~R., {Lindler}, D.~J., {Lanz}, T.~M., {et~al.} 2000, \apj, 539, 435,
  \dodoi{10.1086/309188}

\bibitem[{{Hines} {et~al.}(2007){Hines}, {Schneider}, {Hollenbach}, {Mamajek},
  {Hillenbrand}, {Metchev}, {Meyer}, {Carpenter}, {Moro-Mart{\'\i}n},
  {Silverstone}, {Kim}, {Henning}, {Bouwman}, \& {Wolf}}]{Hines07}
{Hines}, D.~C., {Schneider}, G., {Hollenbach}, D., {et~al.} 2007, \apjl, 671,
  L165, \dodoi{10.1086/525016}

\bibitem[{{Hom} {et~al.}(2023, in prep){Hom}, {Patience}, {Duch{\^e}ne},
  {Chen}, {Mazoyer}, {Millar-Blanchaer}, {Esposito}, \& {Kalas}}]{Hom23}
{Hom}, J., {Patience}, J., {Duch{\^e}ne}, G., {et~al.} 2023, in prep

\bibitem[{{Hom} {et~al.}(2020){Hom}, {Patience}, {Esposito}, {Duch{\^e}ne},
  {Worthen}, {Kalas}, {Jang-Condell}, {Saboi}, {Arriaga}, {Mazoyer}, {Wolff},
  {Millar-Blanchaer}, {Fitzgerald}, {Perrin}, {Chen}, {Macintosh}, {Matthews},
  {Wang}, {Graham}, {Marchis}, {Ammons}, {Bailey}, {Barman}, {Bulger},
  {Chilcote}, {Cotten}, {De Rosa}, {Doyon}, {Follette}, {Goodsell},
  {Greenbaum}, {Hibon}, {Ingraham}, {Konopacky}, {Larkin}, {Maire}, {Marley},
  {Marois}, {Matthews}, {Metchev}, {Nielsen}, {Oppenheimer}, {Palmer},
  {Poyneer}, {Pueyo}, {Rajan}, {Rameau}, {Rantakyr{\"o}}, {Ren}, {Savransky},
  {Schneider}, {Sivaramakrishnan}, {Song}, {Soummer}, {Tallis}, {Thomas},
  {Wallace}, {Ward-Duong}, {Wiktorowicz}, \& {Zuckerman}}]{Hom20}
{Hom}, J., {Patience}, J., {Esposito}, T.~M., {et~al.} 2020, \aj, 159, 31,
  \dodoi{10.3847/1538-3881/ab5af2}

\bibitem[{{Hughes} {et~al.}(2018){Hughes}, {Duch{\^e}ne}, \&
  {Matthews}}]{Hughes18}
{Hughes}, A.~M., {Duch{\^e}ne}, G., \& {Matthews}, B.~C. 2018, \araa, 56, 541,
  \dodoi{10.1146/annurev-astro-081817-052035}

\bibitem[{{Hung} {et~al.}(2015{\natexlab{a}}){Hung}, {Fitzgerald}, {Chen},
  {Mittal}, {Kalas}, \& {Graham}}]{Hung15a}
{Hung}, L.-W., {Fitzgerald}, M.~P., {Chen}, C.~H., {et~al.} 2015{\natexlab{a}},
  \apj, 802, 138, \dodoi{10.1088/0004-637X/802/2/138}

\bibitem[{{Hung} {et~al.}(2015{\natexlab{b}}){Hung}, {Duch{\^e}ne}, {Arriaga},
  {Fitzgerald}, {Maire}, {Marois}, {Millar-Blanchaer}, {Bruzzone}, {Rajan},
  {Pueyo}, {Kalas}, {De Rosa}, {Graham}, {Konopacky}, {Wolff}, {Ammons},
  {Chen}, {Chilcote}, {Draper}, {Esposito}, {Gerard}, {Goodsell}, {Greenbaum},
  {Hibon}, {Hinkley}, {Macintosh}, {Marchis}, {Metchev}, {Nielsen},
  {Oppenheimer}, {Patience}, {Perrin}, {Rantakyr{\"o}}, {Sivaramakrishnan},
  {Wang}, {Ward-Duong}, \& {Wiktorowicz}}]{Hung15b}
{Hung}, L.-W., {Duch{\^e}ne}, G., {Arriaga}, P., {et~al.} 2015{\natexlab{b}},
  \apjl, 815, L14, \dodoi{10.1088/2041-8205/815/1/L14}

\bibitem[{Hunter(2007)}]{Hunter07}
Hunter, J.~D. 2007, Computing in Science Engineering, 9, 90-95

\bibitem[{{Ingraham} {et~al.}(2014){Ingraham}, {Ruffio}, {Perrin}, {Wolff},
  {Draper}, {Maire}, {Marchis}, \& {Fesquet}}]{Ingraham14}
{Ingraham}, P., {Ruffio}, J.-B., {Perrin}, M.~D., {et~al.} 2014, in Society of
  Photo-Optical Instrumentation Engineers (SPIE) Conference Series, Vol. 9147,
  Ground-based and Airborne Instrumentation for Astronomy V, ed. S.~K.
  {Ramsay}, I.~S. {McLean}, \& H.~{Takami}, 91477K, \dodoi{10.1117/12.2055283}

\bibitem[{{Jackson} {et~al.}(2014){Jackson}, {Wyatt}, {Bonsor}, \&
  {Veras}}]{Jackson14}
{Jackson}, A.~P., {Wyatt}, M.~C., {Bonsor}, A., \& {Veras}, D. 2014, \mnras,
  440, 3757, \dodoi{10.1093/mnras/stu476}

\bibitem[{{Janson} {et~al.}(2021){Janson}, {Brandeker}, {Olofsson}, \&
  {Liseau}}]{Janson21}
{Janson}, M., {Brandeker}, A., {Olofsson}, G., \& {Liseau}, R. 2021, \aap, 646,
  A132, \dodoi{10.1051/0004-6361/202039990}

\bibitem[{{Johnson} {et~al.}(2012){Johnson}, {Lisse}, {Chen}, {Melosh},
  {Wyatt}, {Thebault}, {Henning}, {Gaidos}, {Elkins-Tanton}, {Bridges}, \&
  {Morlok}}]{Johnson12}
{Johnson}, B.~C., {Lisse}, C.~M., {Chen}, C.~H., {et~al.} 2012, \apj, 761, 45,
  \dodoi{10.1088/0004-637X/761/1/45}

\bibitem[{{Jones} {et~al.}(2023){Jones}, {Chiang}, {Duchene}, {Kalas}, \&
  {Esposito}}]{Jones23}
{Jones}, J.~W., {Chiang}, E., {Duchene}, G., {Kalas}, P., \& {Esposito}, T.~M.
  2023, arXiv e-prints, arXiv:2303.10189, \dodoi{10.48550/arXiv.2303.10189}

\bibitem[{{Kalas} \& {Jewitt}(1995)}]{Kalas95}
{Kalas}, P., \& {Jewitt}, D. 1995, \aj, 110, 794, \dodoi{10.1086/117565}

\bibitem[{{Kalas} {et~al.}(2015){Kalas}, {Rajan}, {Wang}, {Millar-Blanchaer},
  {Duchene}, {Chen}, {Fitzgerald}, {Dong}, {Graham}, {Patience}, {Macintosh},
  {Murray-Clay}, {Matthews}, {Rameau}, {Marois}, {Chilcote}, {De Rosa},
  {Doyon}, {Draper}, {Lawler}, {Ammons}, {Arriaga}, {Bulger}, {Cotten},
  {Follette}, {Goodsell}, {Greenbaum}, {Hibon}, {Hinkley}, {Hung}, {Ingraham},
  {Konapacky}, {Lafreniere}, {Larkin}, {Long}, {Maire}, {Marchis}, {Metchev},
  {Morzinski}, {Nielsen}, {Oppenheimer}, {Perrin}, {Pueyo}, {Rantakyr{\"o}},
  {Ruffio}, {Saddlemyer}, {Savransky}, {Schneider}, {Sivaramakrishnan},
  {Soummer}, {Song}, {Thomas}, {Vasisht}, {Ward-Duong}, {Wiktorowicz}, \&
  {Wolff}}]{Kalas15}
{Kalas}, P.~G., {Rajan}, A., {Wang}, J.~J., {et~al.} 2015, \apj, 814, 32,
  \dodoi{10.1088/0004-637X/814/1/32}

\bibitem[{{Kasper} {et~al.}(2015){Kasper}, {Apai}, {Wagner}, \&
  {Robberto}}]{Kasper15}
{Kasper}, M., {Apai}, D., {Wagner}, K., \& {Robberto}, M. 2015, \apjl, 812,
  L33, \dodoi{10.1088/2041-8205/812/2/L33}

\bibitem[{{Kiefer} {et~al.}(2014){Kiefer}, {Lecavelier des Etangs}, {Augereau},
  {Vidal-Madjar}, {Lagrange}, \& {Beust}}]{Kiefer14}
{Kiefer}, F., {Lecavelier des Etangs}, A., {Augereau}, J.~C., {et~al.} 2014,
  \aap, 561, L10, \dodoi{10.1051/0004-6361/201323128}

\bibitem[{{Kiefer} {et~al.}(2023){Kiefer}, {Van Grootel}, {Lecavelier des
  Etangs}, {Szab{\'o}}, {Brandeker}, {Broeg}, {Collier Cameron}, {Deline},
  {Olofsson}, {Wilson}, {Sousa}, {Gandolfi}, {H{\'e}brard}, {Alibert},
  {Alonso}, {Anglada}, {B{\'a}rczy}, {Barrado}, {Barros}, {Baumjohann}, {Beck},
  {Beck}, {Benz}, {Billot}, {Bonfils}, {Cabrera}, {Charnoz}, {Csizmadia},
  {Davies}, {Deleuil}, {Delrez}, {Demangeon}, {Demory}, {Ehrenreich},
  {Erikson}, {Fortier}, {Fossati}, {Fridlund}, {Gillon}, {G{\"u}del}, {Heng},
  {Hoyer}, {Isaak}, {Kiss}, {Laskar}, {Lendl}, {Lovis}, {Magrin}, {Maxted},
  {Munari}, {Nascimbeni}, {Ottensamer}, {Pagano}, {Pall{\'e}}, {Peter},
  {Piazza}, {Piotto}, {Pollacco}, {Queloz}, {Ragazzoni}, {Rando}, {Ratti},
  {Rauer}, {Reimers}, {Ribas}, {Santos}, {Scandariato}, {S{\'e}gransan},
  {Simon}, {Smith}, {Steller}, {Thomas}, {Udry}, {Walter}, \&
  {Walton}}]{Kiefer23}
{Kiefer}, F., {Van Grootel}, V., {Lecavelier des Etangs}, A., {et~al.} 2023,
  arXiv e-prints, arXiv:2301.07418, \dodoi{10.48550/arXiv.2301.07418}

\bibitem[{{K{\'o}sp{\'a}l} {et~al.}(2013){K{\'o}sp{\'a}l}, {Mo{\'o}r},
  {Juh{\'a}sz}, {{\'A}brah{\'a}m}, {Apai}, {Csengeri}, {Grady}, {Henning},
  {Hughes}, {Kiss}, {Pascucci}, \& {Schmalzl}}]{Kospal13}
{K{\'o}sp{\'a}l}, {\'A}., {Mo{\'o}r}, A., {Juh{\'a}sz}, A., {et~al.} 2013,
  \apj, 776, 77, \dodoi{10.1088/0004-637X/776/2/77}

\bibitem[{{Kral} {et~al.}(2019){Kral}, {Marino}, {Wyatt}, {Kama}, \&
  {Matr{\`a}}}]{Kral19}
{Kral}, Q., {Marino}, S., {Wyatt}, M.~C., {Kama}, M., \& {Matr{\`a}}, L. 2019,
  \mnras, 489, 3670, \dodoi{10.1093/mnras/sty2923}

\bibitem[{{Kral} {et~al.}(2020){Kral}, {Matr{\`a}}, {Kennedy}, {Marino}, \&
  {Wyatt}}]{Kral20}
{Kral}, Q., {Matr{\`a}}, L., {Kennedy}, G.~M., {Marino}, S., \& {Wyatt}, M.~C.
  2020, \mnras, 497, 2811, \dodoi{10.1093/mnras/staa2038}

\bibitem[{{Lagrange} {et~al.}(2009){Lagrange}, {Gratadour}, {Chauvin}, {Fusco},
  {Ehrenreich}, {Mouillet}, {Rousset}, {Rouan}, {Allard}, {Gendron}, {Charton},
  {Mugnier}, {Rabou}, {Montri}, \& {Lacombe}}]{Lagrange09}
{Lagrange}, A.~M., {Gratadour}, D., {Chauvin}, G., {et~al.} 2009, \aap, 493,
  L21, \dodoi{10.1051/0004-6361:200811325}

\bibitem[{{Lagrange} {et~al.}(2010){Lagrange}, {Bonnefoy}, {Chauvin}, {Apai},
  {Ehrenreich}, {Boccaletti}, {Gratadour}, {Rouan}, {Mouillet}, {Lacour}, \&
  {Kasper}}]{Lagrange10}
{Lagrange}, A.~M., {Bonnefoy}, M., {Chauvin}, G., {et~al.} 2010, Science, 329,
  57, \dodoi{10.1126/science.1187187}

\bibitem[{{Lagrange} {et~al.}(2016){Lagrange}, {Langlois}, {Gratton}, {Maire},
  {Milli}, {Olofsson}, {Vigan}, {Bailey}, {Mesa}, {Chauvin}, {Boccaletti},
  {Galicher}, {Girard}, {Bonnefoy}, {Samland}, {Menard}, {Henning},
  {Kenworthy}, {Thalmann}, {Beust}, {Beuzit}, {Brandner}, {Buenzli},
  {Cheetham}, {Janson}, {le Coroller}, {Lannier}, {Mouillet}, {Peretti},
  {Perrot}, {Salter}, {Sissa}, {Wahhaj}, {Abe}, {Desidera}, {Feldt}, {Madec},
  {Perret}, {Petit}, {Rabou}, {Soenke}, \& {Weber}}]{Lagrange16}
{Lagrange}, A.~M., {Langlois}, M., {Gratton}, R., {et~al.} 2016, \aap, 586, L8,
  \dodoi{10.1051/0004-6361/201527264}

\bibitem[{{Lagrange} {et~al.}(2019){Lagrange}, {Meunier}, {Rubini}, {Keppler},
  {Galland}, {Chapellier}, {Michel}, {Balona}, {Beust}, {Guillot}, {Grandjean},
  {Borgniet}, {M{\'e}karnia}, {Wilson}, {Kiefer}, {Bonnefoy}, {Lillo-Box},
  {Pantoja}, {Jones}, {Iglesias}, {Rodet}, {Diaz}, {Zapata}, {Abe}, \&
  {Schmider}}]{Lagrange19}
{Lagrange}, A.~M., {Meunier}, N., {Rubini}, P., {et~al.} 2019, Nature
  Astronomy, 3, 1135, \dodoi{10.1038/s41550-019-0857-1}

\bibitem[{{Langlois} {et~al.}(2021){Langlois}, {Gratton}, {Lagrange},
  {Delorme}, {Boccaletti}, {Bonnefoy}, {Maire}, {Mesa}, {Chauvin}, {Desidera},
  {Vigan}, {Cheetham}, {Hagelberg}, {Feldt}, {Meyer}, {Rubini}, {Le Coroller},
  {Cantalloube}, {Biller}, {Bonavita}, {Bhowmik}, {Brandner}, {Daemgen},
  {D'Orazi}, {Flasseur}, {Fontanive}, {Galicher}, {Girard}, {Janin-Potiron},
  {Janson}, {Keppler}, {Kopytova}, {Lagadec}, {Lannier}, {Lazzoni}, {Ligi},
  {Meunier}, {Perreti}, {Perrot}, {Rodet}, {Romero}, {Rouan}, {Samland},
  {Salter}, {Sissa}, {Schmidt}, {Zurlo}, {Mouillet}, {Denis}, {Thi{\'e}baut},
  {Milli}, {Wahhaj}, {Beuzit}, {Dominik}, {Henning}, {M{\'e}nard},
  {M{\"u}ller}, {Schmid}, {Turatto}, {Udry}, {Abe}, {Antichi}, {Allard},
  {Baruffolo}, {Baudoz}, {Baudrand}, {Bazzon}, {Blanchard}, {Carbillet},
  {Carle}, {Cascone}, {Charton}, {Claudi}, {Costille}, {De Caprio},
  {Delboulb{\'e}}, {Dohlen}, {Fantinel}, {Feautrier}, {Fusco}, {Gigan}, {Giro},
  {Gisler}, {Gluck}, {Gry}, {Hubin}, {Hugot}, {Jaquet}, {Kasper}, {Le Mignant},
  {Llored}, {Madec}, {Magnard}, {Martinez}, {Maurel}, {Messina},
  {M{\"o}ller-Nilsson}, {Mugnier}, {Moulin}, {Orign{\'e}}, {Pavlov}, {Perret},
  {Petit}, {Pragt}, {Puget}, {Rabou}, {Ramos}, {Rigal}, {Rochat}, {Roelfsema},
  {Rousset}, {Roux}, {Salasnich}, {Sauvage}, {Sevin}, {Soenke}, {Stadler},
  {Suarez}, {Weber}, {Wildi}, \& {Rickman}}]{Langlois21}
{Langlois}, M., {Gratton}, R., {Lagrange}, A.~M., {et~al.} 2021, \aap, 651,
  A71, \dodoi{10.1051/0004-6361/202039753}

\bibitem[{{Lee} \& {Chiang}(2016)}]{LC16}
{Lee}, E.~J., \& {Chiang}, E. 2016, \apj, 827, 125,
  \dodoi{10.3847/0004-637X/827/2/125}

\bibitem[{{Lin} \& {Chiang}(2019)}]{LC19}
{Lin}, J.~W., \& {Chiang}, E. 2019, \apj, 883, 68,
  \dodoi{10.3847/1538-4357/ab35da}

\bibitem[{{Lisse} {et~al.}(2008){Lisse}, {Chen}, {Wyatt}, \&
  {Morlok}}]{Lisse08}
{Lisse}, C.~M., {Chen}, C.~H., {Wyatt}, M.~C., \& {Morlok}, A. 2008, in 39th
  Annual Lunar and Planetary Science Conference, Lunar and Planetary Science
  Conference, 2119

\bibitem[{{L{\"o}hne}(2020)}]{Lohne20}
{L{\"o}hne}, T. 2020, \aap, 641, A75, \dodoi{10.1051/0004-6361/202037858}

\bibitem[{{MacGregor} {et~al.}(2016){MacGregor}, {Wilner}, {Chandler}, {Ricci},
  {Maddison}, {Cranmer}, {Andrews}, {Hughes}, \& {Steele}}]{MacGregor16}
{MacGregor}, M.~A., {Wilner}, D.~J., {Chandler}, C., {et~al.} 2016, \apj, 823,
  79, \dodoi{10.3847/0004-637X/823/2/79}

\bibitem[{{MacGregor} {et~al.}(2018){MacGregor}, {Weinberger}, {Hughes},
  {Wilner}, {Currie}, {Debes}, {Donaldson}, {Redfield}, {Roberge}, \&
  {Schneider}}]{MacGregor18}
{MacGregor}, M.~A., {Weinberger}, A.~J., {Hughes}, A.~M., {et~al.} 2018, \apj,
  869, 75, \dodoi{10.3847/1538-4357/aaec71}

\bibitem[{{Macintosh} {et~al.}(2014){Macintosh}, {Graham}, {Ingraham},
  {Konopacky}, {Marois}, {Perrin}, {Poyneer}, {Bauman}, {Barman}, {Burrows},
  {Cardwell}, {Chilcote}, {De Rosa}, {Dillon}, {Doyon}, {Dunn}, {Erikson},
  {Fitzgerald}, {Gavel}, {Goodsell}, {Hartung}, {Hibon}, {Kalas}, {Larkin},
  {Maire}, {Marchis}, {Marley}, {McBride}, {Millar-Blanchaer}, {Morzinski},
  {Norton}, {Oppenheimer}, {Palmer}, {Patience}, {Pueyo}, {Rantakyro},
  {Sadakuni}, {Saddlemyer}, {Savransky}, {Serio}, {Soummer},
  {Sivaramakrishnan}, {Song}, {Thomas}, {Wallace}, {Wiktorowicz}, \&
  {Wolff}}]{Macintosh14}
{Macintosh}, B., {Graham}, J.~R., {Ingraham}, P., {et~al.} 2014, Proceedings of
  the National Academy of Science, 111, 12661, \dodoi{10.1073/pnas.1304215111}

\bibitem[{{Macintosh} {et~al.}(2018){Macintosh}, {Chilcote}, {Bailey}, {de
  Rosa}, {Nielsen}, {Norton}, {Poyneer}, {Wang}, {Ruffio}, {Graham}, {Marois},
  {Savransky}, \& {Veran}}]{Macintosh18}
{Macintosh}, B., {Chilcote}, J.~K., {Bailey}, V.~P., {et~al.} 2018, in Society
  of Photo-Optical Instrumentation Engineers (SPIE) Conference Series, Vol.
  10703, Adaptive Optics Systems VI, ed. L.~M. {Close}, L.~{Schreiber}, \&
  D.~{Schmidt}, 107030K, \dodoi{10.1117/12.2314253}

\bibitem[{{Macintosh} {et~al.}(2008){Macintosh}, {Graham}, {Palmer}, {Doyon},
  {Dunn}, {Gavel}, {Larkin}, {Oppenheimer}, {Saddlemyer}, {Sivaramakrishnan},
  {Wallace}, {Bauman}, {Erickson}, {Marois}, {Poyneer}, \&
  {Soummer}}]{Macintosh08}
{Macintosh}, B.~A., {Graham}, J.~R., {Palmer}, D.~W., {et~al.} 2008, in Society
  of Photo-Optical Instrumentation Engineers (SPIE) Conference Series, Vol.
  7015, Adaptive Optics Systems, ed. N.~{Hubin}, C.~E. {Max}, \& P.~L.
  {Wizinowich}, 701518, \dodoi{10.1117/12.788083}

\bibitem[{{Maness} {et~al.}(2009){Maness}, {Kalas}, {Peek}, {Chiang},
  {Scherer}, {Fitzgerald}, {Graham}, {Hines}, {Schneider}, \&
  {Metchev}}]{Maness09}
{Maness}, H.~L., {Kalas}, P., {Peek}, K.~M.~G., {et~al.} 2009, \apj, 707, 1098,
  \dodoi{10.1088/0004-637X/707/2/1098}

\bibitem[{{Marino}(2021)}]{Marino21}
{Marino}, S. 2021, \mnras, 503, 5100, \dodoi{10.1093/mnras/stab771}

\bibitem[{{Martioli} {et~al.}(2021){Martioli}, {H{\'e}brard}, {Correia},
  {Laskar}, \& {Lecavelier des Etangs}}]{Martioli21}
{Martioli}, E., {H{\'e}brard}, G., {Correia}, A.~C.~M., {Laskar}, J., \&
  {Lecavelier des Etangs}, A. 2021, \aap, 649, A177,
  \dodoi{10.1051/0004-6361/202040235}

\bibitem[{{Matr{\`a}} {et~al.}(2018){Matr{\`a}}, {Marino}, {Kennedy}, {Wyatt},
  {{\"O}berg}, \& {Wilner}}]{Matra18}
{Matr{\`a}}, L., {Marino}, S., {Kennedy}, G.~M., {et~al.} 2018, \apj, 859, 72,
  \dodoi{10.3847/1538-4357/aabcc4}

\bibitem[{{Matr{\`a}} {et~al.}(2019){Matr{\`a}}, {Wyatt}, {Wilner}, {Dent},
  {Marino}, {Kennedy}, \& {Milli}}]{Matra19}
{Matr{\`a}}, L., {Wyatt}, M.~C., {Wilner}, D.~J., {et~al.} 2019, \aj, 157, 135,
  \dodoi{10.3847/1538-3881/ab06c0}

\bibitem[{{Matthews} {et~al.}(2014){Matthews}, {Krivov}, {Wyatt}, {Bryden}, \&
  {Eiroa}}]{Matthews14}
{Matthews}, B.~C., {Krivov}, A.~V., {Wyatt}, M.~C., {Bryden}, G., \& {Eiroa},
  C. 2014, in Protostars and Planets VI, ed. H.~{Beuther}, R.~S. {Klessen},
  C.~P. {Dullemond}, \& T.~{Henning}, 521--544,
  \dodoi{10.2458/azu_uapress_9780816531240-ch023}

\bibitem[{{Matthews} {et~al.}(2017){Matthews}, {Hinkley}, {Vigan}, {Kennedy},
  {Rizzuto}, {Stapelfeldt}, {Mawet}, {Booth}, {Chen}, \&
  {Jang-Condell}}]{Matthews17}
{Matthews}, E., {Hinkley}, S., {Vigan}, A., {et~al.} 2017, \apjl, 843, L12,
  \dodoi{10.3847/2041-8213/aa7943}

\bibitem[{{Michel} {et~al.}(2021){Michel}, {van der Marel}, \&
  {Matthews}}]{Michel21}
{Michel}, A., {van der Marel}, N., \& {Matthews}, B.~C. 2021, \apj, 921, 72,
  \dodoi{10.3847/1538-4357/ac1bbb}

\bibitem[{{Millar-Blanchaer} {et~al.}(2016{\natexlab{a}}){Millar-Blanchaer},
  {Perrin}, {Hung}, {Fitzgerald}, {Wang}, {Chilcote}, {Graham}, {Bruzzone}, \&
  {Kalas}}]{MB16b}
{Millar-Blanchaer}, M.~A., {Perrin}, M.~D., {Hung}, L.-W., {et~al.}
  2016{\natexlab{a}}, in Society of Photo-Optical Instrumentation Engineers
  (SPIE) Conference Series, Vol. 9908, Ground-based and Airborne
  Instrumentation for Astronomy VI, ed. C.~J. {Evans}, L.~{Simard}, \&
  H.~{Takami}, 990836, \dodoi{10.1117/12.2233071}

\bibitem[{{Millar-Blanchaer} {et~al.}(2016{\natexlab{b}}){Millar-Blanchaer},
  {Wang}, {Kalas}, {Graham}, {Duch{\^e}ne}, {Nielsen}, {Perrin}, {Moon},
  {Padgett}, {Metchev}, {Ammons}, {Bailey}, {Barman}, {Bruzzone}, {Bulger},
  {Chen}, {Chilcote}, {Cotten}, {De Rosa}, {Doyon}, {Draper}, {Esposito},
  {Fitzgerald}, {Follette}, {Gerard}, {Greenbaum}, {Hibon}, {Hinkley}, {Hung},
  {Ingraham}, {Johnson-Groh}, {Konopacky}, {Larkin}, {Macintosh}, {Maire},
  {Marchis}, {Marley}, {Marois}, {Matthews}, {Oppenheimer}, {Palmer},
  {Patience}, {Poyneer}, {Pueyo}, {Rajan}, {Rameau}, {Rantakyr{\"o}},
  {Savransky}, {Schneider}, {Sivaramakrishnan}, {Song}, {Soummer}, {Thomas},
  {Vega}, {Wallace}, {Ward-Duong}, {Wiktorowicz}, \& {Wolff}}]{MB16}
{Millar-Blanchaer}, M.~A., {Wang}, J.~J., {Kalas}, P., {et~al.}
  2016{\natexlab{b}}, \aj, 152, 128, \dodoi{10.3847/0004-6256/152/5/128}

\bibitem[{{Milli} {et~al.}(2017){Milli}, {Vigan}, {Mouillet}, {Lagrange},
  {Augereau}, {Pinte}, {Mawet}, {Schmid}, {Boccaletti}, {Matr{\`a}}, {Kral},
  {Ertel}, {Chauvin}, {Bazzon}, {M{\'e}nard}, {Beuzit}, {Thalmann}, {Dominik},
  {Feldt}, {Henning}, {Min}, {Girard}, {Galicher}, {Bonnefoy}, {Fusco}, {de
  Boer}, {Janson}, {Maire}, {Mesa}, {Schlieder}, \& {SPHERE
  Consortium}}]{Milli17}
{Milli}, J., {Vigan}, A., {Mouillet}, D., {et~al.} 2017, \aap, 599, A108,
  \dodoi{10.1051/0004-6361/201527838}

\bibitem[{{Milli} {et~al.}(2019){Milli}, {Engler}, {Schmid}, {Olofsson},
  {M{\'e}nard}, {Kral}, {Boccaletti}, {Th{\'e}bault}, {Choquet}, {Mouillet},
  {Lagrange}, {Augereau}, {Pinte}, {Chauvin}, {Dominik}, {Perrot}, {Zurlo},
  {Henning}, {Beuzit}, {Avenhaus}, {Bazzon}, {Moulin}, {Llored},
  {Moeller-Nilsson}, {Roelfsema}, \& {Pragt}}]{Milli19}
{Milli}, J., {Engler}, N., {Schmid}, H.~M., {et~al.} 2019, \aap, 626, A54,
  \dodoi{10.1051/0004-6361/201935363}

\bibitem[{{Moore} {et~al.}(2023){Moore}, {Li}, {Hassenzahl}, {Nesvold}, {Naoz},
  \& {Adams}}]{Moore23}
{Moore}, N. W.~H., {Li}, G., {Hassenzahl}, L., {et~al.} 2023, \apj, 943, 6,
  \dodoi{10.3847/1538-4357/aca766}

\bibitem[{{Mouillet} {et~al.}(1997){Mouillet}, {Larwood}, {Papaloizou}, \&
  {Lagrange}}]{Mouillet97}
{Mouillet}, D., {Larwood}, J.~D., {Papaloizou}, J.~C.~B., \& {Lagrange}, A.~M.
  1997, \mnras, 292, 896, \dodoi{10.1093/mnras/292.4.896}

\bibitem[{{Nesvold} {et~al.}(2017){Nesvold}, {Naoz}, \&
  {Fitzgerald}}]{Nesvold17}
{Nesvold}, E.~R., {Naoz}, S., \& {Fitzgerald}, M.~P. 2017, \apjl, 837, L6,
  \dodoi{10.3847/2041-8213/aa61a7}

\bibitem[{{Nguyen} {et~al.}(2021){Nguyen}, {De Rosa}, \& {Kalas}}]{Nguyen21}
{Nguyen}, M.~M., {De Rosa}, R.~J., \& {Kalas}, P. 2021, \aj, 161, 22,
  \dodoi{10.3847/1538-3881/abc012}

\bibitem[{{Nielsen} {et~al.}(2016){Nielsen}, {De Rosa}, {Wang}, {Rameau},
  {Song}, {Graham}, {Macintosh}, {Ammons}, {Bailey}, {Barman}, {Bulger},
  {Chilcote}, {Cotten}, {Doyon}, {Duch{\^e}ne}, {Fitzgerald}, {Follette},
  {Greenbaum}, {Hibon}, {Hung}, {Ingraham}, {Kalas}, {Konopacky}, {Larkin},
  {Maire}, {Marchis}, {Marley}, {Marois}, {Metchev}, {Millar-Blanchaer},
  {Oppenheimer}, {Palmer}, {Patience}, {Perrin}, {Poyneer}, {Pueyo}, {Rajan},
  {Rantakyr{\"o}}, {Savransky}, {Schneider}, {Sivaramakrishnan}, {Soummer},
  {Thomas}, {Wallace}, {Ward-Duong}, {Wiktorowicz}, \& {Wolff}}]{Nielson16}
{Nielsen}, E.~L., {De Rosa}, R.~J., {Wang}, J., {et~al.} 2016, \aj, 152, 175,
  \dodoi{10.3847/0004-6256/152/6/175}

\bibitem[{{Nielsen} {et~al.}(2019){Nielsen}, {De Rosa}, {Macintosh}, {Wang},
  {Ruffio}, {Chiang}, {Marley}, {Saumon}, {Savransky}, {Ammons}, {Bailey},
  {Barman}, {Blain}, {Bulger}, {Burrows}, {Chilcote}, {Cotten}, {Czekala},
  {Doyon}, {Duch{\^e}ne}, {Esposito}, {Fabrycky}, {Fitzgerald}, {Follette},
  {Fortney}, {Gerard}, {Goodsell}, {Graham}, {Greenbaum}, {Hibon}, {Hinkley},
  {Hirsch}, {Hom}, {Hung}, {Dawson}, {Ingraham}, {Kalas}, {Konopacky},
  {Larkin}, {Lee}, {Lin}, {Maire}, {Marchis}, {Marois}, {Metchev},
  {Millar-Blanchaer}, {Morzinski}, {Oppenheimer}, {Palmer}, {Patience},
  {Perrin}, {Poyneer}, {Pueyo}, {Rafikov}, {Rajan}, {Rameau}, {Rantakyr{\"o}},
  {Ren}, {Schneider}, {Sivaramakrishnan}, {Song}, {Soummer}, {Tallis},
  {Thomas}, {Ward-Duong}, \& {Wolff}}]{Nielson19}
{Nielsen}, E.~L., {De Rosa}, R.~J., {Macintosh}, B., {et~al.} 2019, \aj, 158,
  13, \dodoi{10.3847/1538-3881/ab16e9}

\bibitem[{{Nielsen} {et~al.}(2020){Nielsen}, {De Rosa}, {Wang}, {Sahlmann},
  {Kalas}, {Duch{\^e}ne}, {Rameau}, {Marley}, {Saumon}, {Macintosh},
  {Millar-Blanchaer}, {Nguyen}, {Ammons}, {Bailey}, {Barman}, {Bulger},
  {Chilcote}, {Cotten}, {Doyon}, {Esposito}, {Fitzgerald}, {Follette},
  {Gerard}, {Goodsell}, {Graham}, {Greenbaum}, {Hibon}, {Hung}, {Ingraham},
  {Konopacky}, {Larkin}, {Maire}, {Marchis}, {Marois}, {Metchev},
  {Oppenheimer}, {Palmer}, {Patience}, {Perrin}, {Poyneer}, {Pueyo}, {Rajan},
  {Rantakyr{\"o}}, {Ruffio}, {Savransky}, {Schneider}, {Sivaramakrishnan},
  {Song}, {Soummer}, {Thomas}, {Wallace}, {Ward-Duong}, {Wiktorowicz}, \&
  {Wolff}}]{Nielson20}
{Nielsen}, E.~L., {De Rosa}, R.~J., {Wang}, J.~J., {et~al.} 2020, \aj, 159, 71,
  \dodoi{10.3847/1538-3881/ab5b92}

\bibitem[{{Norfolk} {et~al.}(2021){Norfolk}, {Maddison}, {Marshall}, {Kennedy},
  {Duch{\^e}ne}, {Wilner}, {Pinte}, {Mo{\'o}r}, {Matthews}, {{\'A}brah{\'a}m},
  {K{\'o}sp{\'a}l}, \& {van der Marel}}]{Norfolk21}
{Norfolk}, B.~J., {Maddison}, S.~T., {Marshall}, J.~P., {et~al.} 2021, \mnras,
  507, 3139, \dodoi{10.1093/mnras/stab1901}

\bibitem[{Oliphant(2006)}]{Oliphant_06}
Oliphant, T.~E. 2006, A Guide to NumPy, Vol. 1 (Spanish Fork, UT: Trelgol
  Publishing)

\bibitem[{{Olofsson} {et~al.}(2020){Olofsson}, {Milli}, {Bayo}, {Henning}, \&
  {Engler}}]{Olofsson20}
{Olofsson}, J., {Milli}, J., {Bayo}, A., {Henning}, T., \& {Engler}, N. 2020,
  \aap, 640, A12, \dodoi{10.1051/0004-6361/202038237}

\bibitem[{{Olofsson} {et~al.}(2016){Olofsson}, {Samland}, {Avenhaus},
  {Caceres}, {Henning}, {Mo{\'o}r}, {Milli}, {Canovas}, {Quanz}, {Schreiber},
  {Augereau}, {Bayo}, {Bazzon}, {Beuzit}, {Boccaletti}, {Buenzli}, {Casassus},
  {Chauvin}, {Dominik}, {Desidera}, {Feldt}, {Gratton}, {Janson}, {Lagrange},
  {Langlois}, {Lannier}, {Maire}, {Mesa}, {Pinte}, {Rouan}, {Salter},
  {Thalmann}, \& {Vigan}}]{Olofsson16}
{Olofsson}, J., {Samland}, M., {Avenhaus}, H., {et~al.} 2016, \aap, 591, A108,
  \dodoi{10.1051/0004-6361/201628196}

\bibitem[{{Olofsson} {et~al.}(2018){Olofsson}, {van Holstein}, {Boccaletti},
  {Janson}, {Th{\'e}bault}, {Gratton}, {Lazzoni}, {Kral}, {Bayo}, {Canovas},
  {Caceres}, {Ginski}, {Pinte}, {Asensio-Torres}, {Chauvin}, {Desidera},
  {Henning}, {Langlois}, {Milli}, {Schlieder}, {Schreiber}, {Augereau},
  {Bonnefoy}, {Buenzli}, {Brandner}, {Durkan}, {Engler}, {Feldt}, {Godoy},
  {Grady}, {Hagelberg}, {Lagrange}, {Lannier}, {Ligi}, {Maire}, {Mawet},
  {M{\'e}nard}, {Mesa}, {Mouillet}, {Peretti}, {Perrot}, {Salter}, {Schmidt},
  {Sissa}, {Thalmann}, {Vigan}, {Abe}, {Feautrier}, {Le Mignant}, {Moulin},
  {Pavlov}, {Rabou}, {Rousset}, \& {Roux}}]{Olofsson18}
{Olofsson}, J., {van Holstein}, R.~G., {Boccaletti}, A., {et~al.} 2018, \aap,
  617, A109, \dodoi{10.1051/0004-6361/201832583}

\bibitem[{{Olofsson} {et~al.}(2019){Olofsson}, {Milli}, {Th{\'e}bault}, {Kral},
  {M{\'e}nard}, {Janson}, {Augereau}, {Bayo}, {Beam{\'\i}n}, {Henning},
  {Iglesias}, {Kennedy}, {Montesinos}, {Pawellek}, {Schreiber}, {Zamora},
  {Carbillet}, {Feautrier}, {Fusco}, {Madec}, {Rabou}, {Sevin}, {Szul{\'a}gyi},
  \& {Zurlo}}]{Olofsson19}
{Olofsson}, J., {Milli}, J., {Th{\'e}bault}, P., {et~al.} 2019, \aap, 630,
  A142, \dodoi{10.1051/0004-6361/201935998}

\bibitem[{{Olofsson} {et~al.}(2022){Olofsson}, {Th{\'e}bault}, {Kral}, {Bayo},
  {Boccaletti}, {Godoy}, {Henning}, {van Holstein}, {Mauc{\'o}}, {Milli},
  {Montesinos}, {Rein}, \& {Sefilian}}]{Olofsson22}
{Olofsson}, J., {Th{\'e}bault}, P., {Kral}, Q., {et~al.} 2022, \mnras, 513,
  713, \dodoi{10.1093/mnras/stac455}

\bibitem[{Padgett \& Stapelfeldt(2015)}]{padgett_stapelfeldt_2015}
Padgett, D., \& Stapelfeldt, K. 2015, Proceedings of the International
  Astronomical Union, 10, 175–178, \dodoi{10.1017/S1743921315006456}

\bibitem[{{Pan} {et~al.}(2016){Pan}, {Nesvold}, \& {Kuchner}}]{Pan16}
{Pan}, M., {Nesvold}, E.~R., \& {Kuchner}, M.~J. 2016, \apj, 832, 81,
  \dodoi{10.3847/0004-637X/832/1/81}

\bibitem[{{Pan} \& {Schlichting}(2012)}]{Pan12}
{Pan}, M., \& {Schlichting}, H.~E. 2012, \apj, 747, 113,
  \dodoi{10.1088/0004-637X/747/2/113}

\bibitem[{{Pearce} \& {Wyatt}(2014)}]{PW14}
{Pearce}, T.~D., \& {Wyatt}, M.~C. 2014, \mnras, 443, 2541,
  \dodoi{10.1093/mnras/stu1302}

\bibitem[{{Pearce} {et~al.}(2022){Pearce}, {Launhardt}, {Ostermann}, {Kennedy},
  {Gennaro}, {Booth}, {Krivov}, {Cugno}, {Henning}, {Quirrenbach}, {Barcucci},
  {Matthews}, {Ruh}, \& {Stone}}]{Pearce22}
{Pearce}, T.~D., {Launhardt}, R., {Ostermann}, R., {et~al.} 2022, \aap, 659,
  A135, \dodoi{10.1051/0004-6361/202142720}

\bibitem[{{Pecaut} \& {Mamajek}(2016)}]{Pecaut16}
{Pecaut}, M.~J., \& {Mamajek}, E.~E. 2016, \mnras, 461, 794,
  \dodoi{10.1093/mnras/stw1300}

\bibitem[{Perez \& Granger(2007)}]{perez07}
Perez, F., \& Granger, B.~E. 2007, IPython: A System for Interactive Scientific
  Computing, 3, 21-29, 9

\bibitem[{{Perrin} {et~al.}(2014){Perrin}, {Maire}, {Ingraham}, {Savransky},
  {Millar-Blanchaer}, {Wolff}, {Ruffio}, {Wang}, {Draper}, {Sadakuni},
  {Marois}, {Rajan}, {Fitzgerald}, {Macintosh}, {Graham}, {Doyon}, {Larkin},
  {Chilcote}, {Goodsell}, {Palmer}, {Labrie}, {Beaulieu}, {De Rosa},
  {Greenbaum}, {Hartung}, {Hibon}, {Konopacky}, {Lafreniere}, {Lavigne},
  {Marchis}, {Patience}, {Pueyo}, {Rantakyr{\"o}}, {Soummer},
  {Sivaramakrishnan}, {Thomas}, {Ward-Duong}, \& {Wiktorowicz}}]{Perrin14}
{Perrin}, M.~D., {Maire}, J., {Ingraham}, P., {et~al.} 2014, in Society of
  Photo-Optical Instrumentation Engineers (SPIE) Conference Series, Vol. 9147,
  Ground-based and Airborne Instrumentation for Astronomy V, ed. S.~K.
  {Ramsay}, I.~S. {McLean}, \& H.~{Takami}, 91473J, \dodoi{10.1117/12.2055246}

\bibitem[{{Perrin} {et~al.}(2015){Perrin}, {Duchene}, {Millar-Blanchaer},
  {Fitzgerald}, {Graham}, {Wiktorowicz}, {Kalas}, {Macintosh}, {Bauman},
  {Cardwell}, {Chilcote}, {De Rosa}, {Dillon}, {Doyon}, {Dunn}, {Erikson},
  {Gavel}, {Goodsell}, {Hartung}, {Hibon}, {Ingraham}, {Kerley}, {Konapacky},
  {Larkin}, {Maire}, {Marchis}, {Marois}, {Mittal}, {Morzinski}, {Oppenheimer},
  {Palmer}, {Patience}, {Poyneer}, {Pueyo}, {Rantakyr{\"o}}, {Sadakuni},
  {Saddlemyer}, {Savransky}, {Soummer}, {Sivaramakrishnan}, {Song}, {Thomas},
  {Wallace}, {Wang}, \& {Wolff}}]{Perrin15}
{Perrin}, M.~D., {Duchene}, G., {Millar-Blanchaer}, M., {et~al.} 2015, \apj,
  799, 182, \dodoi{10.1088/0004-637X/799/2/182}

\bibitem[{{Plavchan} {et~al.}(2020){Plavchan}, {Barclay}, {Gagn{\'e}}, {Gao},
  {Cale}, {Matzko}, {Dragomir}, {Quinn}, {Feliz}, {Stassun}, {Crossfield},
  {Berardo}, {Latham}, {Tieu}, {Anglada-Escud{\'e}}, {Ricker}, {Vanderspek},
  {Seager}, {Winn}, {Jenkins}, {Rinehart}, {Krishnamurthy}, {Dynes}, {Doty},
  {Adams}, {Afanasev}, {Beichman}, {Bottom}, {Bowler}, {Brinkworth}, {Brown},
  {Cancino}, {Ciardi}, {Clampin}, {Clark}, {Collins}, {Davison},
  {Foreman-Mackey}, {Furlan}, {Gaidos}, {Geneser}, {Giddens}, {Gilbert},
  {Hall}, {Hellier}, {Henry}, {Horner}, {Howard}, {Huang}, {Huber}, {Kane},
  {Kenworthy}, {Kielkopf}, {Kipping}, {Klenke}, {Kruse}, {Latouf}, {Lowrance},
  {Mennesson}, {Mengel}, {Mills}, {Morton}, {Narita}, {Newton}, {Nishimoto},
  {Okumura}, {Palle}, {Pepper}, {Quintana}, {Roberge}, {Roccatagliata},
  {Schlieder}, {Tanner}, {Teske}, {Tinney}, {Vanderburg}, {von Braun}, {Walp},
  {Wang}, {Wang}, {Weigand}, {White}, {Wittenmyer}, {Wright}, {Youngblood},
  {Zhang}, \& {Zilberman}}]{Plavchan20}
{Plavchan}, P., {Barclay}, T., {Gagn{\'e}}, J., {et~al.} 2020, \nat, 582, 497,
  \dodoi{10.1038/s41586-020-2400-z}

\bibitem[{{Ren} {et~al.}(2019){Ren}, {Choquet}, {Perrin}, {Duch{\^e}ne},
  {Debes}, {Pueyo}, {Rice}, {Chen}, {Schneider}, {Esposito}, {Poteet}, {Wang},
  {Ammons}, {Ansdell}, {Arriaga}, {Bailey}, {Barman}, {Sebasti{\'a}n Bruzzone},
  {Bulger}, {Chilcote}, {Cotten}, {De Rosa}, {Doyon}, {Fitzgerald}, {Follette},
  {Goodsell}, {Gerard}, {Graham}, {Greenbaum}, {Hagan}, {Hibon}, {Hines},
  {Hung}, {Ingraham}, {Kalas}, {Konopacky}, {Larkin}, {Macintosh}, {Maire},
  {Marchis}, {Marois}, {Mazoyer}, {M{\'e}nard}, {Metchev}, {Millar-Blanchaer},
  {Mittal}, {Moerchen}, {Nielsen}, {N'Diaye}, {Oppenheimer}, {Palmer},
  {Patience}, {Pinte}, {Poyneer}, {Rajan}, {Rameau}, {Rantakyr{\"o}}, {Ruffio},
  {Ryan}, {Savransky}, {Schneider}, {Sivaramakrishnan}, {Song}, {Soummer},
  {Stark}, {Thomas}, {Vigan}, {Wallace}, {Ward-Duong}, {Wiktorowicz}, {Wolff},
  {Ygouf}, \& {Norman}}]{Ren19}
{Ren}, B., {Choquet}, {\'E}., {Perrin}, M.~D., {et~al.} 2019, \apj, 882, 64,
  \dodoi{10.3847/1538-4357/ab3403}

\bibitem[{{Ren} {et~al.}(2021){Ren}, {Choquet}, {Perrin}, {Mawet}, {Chen},
  {Milli}, {Debes}, {Rebollido}, {Stark}, {Hagan}, {Hines}, {Millar-Blanchaer},
  {Pueyo}, {Roberge}, {Schneider}, {Serabyn}, {Soummer}, \& {Wolff}}]{Ren21}
---. 2021, \apj, 914, 95, \dodoi{10.3847/1538-4357/ac03b9}

\bibitem[{{Ren} {et~al.}(2023){Ren}, {Rebollido}, {Choquet}, {Zhou}, {Perrin},
  {Schneider}, {Milli}, {Wolff}, {Chen}, {Debes}, {Hagan}, {Hines},
  {Millar-Blanchaer}, {Pueyo}, {Roberge}, {Serabyn}, \& {Soummer}}]{Ren23}
{Ren}, B.~B., {Rebollido}, I., {Choquet}, {\'E}., {et~al.} 2023, arXiv
  e-prints, arXiv:2302.04273, \dodoi{10.48550/arXiv.2302.04273}

\bibitem[{{Schneider} {et~al.}(2014){Schneider}, {Grady}, {Hines}, {Stark},
  {Debes}, {Carson}, {Kuchner}, {Perrin}, {Weinberger}, {Wisniewski},
  {Silverstone}, {Jang-Condell}, {Henning}, {Woodgate}, {Serabyn},
  {Moro-Martin}, {Tamura}, {Hinz}, \& {Rodigas}}]{Schneider14}
{Schneider}, G., {Grady}, C.~A., {Hines}, D.~C., {et~al.} 2014, \aj, 148, 59,
  \dodoi{10.1088/0004-6256/148/4/59}

\bibitem[{{Schneiderman} {et~al.}(2021){Schneiderman}, {Matr{\`a}}, {Jackson},
  {Kennedy}, {Kral}, {Marino}, {{\"O}berg}, {Su}, {Wilner}, \&
  {Wyatt}}]{Schneiderman21}
{Schneiderman}, T., {Matr{\`a}}, L., {Jackson}, A.~P., {et~al.} 2021, \nat,
  598, 425, \dodoi{10.1038/s41586-021-03872-x}

\bibitem[{{Smirnov-Pinchukov} {et~al.}(2022){Smirnov-Pinchukov}, {Mo{\'o}r},
  {Semenov}, {{\'A}brah{\'a}m}, {Henning}, {K{\'o}sp{\'a}l}, {Hughes}, \& {di
  Folco}}]{Smirnov22}
{Smirnov-Pinchukov}, G.~V., {Mo{\'o}r}, A., {Semenov}, D.~A., {et~al.} 2022,
  \mnras, 510, 1148, \dodoi{10.1093/mnras/stab3146}

\bibitem[{{Smith} \& {Terrile}(1984)}]{Smith84}
{Smith}, B.~A., \& {Terrile}, R.~J. 1984, Science, 226, 1421,
  \dodoi{10.1126/science.226.4681.1421}

\bibitem[{{Soummer} {et~al.}(2014){Soummer}, {Perrin}, {Pueyo}, {Choquet},
  {Chen}, {Golimowski}, {Hagan}, {Mittal}, {Moerchen}, {N'Diaye}, {Rajan},
  {Wolff}, {Debes}, {Hines}, \& {Schneider}}]{Soummer14}
{Soummer}, R., {Perrin}, M.~D., {Pueyo}, L., {et~al.} 2014, \apjl, 786, L23,
  \dodoi{10.1088/2041-8205/786/2/L23}

\bibitem[{{Stasevic} {et~al.}(2023){Stasevic}, {Milli}, {Mazoyer}, {Lagrange},
  {Bonnefoy}, {Faramaz-Gorka}, {M{\'e}nard}, {Boccaletti}, {Choquet}, {Shuai},
  {Olofsson}, {Chomez}, {Ren}, {Rubini}, {Desgrange}, {Gratton}, {Chauvin},
  {Vigan}, \& {Matthews}}]{stasevic23}
{Stasevic}, S., {Milli}, J., {Mazoyer}, J., {et~al.} 2023, \aap, 678, A8,
  \dodoi{10.1051/0004-6361/202346720}

\bibitem[{{Takasawa} {et~al.}(2011){Takasawa}, {Nakamura}, {Kadono}, {Arakawa},
  {Dohi}, {Ohno}, {Seto}, {Maeda}, {Shigemori}, {Hironaka}, {Sakaiya},
  {Fujioka}, {Sano}, {Otani}, {Watari}, {Sangen}, {Setoh}, {Machii}, \&
  {Takeuchi}}]{Takasawa11}
{Takasawa}, S., {Nakamura}, A.~M., {Kadono}, T., {et~al.} 2011, \apjl, 733,
  L39, \dodoi{10.1088/2041-8205/733/2/L39}

\bibitem[{{Telesco} {et~al.}(2005){Telesco}, {Fisher}, {Wyatt}, {Dermott},
  {Kehoe}, {Novotny}, {Mari{\~n}as}, {Radomski}, {Packham}, {De Buizer}, \&
  {Hayward}}]{Telesco05}
{Telesco}, C.~M., {Fisher}, R.~S., {Wyatt}, M.~C., {et~al.} 2005, \nat, 433,
  133, \dodoi{10.1038/nature03255}

\bibitem[{{Terrill} {et~al.}(2023){Terrill}, {Marino}, {Booth}, {Han},
  {Jennings}, \& {Wyatt}}]{Terrill23}
{Terrill}, J., {Marino}, S., {Booth}, R.~A., {et~al.} 2023, \mnras,
  \dodoi{10.1093/mnras/stad1847}

\bibitem[{{Thalmann} {et~al.}(2013){Thalmann}, {Janson}, {Buenzli}, {Brandt},
  {Wisniewski}, {Dominik}, {Carson}, {McElwain}, {Currie}, {Knapp},
  {Moro-Mart{\'\i}n}, {Usuda}, {Abe}, {Brandner}, {Egner}, {Feldt}, {Golota},
  {Goto}, {Guyon}, {Hashimoto}, {Hayano}, {Hayashi}, {Hayashi}, {Henning},
  {Hodapp}, {Ishii}, {Iye}, {Kandori}, {Kudo}, {Kusakabe}, {Kuzuhara}, {Kwon},
  {Matsuo}, {Mayama}, {Miyama}, {Morino}, {Nishimura}, {Pyo}, {Serabyn},
  {Suto}, {Suzuki}, {Takami}, {Takato}, {Terada}, {Tomono}, {Turner},
  {Watanabe}, {Yamada}, {Takami}, \& {Tamura}}]{Thalmann13}
{Thalmann}, C., {Janson}, M., {Buenzli}, E., {et~al.} 2013, \apjl, 763, L29,
  \dodoi{10.1088/2041-8205/763/2/L29}

\bibitem[{{The Astropy Collaboration} {et~al.}(2018){The Astropy
  Collaboration}, Price-Whelan, Sip{\H o}cz, G{\"u}nther, Lim, Crawford,
  Conseil, Shupe, Craig, Dencheva, Ginsburg, VanderPlas, Bradley,
  P{\'e}rez-Su{\'a}rez, de~Val-Borro, Aldcroft, Cruz, Robitaille, Tollerud,
  Ardelean, Babej, Bachetti, Bakanov, Bamford, \&
  Barentsen}]{Collaboration:2018ab}
{The Astropy Collaboration}, Price-Whelan, A.~M., Sip{\H o}cz, B.~M., {et~al.}
  2018, \dodoi{10.3847/1538-3881/aabc4f}

\bibitem[{{Thebault} \& {Kral}(2019)}]{TK19}
{Thebault}, P., \& {Kral}, Q. 2019, \aap, 626, A24,
  \dodoi{10.1051/0004-6361/201935341}

\bibitem[{{Thilliez} \& {Maddison}(2017)}]{Thilliez17}
{Thilliez}, E., \& {Maddison}, S.~T. 2017, \mnras, 464, 1434,
  \dodoi{10.1093/mnras/stw2427}

\bibitem[{{Torres} {et~al.}(2006){Torres}, {Quast}, {da Silva}, {de La Reza},
  {Melo}, \& {Sterzik}}]{Torres06}
{Torres}, C.~A.~O., {Quast}, G.~R., {da Silva}, L., {et~al.} 2006, \aap, 460,
  695, \dodoi{10.1051/0004-6361:20065602}

\bibitem[{Virtanen {et~al.}(2020)Virtanen, Gommers, Oliphant, Haberland, Reddy,
  Cournapeau, Burovski, \& Peterson}]{Virtanen_20}
Virtanen, P., Gommers, R., Oliphant, T.~E., {et~al.} 2020, Nature Methods, 17,
  261-272

\bibitem[{{Vizgan} {et~al.}(2022){Vizgan}, {Hughes}, {Carter}, {Flaherty},
  {Pan}, {Chiang}, {Schlichting}, {Wilner}, {Andrews}, {Carpenter}, {Mo{\'o}r},
  \& {MacGregor}}]{Vizgan22}
{Vizgan}, D., {Hughes}, A.~M., {Carter}, E.~S., {et~al.} 2022, \apj, 935, 131,
  \dodoi{10.3847/1538-4357/ac80b8}

\bibitem[{{Wahhaj} {et~al.}(2016){Wahhaj}, {Milli}, {Kennedy}, {Ertel},
  {Matr{\`a}}, {Boccaletti}, {del Burgo}, {Wyatt}, {Pinte}, {Lagrange},
  {Absil}, {Choquet}, {G{\'o}mez Gonz{\'a}lez}, {Kobayashi}, {Mawet},
  {Mouillet}, {Pueyo}, {Dent}, {Augereau}, \& {Girard}}]{Wahhaj16}
{Wahhaj}, Z., {Milli}, J., {Kennedy}, G., {et~al.} 2016, \aap, 596, L4,
  \dodoi{10.1051/0004-6361/201629769}

\bibitem[{{Wang} {et~al.}(2014){Wang}, {Rajan}, {Graham}, {Savransky},
  {Ingraham}, {Ward-Duong}, {Patience}, {De Rosa}, {Bulger},
  {Sivaramakrishnan}, {Perrin}, {Thomas}, {Sadakuni}, {Greenbaum}, {Pueyo},
  {Marois}, {Oppenheimer}, {Kalas}, {Cardwell}, {Goodsell}, {Hibon}, \&
  {Rantakyr{\"o}}}]{Wang14}
{Wang}, J.~J., {Rajan}, A., {Graham}, J.~R., {et~al.} 2014, in Society of
  Photo-Optical Instrumentation Engineers (SPIE) Conference Series, Vol. 9147,
  Ground-based and Airborne Instrumentation for Astronomy V, ed. S.~K.
  {Ramsay}, I.~S. {McLean}, \& H.~{Takami}, 914755, \dodoi{10.1117/12.2055753}

\bibitem[{{Wyatt}(2008)}]{Wyatt08}
{Wyatt}, M.~C. 2008, \araa, 46, 339,
  \dodoi{10.1146/annurev.astro.45.051806.110525}

\bibitem[{{Wyatt} {et~al.}(2011){Wyatt}, {Clarke}, \& {Booth}}]{Wyatt11}
{Wyatt}, M.~C., {Clarke}, C.~J., \& {Booth}, M. 2011, Celestial Mechanics and
  Dynamical Astronomy, 111, 1, \dodoi{10.1007/s10569-011-9345-3}

\bibitem[{{Wyatt} {et~al.}(1999){Wyatt}, {Dermott}, {Telesco}, {Fisher},
  {Grogan}, {Holmes}, \& {Pi{\~n}a}}]{Wyatt99}
{Wyatt}, M.~C., {Dermott}, S.~F., {Telesco}, C.~M., {et~al.} 1999, \apj, 527,
  918, \dodoi{10.1086/308093}

\bibitem[{{Zakhozhay} {et~al.}(2022){Zakhozhay}, {Launhardt}, {Trifonov},
  {K{\"u}rster}, {Reffert}, {Henning}, {Brahm}, {Vin{\'e}s}, {Marleau}, \&
  {Patel}}]{Zakhozhay22}
{Zakhozhay}, O.~V., {Launhardt}, R., {Trifonov}, T., {et~al.} 2022, \aap, 667,
  L14, \dodoi{10.1051/0004-6361/202244747}

\bibitem[{{Zuckerman}(2019)}]{Zuckerman19}
{Zuckerman}, B. 2019, \apj, 870, 27, \dodoi{10.3847/1538-4357/aaee66}

\end{thebibliography}
\bibliographystyle{aasjournal}

\appendix
\renewcommand{\thesubsection}{\Alph{subsection}}


\begin{figure*}
        \centering
	\caption{\label{Fig:SN_H} \textbf{Top:} Signal to Noise maps for each disk with multiwavelength observations. The circles represent the size of the FPM in $J$, $H$, and $K1$ ($0.09''$, $0.12''$ and $0.15''$ respectively), and the crosses represent the location of the star. \textbf{Bottom:} Signal to Noise maps for the remaining disks with observations in the $H$ band only. The circles represent the size of the FPM in $H$ band ($0.12''$), and the crosses represent the location of the star.}
	\includegraphics[width=0.85\textwidth]{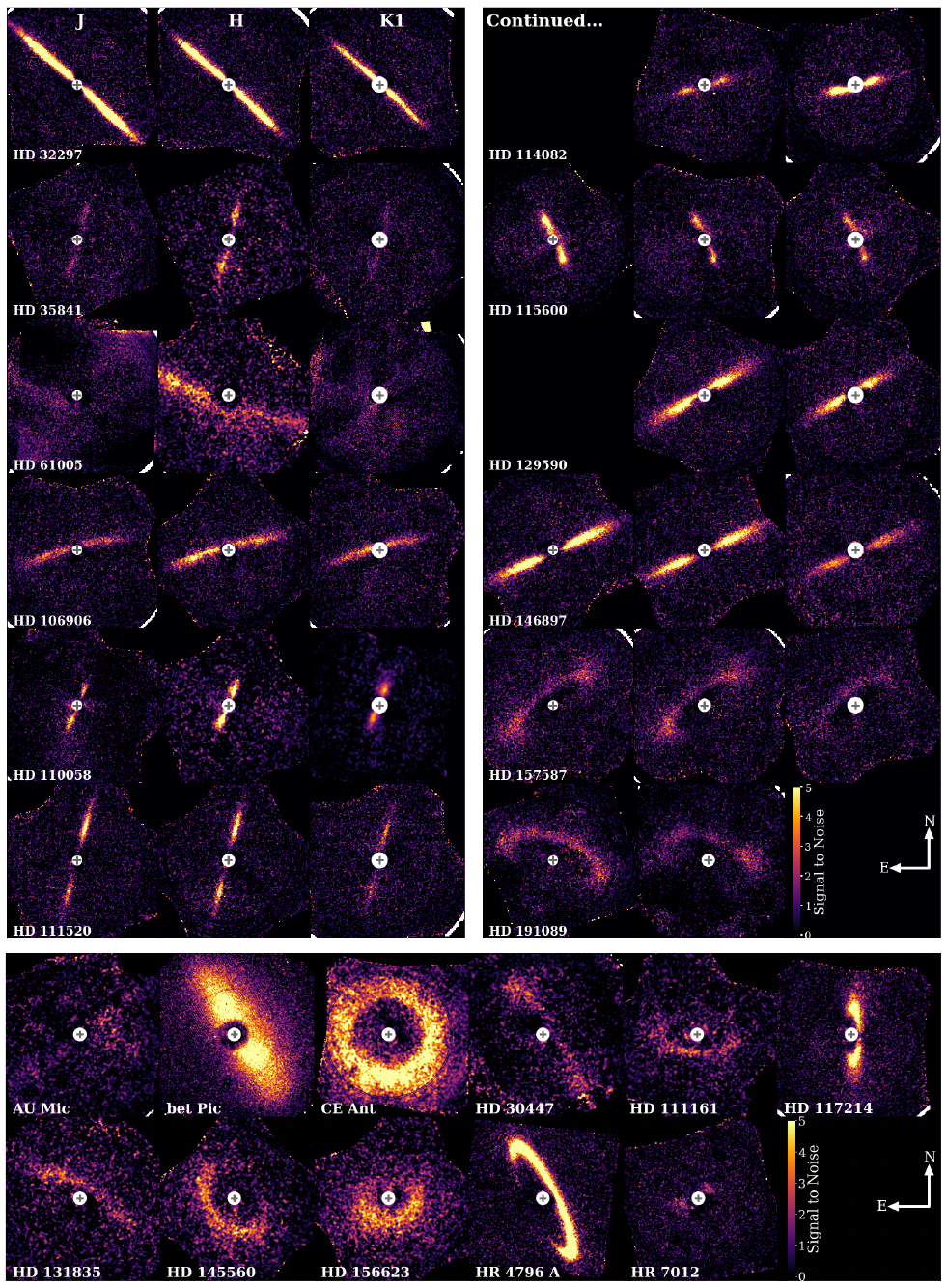}
\end{figure*}

\begin{figure*}
\centering
	\caption{\label{Fig:spine_H_data} $H$-band observations rotated by their $PA - 90^{\circ}$ and overlaid with their best fitting ring model (orange curves).}
	\includegraphics[width=\textwidth]{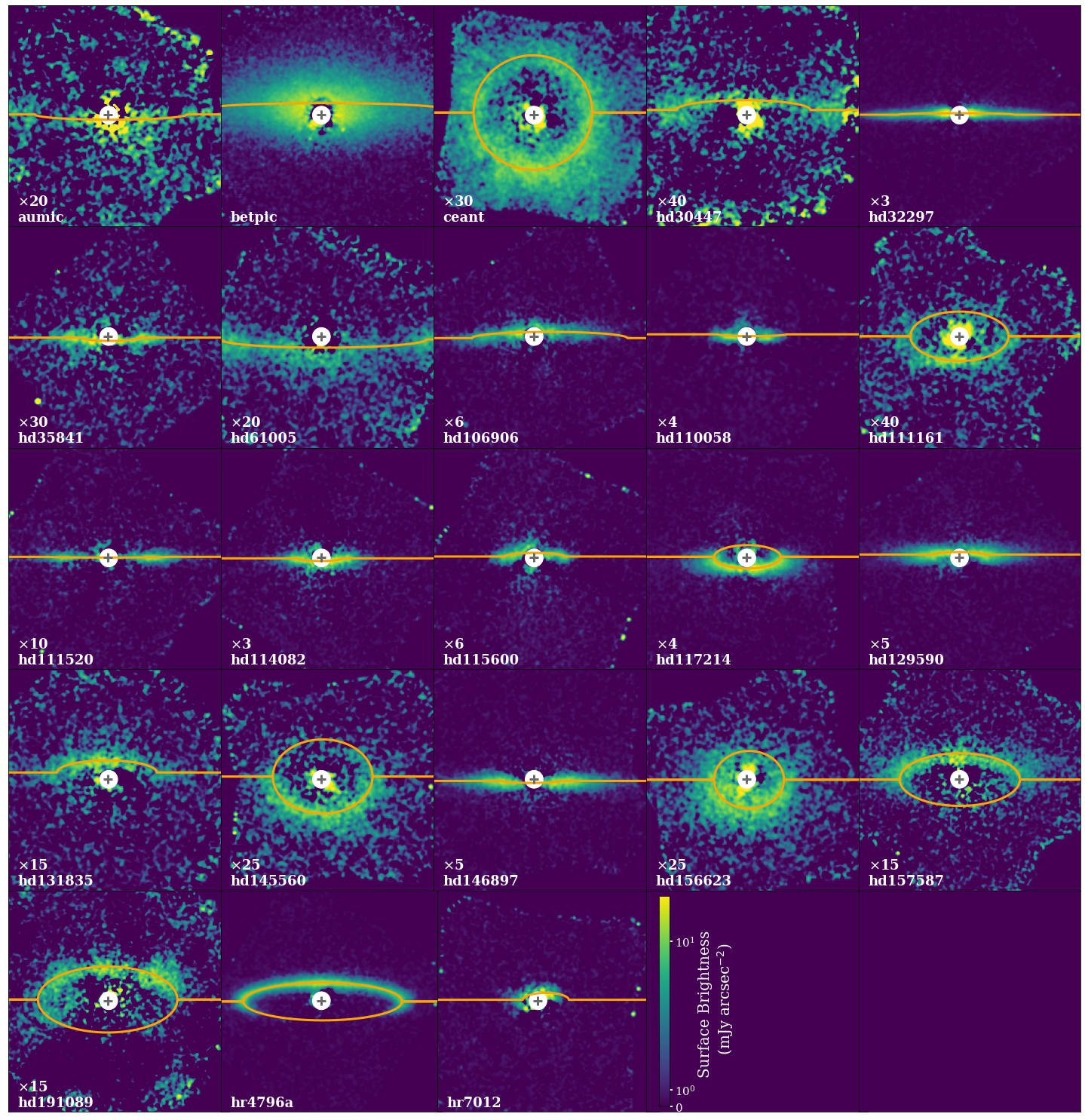}
\end{figure*}

\subsection{S/N Maps and Vertical/Radial Offset Profiles}
We present here additional figures of the debris disks in our sample. Figure \ref{Fig:SN_H} shows S/N maps of each disk in the $J$, $H$ and $K1$ bands. These are created by dividing the noise maps derived from $U_{\phi}$ from our $Q_{\phi}$ images (see Section \ref{sec:observations}. Every disk is scaled between a S/N of 0 to 5. Figure \ref{Fig:spine_H_data} shows the best fitting ring models overlaid on top of the $H$-band data. Each disk is rotated by its measured $PA - 90^{\circ}$, so that the disk major-axis is horizontal in the image.

\subsection{East vs. West Frame of Reference} 
Throughout this paper, we refer to the two extensions of each disk as the East and West sides/extensions. These definitions are based on a consistent frame of reference, rather than the original cardinal directions. To create the new frame of reference, we simply rotate each disk clockwise/counterclockwise so that the disk major-axis is horizontal in the image. The degrees rotated is equal to the disk $PA - 90^{\circ}$. In Table \ref{tab:rotate}, we list the angle each disk is rotated, as well as the change in cardinal directions from the original frame of reference to the new frame of reference.

\begin{table}
\centering
	\caption{\label{tab:rotate} Degrees each disk is rotated to create a consistent disk orientation and frame of reference between all 23 disks, i.e. disk emission left of the star = East side, disk emission right of the star = West side. Positive values represent clockwise rotation, while negative values represent counter-clockwise rotation. Column three shows the change in cardinal directions to East and West for both sides of each disk.}
	\begin{tabular}{ccc}
	    \hline
	    \hline
		Disk & Degrees Rotated & Cardinal Change \\
		\hline
            AU Mic & 36.7 & SE $\rightarrow$ E, NW $\rightarrow$ W \\
            $\beta$ Pic & -57.8 & NE $\rightarrow$ E, SW $\rightarrow$ W \\
            CE Ant & 1.02 & E $\rightarrow$ E, W $\rightarrow$ W \\
            HD 30447 & -56.4 & NE $\rightarrow$ E, SW $\rightarrow$ W \\
            HD 32297 & -42.4 & NE $\rightarrow$ E, SW $\rightarrow$ W \\
            HD 35841 & 77.5 & SE $\rightarrow$ E, NW $\rightarrow$ W \\
            HD 61005 & -19.2 & NE $\rightarrow$ E, SW $\rightarrow$ W \\
            HD 106906 & 14.0 & SE $\rightarrow$ E, NW $\rightarrow$ W \\
            HD 110058 & 68.6 & SE $\rightarrow$ E, NW $\rightarrow$ W \\
            HD 111161 & -6.7 & E $\rightarrow$ E, W $\rightarrow$ W \\
            HD 111520 & 75.7 & SE $\rightarrow$ E, NW $\rightarrow$ W \\
            HD 114082 & 15.0 & SE $\rightarrow$ E, NW $\rightarrow$ W \\
            HD 115600 & -65.8 & NE $\rightarrow$ E, SW $\rightarrow$ W \\
            HD 117214 & 90.5 & S $\rightarrow$ E, N $\rightarrow$ W \\
            HD 129590 & 30.3 & SE $\rightarrow$ E, NW $\rightarrow$ W \\
            HD 131835 & -29.2 & NE $\rightarrow$ E, SW $\rightarrow$ W \\
            HD 145560 & -50.5 & NE $\rightarrow$ E, SW $\rightarrow$ W \\
            HD 146897 & 24.6 & SE $\rightarrow$ E, NW $\rightarrow$ W \\
            HD 156623 & 12.9 & E $\rightarrow$ E, W $\rightarrow$ W \\
            HD 157587 & 37.7 & SE $\rightarrow$ E, NW $\rightarrow$ W \\
            HD 191089 & -18.2 & NE $\rightarrow$ E, SW $\rightarrow$ W \\
            HR 4796 A & -63.6 & NE $\rightarrow$ E, SW $\rightarrow$ W \\
            HR 7012 & 23.8 & SE $\rightarrow$ E, NW $\rightarrow$ W \\
            \hline
            \hline
        \end{tabular}
\end{table}

\subsection{Individual Disk Results} \label{sec:summary}
In the previous Sections, we describe the methods used in this study, as well as the results of these methods when applied to our sample of GPI disks. In this Section, we discuss the summary of our empirical analysis for each disk, and compare our results to those in the literature. We focus mainly on new and/or the most interesting results, while more minor results such as the inclination and $PA$ are not highlighted unless they deviate significantly from previous results or in the cases where the disk is not well studied. 

\subsubsection*{AU Mic}
The AU Mic debris disk is one of only two disks in our sample that resides around an M-type star. Debris disks resolved around M-type stars in general are fairly rare given observational biases. AU Mic is a particularly interesting system, as the disk shows peculiar clumps of dust moving outwards from the star \citep{Boccaletti18}, along with two known planets recently discovered through the transit and radial velocity methods \citep{Plavchan20, Martioli21}. 

The GPI observations for AU Mic are very low S/N, as can be seen in Figure \ref{Fig:SN_H}, and the disk extends beyond GPI's FOV, making it difficult to obtain consistent values for the disk geometry. For example, we obtain a value of $\sim$86$^{\circ}$ for the inclination, when the literature reports an inclination between $88^{\circ}$ and $90^{\circ}$. We also measure a small disk radius of $\sim$10 AU. While this value is consistent within the large uncertainties of previous inner radius measurements, the most recent ALMA data suggests an inner radius around 22 AU, more than twice our measured disk radii measurement \citep{Vizgan22}. While it may simply be that we are unable to probe the true disk radius, given that the disk extends outside GPI's FOV, our derived radius could be a sign of a second disk component. Most recently, by using the code \textit{Frankenstein}, which can deproject disks at any inclination to reveal their radial distribution, \citet{Terrill23} find a second smaller peak in intensity at 10 au using ALMA observations. While this is a tentative detection, the same result has been found in multiple other studies in support of a second disk component around 10 au \citep{Daley19,Marino21,Han22}.

In the $H$ band, we find that the West side of the disk is about 1.4 times brighter than the East side. This brightness asymmetry can also be observed in the most recent SPHERE data \citep{Langlois21, Olofsson22}, in which the disk is better resolved. Currently, no literature reports any disk offsets or eccentricities, consistent with the almost zero offset detected with our ring fitting. However, with the low S/N and high inclination, it is very unlikely we would be able to constrain a disk offset along the major-axis. This does not necessarily mean that no offset exists, especially given that there is a clear brightness asymmetry; however, mm-observations with ALMA show an axisymmetric disk \citep{Vizgan22}, suggesting that this brightness asymmetry and any possible disk offsets are only present in smaller grains. Whether this asymmetry is tied to planets in the system is unclear. The known planets in this system orbit very close to the star, making them dynamically decoupled from the disk, and efforts to search for additional planets farther out have yielded no candidates \citep{Gallenne22}. A much more likely scenario would be that the distribution of dust grains in the disk is being altered. Although we do not have multiwavelength observations to test this, other studies have shown that the outward moving clumps of small dust in the system may be the result of a combination of stellar winds and a catastrophic collision \citep{Chiang17}. This would explain why the disk is highly asymmetric at shorter wavelengths, while being more axisymmetric at longer wavelengths. 

\subsubsection*{$\beta$ Pic}
The $\beta$ Pic debris disk is one of the most well-studied and well-known debris disks to date. Because the system is close (19.44 au), and the disk is particularly bright compared to other debris disks, $\beta$ Pic was the first debris disk to ever be imaged \citep{Smith84}. The disk also hosts multiple interesting features including a warp, brightness asymmetry, radial asymmetry and a clump seen in ALMA observations to name a few \citep{Heap00, Kalas95, Janson21, Telesco05}. On top of this, the disk is one of only a handful of resolved disks with directly imaged planets, $\beta$ Pic b and c \citep{Lagrange10, Lagrange19}, which can be directly linked to the disk's perturbed morphology \citep{Chauvin12}.

While $\beta$ Pic is known to have several asymmetries, the GPI observations show a fairly axisymmetric disk. We detect only a modest brightness asymmetry with a stronger West extension. Our vertical offset fitting also finds no significant offset along the major axis. For other aspects of the disk geometry, we find some discrepancies between our results and previous results in the literature. For one, our obtained inclination of 88.9$^{\circ}$ is several degrees higher than previous estimates of $\sim$85$^{\circ}$. This difference may be the result of the $\beta$ Pic disk being not radially narrow, or that the disk extends farther out than GPI's FOV, preventing us from fitting the entire vertical offset profile. In either case, $\beta$ Pic is an example of limitations of our modelling technique, and likely requires a more complex model to capture the disk's complex morphology.

Although the surface brightness and disk geometry appear mostly axisymmetric, one interesting feature is $\beta$ Pic's vertical width, where it has the highest aspect ratio of our entire sample. This is consistent with previous measurements, where $\beta$ Pic has been found to have a relatively large vertical aspect ratio compared to other debris disks \citep{Olofsson22}. The implications of $\beta$ Pic's high aspect ratio is discussed in Section \ref{sec:fwhm_discussion}.

\subsubsection*{CE Ant}
First imaged by \citet{Choquet16}, CE Ant, also known as TWA 7, is the lowest inclined disk in our sample, as well as the only one where the entire back side of the disk is visible. It is also the second disk that is around an M-type star. The CE Ant disk is a very interesting case as it is one of only a few debris disks with observed multiple rings in scattered light and also exhibits a spiral arm \citep{Olofsson18, Ren21}: features that would otherwise be unobservable if the disk was higher inclined. Due to the FOV of GPI, we can only see the inner ring.

\begin{figure}
	\caption{\label{Fig:ce_ant_spiral} CE Ant (TWA 7) overlaid with surface brightness contours to help highlight the spiral arm first detected in the SPHERE observations \citep{Olofsson18}. The red box defines the location of the spiral arm, which is only marginally detected in our GPI data. The white circle represents the size of the FPM, while the grey cross represents the location of the star.}
	\includegraphics[width=.47\textwidth]{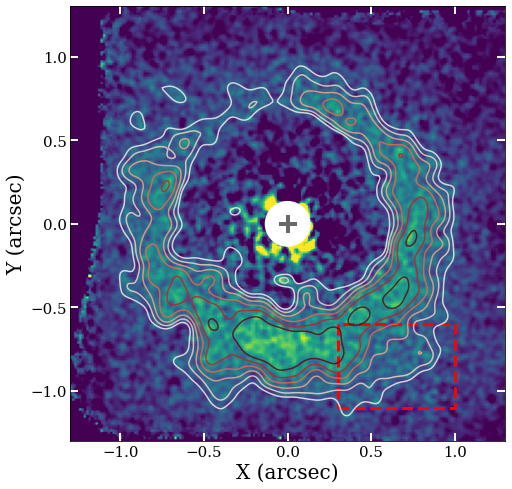}
\end{figure}

Because the entire disk is visible, we are able to fit the full disk geometry. While no eccentricity has been reported, we detect small offsets along both the major and minor axes of $\sim$0.86 au, leading to an eccentricity of 0.03. While this eccentricity is small, we do measure a significant brightness asymmetry, where the West side of the disk is $\sim$1.13 times brighter than the East side. This brightness asymmetry could be the result of an eccentric disk; however, given that the spiral arm is located on the West side, this may also be contributing to the surface brightness of the West extension. Although the spiral arm is not strongly detected in the GPI observations, likely due to being located towards the outer edge of the GPI's FOV, Figure \ref{Fig:ce_ant_spiral} shows the location of the spiral arm which can be made out slightly using surface brightness contours.

While we detect a small eccentricity and modest brightness asymmetry, there are other reasons to believe that there are planets shaping the disk. For one, there is a stark inner clearing within $\sim$0.5$''$ (17 AU). Additionally the disk harbours multiple rings and a spiral arm, all of which are strong indications of one or more planets shaping the disk.

\subsubsection*{HD 30447}
The GPI observations in this study, first published in \citet{Esposito20}, represents one of only two observations total of the HD 30447 debris disk. While these GPI observations are relatively low S/N compared to the rest of the sample, the disk is still better resolved in polarized intensity compared to previous HST observations \citep{Soummer14}. From visual inspection, the disk appears to be highly inclined, with an inner clearing within $\sim$0.8$''$. 

Measuring the disk geometry, we obtain a radius of 75.43 AU, an inclination of $81.47^{\circ}$, and a PA of $213.56^{\circ}$, consistent with measurements done in \citet{Esposito20} within uncertainties. Interestingly, we find a clear disk offset along the major-axis of 6.53 au, bringing the star closer to the West side of the disk in the case of an eccentric disk. However, measuring the surface brightness between the East and West extensions, we find that East extension is 1.13 times brighter than the West extension, consistent with observations of the disk with HST \citep{Soummer14}. If this surface brightness asymmetry is due to a pericenter glow \citep{Wyatt99,Pan16}, we would expect the offset along the major-axis in the opposite direction. When analyzing the surface brightness profile, the polarized intensity peaks between 0.75$''$ and 1.15$''$ from the star in the East extension, while the polarized intensity in the West extension peaks beyond 1.15$''$, suggesting the the East extension may indeed be closer to the star than the West extension. One explanation is that the derived offset could be the result of another geometrical asymmetry rather than an eccentric disk, or the data have too low S/N to properly constrain $\delta_{x}$. Furthermore, more recent HST observations show the East side of the disk halo to be more radially extended than the West side \citep{Ren23}, similar to the disk HD 111520 (discussed later in this Section), where the brighter and possibly closer side of the disk is more radially extended.

The HD 30447 debris disk appears to be perturbed in some manner; while the surface brightness suggests an eccentric disk with the East extension closer to the star, the disk geometry suggests the opposite. To learn more about the source of these asymmetries, multiwavelength and higher S/N observations are essential. A more in depth analysis of the disk halo as observed with HST, may also be helpful to understand the disk structure as a whole.

\subsubsection*{HD 32297}
Like $\beta$ Pic, the HD 32297 debris disk has been studied in great detail over the past two decades. Not only is it bright compared the majority of other disks in our sample, but it also has one of the strongest detections of gas emission \citep{Donaldson13,Greaves16,MacGregor18,Cataldi20}. In the optical, the disk halo can be seen, which extends to at least 1800 au \citep{Schneider14}, and appears to have an interesting curved ``moth"-like morphology. This morphology was originally thought to have been caused by an interaction with a dense portion of the interstellar medium \citep{Debes09}, and is currently thought to be the result of planet-disk interactions \citep{LC16}. Here we introduce the first observations of the disk in the $J$ and $K1$ bands.

We compare our geometrical results with those from \citet{Duchene20}, as they performed a similar ring model fitting to the vertical offset profile. We have, however, included two extra parameters, the $PA$ and $\delta_{y}$. While \citet{Duchene20} found no offset along the major axis, we find a significant offset of 4.6 au towards the East, bringing the West side of the disk closer to the star. To confirm this offset, we also fit a model to $J$-band observations, which has a similarly high S/N, and find that the vertical offset profile also exhibits a $\sim$4.6 au offset. Our results still lie within the 3$\sigma$ upper limit on the eccentricity of 0.05 \citep{Duchene20}, as the derived offset leads to an eccentricity of 0.04. 

Similar to past studies, we find no evidence of a significant brightness asymmetry in any of the three bands. We also find no evidence of an asymmetry in disk color, although the disk appears to be very blue in $J$-$K1$ and $J$-$H$, while being close to neutral in $H$-$K1$, as was similarly found in \citet{Browhmik19}. This is likely caused by the drastic increase in surface brightness in the $J$ band, compared to the $H$ and $K1$ bands. In all, while the HD 32297 debris disk may have a slight offset along the major axis, the eccentricity of the disk is modest at most, and is otherwise axisymmetric. 

\subsubsection*{HD 35841}
The HD 35841 debris disk is a slightly more compact, highly inclined disk that has only been detected so far in the NIR and in the optical with HST \citep{Soummer14}, although newer/higher resolution observations with HST have been presented in \citet{Ren23}. While an in depth study has been done already with the $H$ band data \citep{Esposito18}, we present here the $J$- and $K1$-band data for the first time, allowing for a multiwavelength study.

For the disk geometry, we find a disk radius of 39.12 AU. Interestingly, this radius is within the estimated inner radius of 59.8 au based on radiative-transfer modelling \citep{Esposito18}. Given the high inclination of $83^{\circ}$ (slightly lower than the estimated inclination of $85^{\circ}$ from \citealt{Esposito20}), it may simply be that it is difficult to probe the inner radius. Therefore, it is possible that the minimum radius is actually closer to the star than what is determined with radiative transfer modelling. We also derive an offset along the major axis of $\sim$1 au towards the East extension, although, considering the S/N, this small offset is unlikely to be significant. 

No significant brightness asymmetry is found, consistent with previous measurements \citep{Esposito18}. Additionally, no disk color asymmetry is found between the East and West extensions. While \citet{Esposito18} found a slight blue color between the $H$ band and HST observations, between the $J$, $H$, and $K1$ bands, the disk presents a neutral color in polarized intensity. Overall, the HD 35841 debris disk is found to be axisymmetric.  

\subsubsection*{HD 61005}
HD 61005 is another well studied disk, with multiwavelength observations and an interesting morphology. In the optical, as observed with HST, the disk halo has a swept back morphology, giving it the nickname ``the Moth" \citep{Hines07}. This feature, similar to HD 32297, was originally thought to have been caused by an interaction with the ISM, although later simulations done by \citet{LC16} and \citet{Jones23} show that this morphology can also be created by a planet-disk interaction and a recent giant impact. NIR observations with SPHERE and GPI also show a large brightness asymmetry, with the East side being twice as bright as the West side \citep{Olofsson16, Esposito16}. On the other hand, ALMA data show a millimeter belt that is fairly axisymmetric. Here we discuss the results from our multiwavelength GPI data.

Unfortunately, we do not detect the significant offset along the major axis detected with the SPHERE observations \citep{Olofsson16}, which led to an estimated eccentricity of $\sim$0.1. Given that the SPHERE observations have a higher S/N compared to the GPI observations and that the disk extends beyond GPI's FOV, we may not have the sensitivity to detect this offset. We do, however, detect the brightness asymmetry in all three bands, with the East extension being much brighter than the West extension. We find in the $H$ band, which has the highest S/N out of the three bands, that the East side is $\sim$1.6 times brighter than the West side. This brightness asymmetry is much greater in the $J$ and $K1$ bands, where the East side is 6.3 and 2.6 times brighter, respectively. However, its important to note that these two observations are relatively low S/N, and therefore these brightness asymmetry measurements may not be exact. In addition to a large surface brightness asymmetry, we also find a significant color asymmetry over 3$\sigma$ in $J$-$H$ and $J$-$K1$ between the two sides of the disk. All three measurements show a distinctly blue disk color, which is consistent with past measurements \citep{Esposito16}.

Whether or not these asymmetries are associated with an interaction with the ISM, a planet-disk interaction, or another source has been highly debated. Both an interaction with the ISM and a planet on an eccentric disk could cause the moth-like wings seen in the disk halo \citep{Debes09, Esposito16}. An ISM interaction could also cause the disk color asymmetry, with the East side being more blue than the West side. A recent collision between two large objects may also cause the observed brightness asymmetry and tentative disk color asymmetry, however, ALMA observations do not show any significant clumps, and no gas disk is detected \citep{Olofsson16,MacGregor18}. In this case, planet-disk interactions or an ISM interaction are more likely scenarios, as a combination of the two could cause the majority of asymmetries seen, such as the moth-like halo, brightness asymmetry, eccentricity, and possibly a disk color asymmetry. While, the mm-observations appear to be axisymmetric, residuals in the best fitting models employed by \citet{MacGregor18} suggest that the millimeter sized grains may indeed have some eccentricity, although this would require detection of the star to confirm. 

\subsubsection*{HD 106906}
The HD 106906 debris disk system is the only debris disk in our sample with a massive, directly imaged planet orbiting \textit{outside} of the disk (11 $M_{Jup}$, 735$\pm$5 AU; \citealt{Bailey14, Daemgen17}). The disk itself appears perturbed, with a moderate brightness asymmetry seen in scattered light with GPI, SPHERE, and HST \citep{Kalas15, Lagrange16}, and most recently has been found to have a significant eccentricity \citep{Crotts21}. Additionally, HST observations show that the outer disk halo is radially asymmetric, where the NW extension extends significantly farther than the SE extension in a ``needle"-like fashion \citep{Kalas15}. Although the origin of the planet is still debated, what is clear is that the disk's asymmetries align with being perturbed by the outer planet on an eccentric orbit.

While our geometrical fitting agrees mostly with the analysis done in \citet{Crotts21}, we derive a slightly larger disk radius and $\delta_{x}$ of 107 au and 20 au compared to 104 au and 16 au, respectively. This may be due to the unique shape of HD 106906's disk spine as it has a distinct ``S" shape. This ``S" geometry can be caused by either an eccentric disk or other geometrical asymmetry such as a warp, however, given that there is no detected warp, this led to the conclusion that the disk is rather eccentric. In terms of our simple ring modelling, it is clear that the ``S" shape of the disk geometry can be fit well with multiple different models; however, either way a large offset along the major axis is always required. Keeping a lower limit on the eccentricity of 0.16 as set by \citet{Crotts21} still makes it one of the most eccentric disks in our sample. 

We also confirm the brightness asymmetry in all three bands. This brightness asymmetry is very modest given the large eccentricity; however, as \citet{Crotts21} shows, the scattering phase function (SPF) for an eccentric disk can offset the expected brightness asymmetry based solely on the radial separation of the disk from the star. In terms of the disk color we find the disk has a blue color that becomes increasingly grey at longer wavelengths, consistent with the findings in \citet{Crotts21}. Similarly, we do not find a significant color asymmetry. 

\subsubsection*{HD 110058}
The HD 110058 debris disk is one of the most asymmetric disks in our sample. Along with GPI, HD 110058 has been also imaged with HST, SPHERE, as well as with ALMA \citep{Ren23,Kasper15,stasevic23,Hales22}. In scattered light, a definite warp has been detected towards the outer edges of the disk \citep{Kasper15,stasevic23}, reminiscent of the warp detected in the outer regions of the HD 111520 debris disk \citep{Crotts22}, and also similar to the warp featured in Beta Pic \citep{Heap00}. While perturbation from a planet companion is a strong candidate for this warp, no planets yet have been detected.   

\begin{figure}
\centering
	\caption{\label{Fig:warps} \textbf{Top:} Vertical offset profiles of HD 110058 and HD 115600, which show tentative warps in their vertical offset profiles. \textbf{Bottom:} HD 115600 overlaid with surface brightness contours to help highlight the warp detected in the vertical offset profile beyond 0.4$''$. The red solid line represents the vertical offset profile derived in Section \ref{sec:geom} and plotted above. The red dashed line represents an extension of the warp to show the angle of the warp on both sides of the disk. The white circle represents the size of the FPM, while the grey cross represents the location of the star.}
	\includegraphics[width=.47\textwidth]{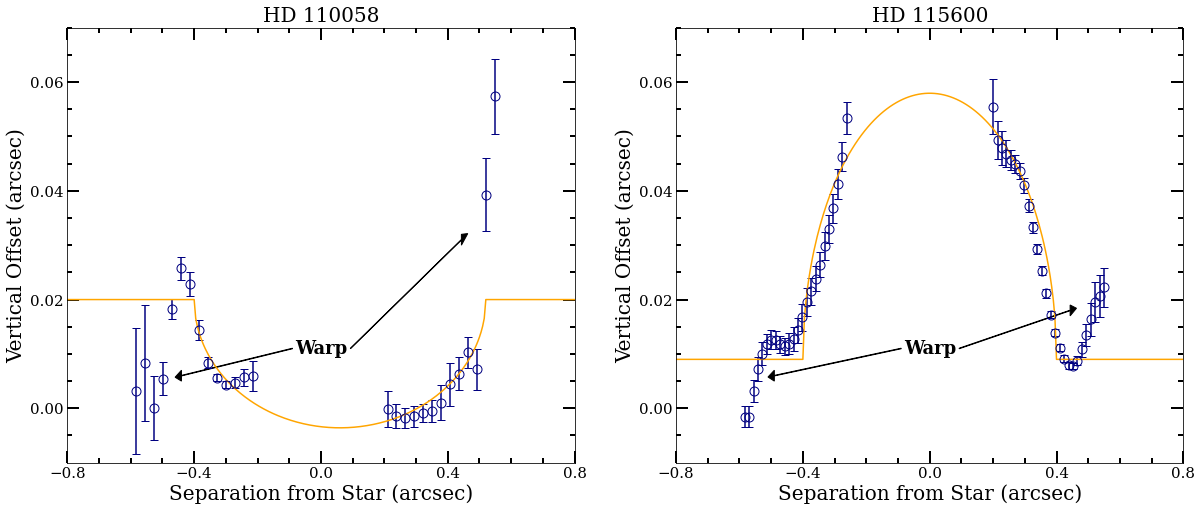}
        \includegraphics[width=.47\textwidth]{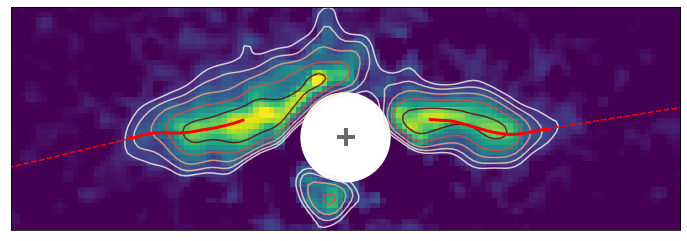}
\end{figure}

While the disk's warp has only been seen so far in total intensity observations, we are also able to detect it in polarized intensity, which we highlight in Figure \ref{Fig:warps}. We find the warp to occur beyond 0.35$''$ or 40 AU, as well as find the South-East warp to have an angle of $\sim$15$^{\circ}$, similar to what has been found in previous studies \citep{Kasper15}. From our geometrical fitting, we find that the MCMC favors two slightly different models: a disk with a 5 au offset towards the East extension and a disk with an 8 au offset towards the West extension. While these are contrasting models, the model with the 8 au offset has a much higher log likelihood and therefore we use this offset to estimate the eccentricity of $\gtrsim$0.13. This is a significantly high eccentricity, and is in contradiction to the low eccentricity ($e < 0.035$) estimated in \citet{Kasper15} based on SPHERE data. It is possible that this offset along the major-axis is a result of the asymmetric geometry due to the warp rather than the disk being eccentric. 

When looking at the brightness asymmetry in all three bands, we find an interesting trend. While no significant brightness asymmetry is seen in the $K1$ band, there is a significant brightness asymmetry in the $J$ and $H$ bands. In the $J$ band, the East extension is 1.7 times brighter than the West extension, however, in the $H$ band, the East extension is only 1.2 times brighter than the West extension, meaning that the brightness asymmetry is most significant at shorter wavelengths. This may be a result of dust grain properties, as we also find a significant disk color asymmetry between the $J$ and $H$ bands, where the East extension is relatively more blue than the West extension. This may suggest that the dust grain properties (such as minimum size, composition and/or porosity) or the distribution of dust grains are in some way being altered. In general, the disk exhibits a strong red color between all three bands, suggesting a larger minimum dust grain size on the order of one to several microns, assuming a porosity of zero \citep{Boccaletti03}. This is consistent with the 2 $\mu$m blowout size for the system.

In summary, the HD 110058 debris disk serves as a very interesting candidate for further investigation. The disk is clearly being perturbed by some mechanism. While a planet is a likely candidate for the observed warp and possible eccentricity, further work is required to understand if perturbation from a planet is enough to create a disk color asymmetry, or if another mechanism is needed. 

\subsubsection*{HD 111161}
The HD 111161 debris disk is one that has not yet been studied in great detail. From visual inspection, the disk appears to be a lower inclined ring that is highly forward scattering, as only the front side of the disk is visible. There is also a cleared gap within the disk's inner radius.

Comparing our disk geometry results to previous measurements, we find a disk radius of $\sim$72.5 au, which is consistent with the estimated inner radius of 71.4$^{+0.5}_{-1.05}$ au \citep{Esposito20}. We find a $PA$ of $\sim$83.3$^{\circ}$, which is also similar with previous measurements done in \citet{Esposito20} using radiative transfer modelling (83.2$^{\circ}$$^{+0.5}_{-0.6}$), while our derived inclination is slightly lower ($\sim$59.8$^{\circ}$ compared to 62.1$^{\circ}$$^{+0.3}_{-0.2}$). Our geometrical fitting does favor a slight disk offset along both the major and minor axes of 1.4 au and 0.66 au, however, these observations are relatively low S/N. 

Estimating the brightness between the East and West extensions, we find no evidence of a significant brightness asymmetry within 3$\sigma$, which is consistent with small to no disk offset. Unfortunately, this disk only has $H$ band data, meaning that we were unable to perform disk color measurements.  

\subsubsection*{HD 111520}
The HD 111520 debris disk is one that presents multiple different asymmetries. Previous studies have shown the disk to have a large brightness asymmetry, radial asymmetry, disk color asymmetry, as well as a warp at 1.7$''$ from the star and a bifurcation feature on the West side of the disk \citep{padgett_stapelfeldt_2015,Draper16,Crotts22}. While we are performing a similar analysis on the GPI $J$-, $H$- and $K1$-band polarimetric observations as \citet{Crotts22}, the analysis presented here allows us to compare the HD 111520 disk to the rest of the disks in our sample.

Comparing our geometrical fitting with the same fitting done in \citet{Crotts22}, we come to similar conclusions. While the disk radius is still difficult to constrain given the high inclination of the disk, we get a consistent result with a disk radius of 91.4 au or $\sim$0.84$''$. Again, similar to \citet{Crotts22}, we find that an offset along the major-axis is also difficult to constrain and is consistent with zero. Our derived inclination of 89.5$^{\circ}$ is slightly higher than what was measured previously for the $H$ band, however, it still is consistent with the disk being less than 2 degrees from edge on.

Within our sample, the disk has one of the highest brightness asymmetries in all three bands, ranging from a right/East brightness ratio of $\sim$1.4:1 to 1.8:1. Similar to \citet{Crotts22}, we find the disk to present a strong blue color between all three bands. While we do measure a disk color asymmetry, this asymmetry is only significant by 2$\sigma$.

\subsubsection*{HD 114082}
HD 114082 is the most recent system to have a resolved debris disk and a known planet \citep{Engler22,Zakhozhay22}. Similar to AU Mic, the planet has been observed via the transit and radial velocity method, where the planet found has a mass of 8 M$_{Jup}$, orbits at a distance of 0.51 au, and has a possible large eccentricity of 0.4 \citep{Zakhozhay22}. The disk lies much farther out from the star compared to the planet, and is fairly compact, similar to the HD 110058 disk.

With the higher S/N $K1$-band data, we find the disk to have a radius of $\sim$28.5 au, which is consistent with the inner radius estimated of 28.7$^{+2.9}_{-3.7}$ \citep{Wahhaj16}. \citealt{Engler22}). We also derive a small offset of 3 au (0.03$''$) along the major axis, bringing the West side of the disk closer to the star, however this is roughly twice as large as the 2$\sigma$ offset placed by \citet{Wahhaj16}. Additionally, no significant offsets are found using SPHERE observations \citep{Engler22}. 

For the surface brightness, we find no significant brightness asymmetry in the $H$ band. However, we do find a small but significant brightness asymmetry in the $K$ band, with the East side being 1.13 times brighter than the West side, in contrast to the derived offset from the geometrical fitting, suggesting that the measured offset may not be due to an eccentric disk. A similar finding was observed in the SPHERE data, where \citet{Engler22} reports a brightness asymmetry in the $K$-band IRDIS observations, but not in the $H$-band IRDIS observations. While this brightness asymmetry is thought to be a result of instrumental noise, the fact that it is also observed with GPI suggests that this feature may be real. Along with the brightness asymmetry, a small color asymmetry is also observed in $H-K1$ where the East side is relatively more red than the West side, however, this asymmetry is only significant within 2$\sigma$.

While this system has a known planet, the planet is too close to the star to be dynamically coupled with the disk (0.5 au compared to 25 au). On the other hand, the disk has one of the highest vertical aspect ratios in our sample, similar to the HD 110058 debris disk which may indicate stirring from another companion closer to the disk.

\subsubsection*{HD 115600}
Previous studies of the HD 115600 disk with GPI and SPHERE have shown the disk to be asymmetric with a moderate to high eccentricity, although this is mainly based on total intensity observations \citep{Currie15, Gibbs19}. 

In polarized intensity we find no disk offset along the major-axis, suggesting that the disk is not eccentric. We do, however, detect a tentative warp in the disk geometry, where the the East extension bends downwards beyond 0.4$''$, while the West extension bends upwards beyond 0.4$''$ (see Figures \ref{Fig:warps}). This is very similar to the HD 110058 debris disk, which hosts a similar warp, while not being necessarily eccentric. This may explain why the disk was found to be highly eccentric in \citet{Currie15}, who performs a similar geometrical analysis, as an asymmetric geometry, such as a warp, can translate into a significant offset that can be interpreted as an eccentric disk. Further observations, such as with HST, can help confirm the existence of this warp. 

We find no surface brightness asymmetry between the East and West extensions in any of the three bands, supporting the findings of a non-eccentric disk. We also find no asymmetry in the disk color between the two sides of the disk. The overall disk color in $J$-$K1$ and $J$-$H$ are strongly blue, with values between -0.6 and -1, while in $H$-$K1$ the disk color jumps to red, somewhat similar to the HD 32297 disk. This large jump in disk color, from strongly blue to red, is discussed in Section \ref{sec:color_vs_star}.

\subsubsection*{HD 117214}
The HD 117214 debris disk has been described as axisymmetric, with no asymmetries currently reported in the literature. While the disk has not been found to be eccentric, we do find a very small offset along the major-axis of $\sim$0.19 au, but it is consistent with 0 au within 2$\sigma$. Overall, the disk geometry is in line with being axisymmetric, as has been observed in \citep{Engler20}. Despite the axisymmetric disk geometry, we do find a significant brightness asymmetry where the West side is $\sim$1.15 times brighter than the East extension. This brightness asymmetry is unlikely to be due to a pericenter glow as we find no significant disk offsets. Multiwavelength observations in the future will be useful to help confirm this brightness asymmetry and better understand what mechanisms are prevalent in the disk.

\subsubsection*{HD 129590}
The HD 129590 debris disk, is one of the few disks around a G-type star that has been found to harbor a detectable amount of gas \citep{Kral20}. Along with low resolution ALMA observations, the disk has also been observed in the $H$ and $YJ$ bands with SPHERE IRDIS and IFS \citep{Matthews17}. Here, we present the first $K1$-band observations, along side $H$-band polarimetric observations with GPI.

Analyzing the geometry, we find a disk radius of 45.5 au, which is smaller than the estimated R$_{0}$ of 66.9 au, and may be closer to the inner radius which is estimated to be $<$40 au \citep{Matthews17}. We also find the inclination is much higher than the estimated inclination of $\sim$75$^{\circ}$ based on total intensity SPHERE data in \citet{Matthews17}. However, modelling done in \citet{Olofsson22} find a more comparable inclination of 82$^{\circ}$. The disk spine fitting does support a small offset along the major axis of $\sim$1.9 au, placing the star closer to the West extension, although such an offset has no precedent in the literature.

Comparing the surface brightness between the $H$ and $K1$ bands, we find the disk to be brighter in the $K1$ band and find the disk to have a red color. While no significant brightness asymmetry is found in the $H$ band between the East and West extensions, we do find that the East side of the disk is about 1.1 times brighter than the West side in the $K1$ band. This brightness asymmetry is contradictory to the offset measured in our geometrical fitting, and we additionally find no significant color asymmetry.

\subsubsection*{HD 131835}
The HD 131835 (HIP 73145) debris disk is another disk in our sample with strong CO detections. This gas disk is co-located with the dust disk, and is found to likely arise from secondary origins \citep{Kral19, Smirnov22}. The dust disk is moderately inclined and appears to have an inner gap within $\sim$75 au, with evidence for two inner/warmer rings \citep{Hung15a, Feldt17}. In this study, we reanalyze the GPI $H$-band observations first presented in \citet{Hung15b}. 

Through the disk geometry, an offset of 4.6 au is detected along the major-axis, bringing the star closer to the West extension and leading to a minimum eccentricity of 0.05. However, such an offset/eccentricity is not reported for other observations, and \citet{Hung15b} ruled out an eccentricity of $>$0.2 at 1$\sigma$. Therefore, if the disk is indeed eccentric, it is not likely to be significantly greater than 0.05. The HD 131835 disk is also reported to be radially broad \citep{Hung15b}, and has relatively low S/N in our GPI observations, meaning that our narrow ring model may not be the best method for deriving disk offsets. Additionally, the disk has been found to possibly consist of three concentric rings \citep{Feldt17}, further complicating the overall disk geometry. See Section \ref{sec:hd131835_rings} at the end of the Appendix for further analysis related to multiple rings in the system. 

In agreement with \citet{Hung15b}, we also find a brightness asymmetry with the East extension being brighter than the West, although we find this asymmetry to be larger at 1.7:1 compared to 1.3:1 when averaging the flux over our selected apertures. This brightness asymmetry appears only in the $GPI$ polarized intensity data, as SPHERE observations do not show a similar brightness asymmetry \citep{Feldt17}, however, this difference may be due to disk self-subtraction, introduced by the PSF subtraction process, as the SPHERE observations are in total intensity. Longer wavelength observations with ALMA also appear axisymmetric, although the disk is not well resolved \citep{Feldt17}. Additionally, a brighter East extension contradicts the measured disk offset which places the star closer to the West extension, assuming the offset is due to eccentricity. Future, higher resolution imaging will be useful to confirm the observed brightness asymmetry.

\subsubsection*{HD 145560}
The HD 145560 system harbors a lower inclined debris disk, which can be described as a narrow ring with an inner clearing within 68 au. As of now, the disk has only been imaged with GPI and with low-resolution ALMA observations, making it one of the less studied disks in our sample. We compare our results with another analysis done using the same GPI $H$-band data \citep{Esposito20, Hom20}.

While other studies used radiative transfer modelling to derive disk geometrical properties, we use our radial offset fitting. We derive a disk radius of 81.2 au, which is located near R$_{0}$ measured in \citet{Esposito20} of 85.3 au. We also derive an inclination of 41.9$^{\circ}$ and a PA of 39.5$^{\circ}$. Both these values are slightly smaller than the measurements derived from radiative transfer modelling \citep{Esposito20, Hom20} of 43.9$^{\circ}$ and 41.5$^{\circ}$, but are still consistent within 2$\sigma$ uncertainties. Our model prefers a small offset along the major-axis of $\sim$0.86 au, leading to a small eccentricity of $>$0.01. However, we do derive a larger offset along the minor-axis of 3.3 au, which brings the estimated eccentricity up to $\sim$0.04. We otherwise find the disk to be axisymmetric, with no brightness asymmetry measured in the $H$ band, which would be expected for the derived small offset along the major-axis.

\subsubsection*{HD 146897}
The HD 146897 system, also well-known as HIP 79977, harbours a highly inclined debris disk that has also been observed with SPHERE and SCExAO on the Subaru telescope \citep{Thalmann13, Engler17, Goebel18}.

In \citet{Engler17}, radiative-transfer modelling was used to determine properties of the HD 146897 disk, comparing two different models: One with a disk radius of 70 au, and one with a disk radius of 40 au. While the disk model with a radius of 70 au was found to be a better fit to the data, we derived a disk radius of $\sim$52 au which is more consistent with the measured R$_{0}$ of 53 au derived in \citet{Goebel18}. Moreover, we find a significant offset along the major-axis of 6.3 au, placing the star closer to the West extension. Considering a disk radius of 52 au, this offset leads to a disk eccentricity of at least 0.12, which is a significant eccentricity compared to the majority of our sample. While previous observations do not report any eccentricity, 0.12 is still consistent with the upper limit of the eccentricity as set by \citet{Thalmann13} of e$\le$0.16. 

Although \citet{Goebel18} found the East extension to be brighter than the West extension in total intensity, our polarized intensity shows the West side to be moderately brighter than the East in the $J$ and $H$ bands with a brightness asymmetry of 1.08 to 1. The reasoning for this difference could be an artifact from disk self-subtraction with total intensity observations. Taking into account the derived disk offset along the major-axis, an eccentric disk with the West side closer to the star is more consistent with the measured brightness asymmetry. While a 1.08:1 brightness asymmetry is small considering an eccentricity of 0.12, one explanation could be similar to HD 106906, where the SPF partially cancels out the brightness asymmetry caused by a 1/r$^{2}$ relationship. 

With our multiwavelength observations, we find that the disk changes color when going from short to longer wavelengths. While a red disk color is measured in $J$-$H$, a neutral color is measured in $J$-$K1$ and a blue color in $H$-$K1$. the HD 146897 disk is the only one in our sample to exhibit this behavior in disk color. When comparing the disk color between the East and West side of the disk, we do not measure a significant disk color asymmetry between any of the three bands.

This analysis reveals an interesting side of the HD 146897 debris disk. While previous studies depicts the disk as being fairly axisymmetric, our results suggest that the disk morphology may actually be more complicated. Fitting the vertical offset or disk spine suggests an eccentric disk, or at the very least, an asymmetrical disk geometry. Measuring the surface brightness also reveals conflicting information with previous observations, suggesting a brighter West side rather than a brighter East side, although this would be more consistent with our derived offset along the major-axis in the case of an eccentric disk. 

\subsubsection*{HD 156623}
HD 156623 is another debris disk system that is rich in gas; however, the high density of gas leads to the speculation that this disk may be a ``hybrid", where the gas may be partially of primordial origin, i.e., a remnant of the protoplanetary disk phase \citep{Kospal13}. In this study, we are analyzing the first scattered light observations of the disk taken in the $H$ band and first presented in \citet{Esposito20}. 

We compare our empirical results for the disk geometry to the results from \citet{Esposito20} who uses radiative-transfer modelling. We derive a disk radius of $\sim$52.6 au, which lies within the derived critical radius, r$_{c}$, of 64.4$\pm$1.8 au \citep{Esposito20}, where r$_{c}$ is the radius where the disk transitions from a dust density power law of $\alpha_{in}$ to $\alpha_{out}$. While our inclination is consistent with previous measurements ($\sim$34.7$^{\circ}$ compared to 34.9$^{\circ}$$^{+3.6}_{-9.5}$), our estimated $PA$ is slightly higher (102.9$^{\circ}$ compared to 100.9$^{\circ}$$^{+1.9}_{-2.2}$); however, these values are still consistent within 2$\sigma$ uncertainties. A small disk offset is measured along the major-axis of 2.1 au, leading to an eccentricity of $\gtrsim$0.04 and bringing the East side of the disk closer to the star. An additional offset is measured along the minor-axis of 1.68 au, which when taken into account, increases the eccentricity to $\sim$0.08. However, these offsets may be exaggerated given that the disk appears radially broad, with no gap observed outside of the FPM, we therefore place an eccentricity of 0.08 as an upper limit.

Measuring the surface brightness reveals a moderate brightness asymmetry, where the East side of the disk is 1.11 times brighter than the West side. This is consistent with the small offset measured, which places the star closer to the East extension, possibly causing a slight pericenter glow \citep{Wyatt99}. Further scattered light observations will be useful to help confirm these asymmetries.

\subsubsection*{HD 157587}

\begin{figure*}
\centering
	\caption{\label{Fig:hd157587} HD 157587 observations in all three bands, overlaid with surface brightness contours to highlight the difference in the vertical width between the East and West extensions at each wavelength. The white circles represent the size of the FPM, while the grey crosses represent the location of the star.}
	\includegraphics[width=\textwidth]{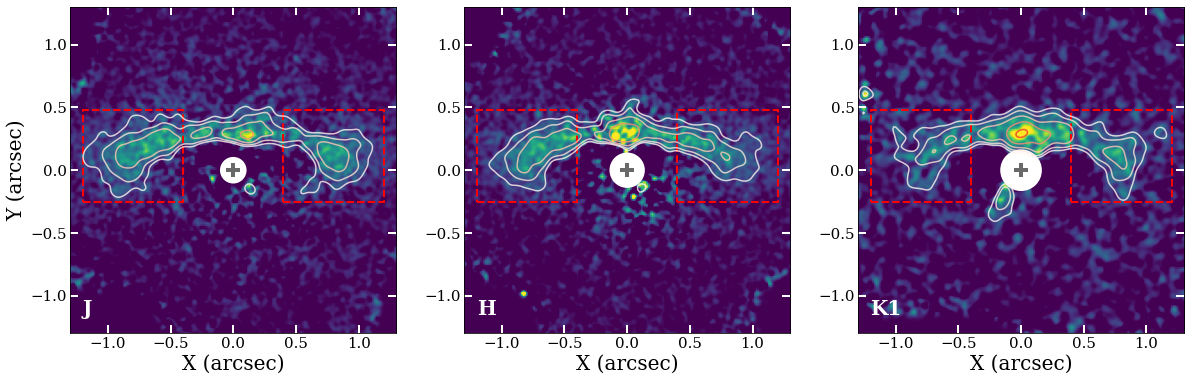}
\end{figure*}

The HD 157587 debris disk is the oldest system in our sample with an estimated age of 165-835 Myr. So far, the disk has only been observed with GPI and HST, where only the $H$-band observations have been fully analyzed \citep{MB16}. In this study we include the $J$- and $K1$-band observations, adding a multiwavelength and disk color analysis.

Through our geometrical fitting, we derive an inclination that is several degrees smaller than found in previous studies (64$^{\circ}$ compared to $\sim$68-72$^{\circ}$; \citet{MB16}). While \citet{MB16} found evidence for an offset along the major-axis placing the East side of the disk closer to the star by $\sim$1.6$\pm$0.6 au, our ring model fitting does not find strong evidence for such an offset (Our results suggest a 0.65 au offset in the opposite direction). The reason for this inconsistency may be an asymmetric disk morphology not related to eccentricity. In the case of HD 157587, we find that the East side of the disk is vertically broader than the West side of the disk, where the weighted average FWHM for the East side is roughly 0.04$''$ (4 au) greater than the weighted average FWHM for the West side in the $H$ band. This discrepancy may have led to an offset along the major-axis in the radial offset profile using our method. We also plot the image of HD 157587 in each band, overlaid with surface brightness contours, to visually show this difference in the vertical FWHM in Figure \ref{Fig:hd157587}.

Similar to \citet{MB16}, we also measure a brightness asymmetry in the disk, with the East side being moderately brighter than the West side. Our brightness asymmetry measurements in the $H$ band of 1.13$\pm$0.05 is consistent to previous measurements of 1.15$\pm$0.02 \citep{MB16}. Conducting the same measurements in the $J$ and $K1$ bands, we find the brightness asymmetry to be even stronger in the $J$ band of 1.22$\pm$0.03, whereas the $K1$ band does not show a significant brightness asymmetry within 2$\sigma$. This brightness asymmetry may partially be due to the difference in vertical width between the East and West extensions, as this feature is most prominent in the $J$ and $H$ bands, while less prominent in the $K1$ band (see Figure \ref{Fig:hd157587}). If the brightness asymmetry is indeed due to an eccentric disk, it is most likely that the offset along the major-axis is towards the opposite direction than what is measured in this study. 

While overall the disk presents a blue to neutral disk color, the East side of the disk is tentatively bluer in $H$-$K1$ and $J$-$H$, while being significantly bluer in $J$-$K1$. If there are asymmetries in the dust grain properties, this may provide an alternate explanation for the brightness asymmetry.  

\subsubsection*{HD 191089}
The HD 191089 debris disk consists of a dust ring from $\sim$26 to $\sim$78 au, and an extended halo out to 640$\pm$130 au as observed with HST \citep{Ren19}. The disk has been observed at multiple wavelengths, from the optical with HST, to the sub-mm with ALMA \citep{Soummer14, Ren19, Churcher11, Kral20}. Along with the already published $H$-band observations \citep{Ren19, Esposito20}, we also include $J$-band observations in our analysis.

We derive a disk radius of $\sim$47 au, which is close to the derived R$_{0}$ from radiative transfer modelling of the GPI $H$-band observations (43.9$\pm$0.3 au; \citealt{Ren19}), as well as the radius derived from mm observations (43.4 au; \citealt{Kral20}). Similarly, we do not detect a significant offset along the major or minor axis, in agreement with the results from \citet{Ren19}, however, a small offset of 1 au is measured along the major-axis. 

We find no significant brightness asymmetry present in either bands. Calculating the disk color shows that the disk presents a strong blue color in $J$-$H$, meaning that dust grains are more efficient at scattering light at shorter wavelengths. We find no disk color asymmetry between the two extensions, further supporting a fairly axisymmetric disk. 

\subsubsection*{HR 4796 A}
The HR 4796 A debris disk is one of the most well studied disks in our sample. The disk is a bright and a distinctly narrow ring, permitting the measurement of a complete SPF compared with other debris disks (e.g. \citealt{Milli17, Milli19}). Given that the disk is already well characterized, we use our polarized $H$-band GPI observations simply as a confirmation of the disk geometry and surface brightness. 

We find a disk radius of $\sim$77.7 au, which is consistent with previous measurements using a similar geometrical fitting (e.g., \citealt{Chen20}). While the disk is known to be eccentric, ranging from 0.01 to $\sim$0.08 depending on the observation and reduction method \citep{Perrin15, Milli17, Milli19, Olofsson19, Olofsson20, Chen20}, with our GPI polarized intensity observations, the derived offset along the major-axis and resulting eccentricity are on the smaller end with an offset of 0.58 au and eccentricity of $\gtrsim$0.01. Including the 1.56 au along the minor-axis leads to an estimated eccentricity of $\sim$0.02st{, which is still on the low end of measured eccentricities for the HR 4796 A disk.}

Measuring the surface brightness of the disk as a function of stellar separation, the surface brightness peaks close to the star, followed by a second peak at the disk ansae before decreasing towards the back side of the disk. Placing several square apertures along the East and West extensions, we confirm a modest brightness asymmetry, where the East extension is $\sim$1.17 times brighter than the West extension, most of which comes from near the East disk ansae.

\subsubsection*{HR 7012}
The HR 7012 (also known as HD 172555) debris disk is one of the warmest and most radially compact disks in our sample, extending only $\sim$0.1$''$ past the FPM. The disk appears to be in a state of heavy bombardment, with strong traces of both SiO and CO \citep{Lisse08, Schneiderman21}, along with indirect and direct detections of exocomet transits \citep{Kiefer14, Kiefer23}. Here, we compare our analysis of the disk morphology using GPI $H$-band observations to previous analysis using SPHERE/ZIMPOL observations \citep{Engler18}.

We derive a disk radius of $8.8$ au, consistent with SPHERE/ZIMPOL measurements of R$_{0}$ within 1$\sigma$ derived from a grid model (10.3$\pm$1.7 au) and within 2$\sigma$ derived from a radiative-transfer model (11.3$\pm$1.7 au; \citealt{Engler18}). This measurement is also consistent with the measured inner radius of 8$\pm2$ au \citep{Engler18}. While \citet{Engler18} find the disk to be axisymmetric, the GPI observations appear to tell a different story. Fitting the vertical offset profile shows a relatively large offset along the major-axis of 2.76 au, which would mean the disk is highly eccentric with e$\gtrsim$0.31. Given that the disk sits very close to the FPM, this asymmetric geometry may simply be due to residual noise close to the star.

This is supported by the surface brightness profile and brightness asymmetry, where the surface brightness profile decreases symmetrically from the star out to $\sim$0.4$''$ within 1$\sigma$ uncertainties. Additionally, averaging the flux over rectangular apertures placed on the highest S/N regions of the disk yields no significant brightness asymmetry within 2$\sigma$. While the disk may not be as asymmetric as it would appear from the polarimetric GPI observations at first glance, the disk does have the 3rd highest vertical aspect ratio in our sample due to the disk being so compact. This may be the result of the stellar companion, CD-64 1208, located $>$2000 au from HR 7012 \citep{Torres06}, which could cause the disk to become truncated depending on its orbit. However, given the large separation of the stellar companion, it would be difficult to confirm if this is the case.

\subsection{HD 131835: Multiple Rings?} \label{sec:hd131835_rings}

\begin{figure*}[t!]
\centering
	\caption{\label{Fig:hd131835_rings} \textbf{Top:} The FWHM profile as a function of stellar separation for the HD 131835 disk. \textbf{Bottom:} Vertical Offset profile for the HD 131835 disk, also shown in Figure \ref{Fig:vert_prof}. The inner two orange shaded regions show the locations of the disk gaps found in \citet{Feldt17}, with the addition of a possible additional gap outside the two already known gaps found in this work.}
	\includegraphics[width=\textwidth]{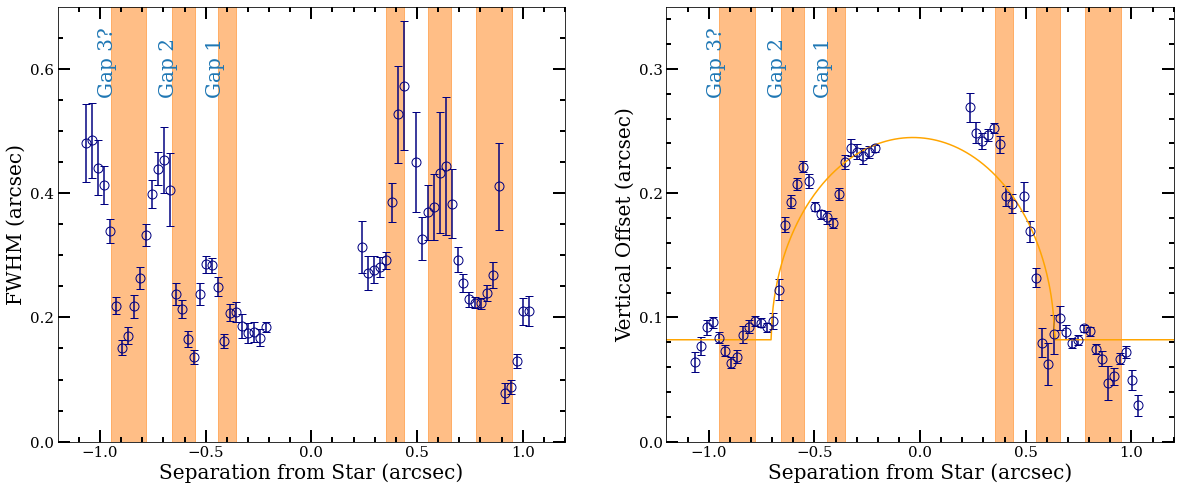}
\end{figure*}

Using total-intensity SPHERE/IRDIS observations in the $H$ band, \citet{Feldt17} discovered that the HD 131835 disk consisted of several concentric rings, features which are often very difficult to detect in higher inclined disks ($i=75-76^{\circ}$). Using the $H$-band polarized intensity GPI observations, we look at the vertical structure to see whether or not these rings are still present in our data.

In Figure \ref{Fig:hd131835_rings}, we re-plot the vertical offset profile, along side the vertical FWHM as a function of stellar separation. We then plotted orange bars to represent the locations of the gaps found in \citet{Feldt17}, which were found at 46-57 au ($\sim$0.36$''$-0.44$''$) and 71-85 au ($\sim$0.55$''$-0.66$''$). Doing so we find that the locations of these gaps strongly co-align with dips in the vertical FWHM, as well as the vertical offset. In addition to the two inner gaps discovered in \citet{Feldt17}, we find a possible third outer gap located between $\sim$101 au and 123 au (0.78$''$-0.95$''$), where another dip in the vertical FWHM is observed. This dip in the vertical FWHM also coincides with a dip in the vertical offset at the same location on either side of the disk. This gap is outside the outer ring observed with SPHERE, although, as these are total intensity observations, it is possible that additional structure outside the outer ring was subtracted during the PSF subtraction process. While we cannot definitively say whether or not this is a physical gap, the fact that the location of the two inner gaps found in \citet{Feldt17} align with dips in both the vertical FWHM and vertical offset, help to confirm that these structures are real.

In terms of other high inclined disks, whether or not the vertical FWHM and vertical offset profiles can be used as probes for multiple rings/gaps is unclear without further evidence. Wavy patterns in either profile could arise from other factors such as low S/N, and therefore may not be indicative of more complex structure. Further analysis is required to explore the connection between the vertical structure and evidence of rings/gaps, although this is outside the scope of our study.

\end{document}